\documentclass[twocolumn]{aastex63}

\usepackage[utf8]{inputenc}

\usepackage{graphicx} 
\usepackage{booktabs} 
\usepackage{physics}
\usepackage{natbib}

\usepackage{longtable}
\usepackage{amssymb}
\usepackage{placeins}

\begin{document} 

\title{A Galactic-scale extreme-precision radial velocity survey}

\author[0000-0002-0786-7307]{Sukanya Chakrabarti}
\affiliation{Department of Physics and Astronomy, University of Alabama \\
Huntsville, Huntsville, Alabama, 35899}

\author[0009-0000-7423-852X]{Jack Wagner}
\affiliation{Department of Physics and Astronomy, University of Alabama \\
Huntsville, Huntsville, Alabama, 35899}

\author[0000-0002-4918-0247]{Robert J. De Rosa}
\affiliation{European Southern Observatory, Alonso de C\'ordova 3107, Vitacura, Casilla 19001, Santiago, Chile}

\author[0009-0000-7423-852X]{Henrique Reggiani}
\affiliation{Gemini South, Gemini Observatory, NSF's NOIRLab, Casilla 603, La Serena, Chile}

\author[0000-0001-6160-5888]{Jason T. Wright}
\affiliation{Department of Astronomy \& Astrophysics,
The Pennsylvania State University,
525 Davey Laboratory, University Park, PA 16802, USA}
\affiliation{Center for Exoplanets and Habitable Worlds,
The Pennsylvania State University,
University Park, PA 16802, USA}

\author[0000-0002-3673-0668]{Peter A. Craig}
\affiliation{Center for Data Intensive and Time Domain Astronomy,
Department of Physics and Astronomy,
Michigan State University,
East Lansing, MI 48824, USA}

\author[0000-0002-4927-9925]{Jacob K. Luhn}
\affiliation{Jet Propulsion Laboratory,
California Institute of Technology,
4800 Oak Grove Drive, Pasadena, CA 91109, USA}

\author[0000-0002-9815-773X]{Francesco Pepe}
\affiliation{Department of Astronomy,
University of Geneva,
Chemin Pegasi 51, 1290 Versoix, Switzerland}

\author[0000-0003-0149-9678]{Paul Robertson}
\affiliation{Department of Physics \& Astronomy, The University of California, Irvine, Irvine, CA 92697, USA}

\author[0000-0002-9332-2011]{Xavier Dumusque}
\affiliation{Department of Astronomy,
University of Geneva,
Chemin Pegasi 51, 1290 Versoix, Switzerland}

\author[0000-0002-8597-9742]{Massimiliano Bonamente}
\affiliation{Department of Physics and Astronomy,
The University of Alabama in Huntsville,
Huntsville, AL 35899, USA}

\begin{abstract}

We describe the first Galactic-scale extreme-precision radial velocity (EPRV) survey and outline our goals of measuring Galactic accelerations and deriving constraints on the dark matter distribution in our Galaxy.  We are targeting quiet subgiant stars within a kiloparsec in radius of the Sun and about 2 kiloparsec in vertical height relative to the Galactic mid-plane. The observations described here were carried out using the Echelle SPectrograph for Rocky Exoplanets and Stable Spectroscopic Observations (ESPRESSO) at the Very Large Telescope (VLT), between 2021-2025.  
We began by operating at a typical RV precision of $\sim$ 70 cm/s to select quiet stars from stellar activity indicators, and are now operating at $\sim$ 50 cm/s.  We  selected stars with low stellar jitter (determined from their RV scatter as a function of epoch) with the goal that they yield a measurement of the Galactic acceleration on approximately a decade timescale.  We refer to this as our ``gold" sample.  We also identify stars with possible companions (either brown dwarfs or planetary companions) or hitherto unknown stellar binaries that we have culled from our survey.  For a few cases, we present initial constraints on the companion mass and orbital period.  We also present the photospheric and fundamental stellar parameters for our ``gold" sample of sub-giant stars.  
This first Galactic-scale EPRV survey provides a pathfinder for next-generation surveys with extremely large telescopes, opening the possibility of directly mapping the gravitational acceleration field across the Milky Way and potentially beyond.
\\\vspace{0.5cm}
\end{abstract}

\section{Introduction}

The last few decades have seen significant advances in precision measurements across astronomy, including increasingly stringent tests of general relativity enabled by pulsar timing \citep{Weisberg2016,Zhuetal2019,Agazie2023}, and in the hunt for and characterization of exoplanets using both transit timing \citep{Batalha2017,Petigura2013,Lissauer2024} and precise spectrographs \citep{Mayor2003,Pepe2010,Fischer2016,Butler2017}.  The effective precision now enabled by pulsar timing observations \citep{Chakrabartietal2021,Donlonetal2024,Donlonetal2025} and by extreme-precision radial velocity (EPRV) observations \citep{Silverwood2019,Chakrabarti2020} as well as by precise long-term timing of eclipses \citep{Chakrabarti2022} is such that we can now directly measure the very small accelerations of stars ($\sim$ 10 cm/s/decade) that live within the gravitational potential of the Milky Way.  Accelerations from pulsar timing have already provided measurements of the Galactic mid-plane density \citep{Chakrabartietal2021}, the slope of the rotation curve and Oort constants, along with evidence of disequilibrium \citep{Donlonetal2024}, and the first measurement of the local dark matter density (at nearly 4-$\sigma$) \citep{Donlonetal2025}.  Very recently, pulsar timing accelerations also give tentative evidence for a dark matter sub-halo near the Sun \citep{Chakrabartietal2026}.  Subsequent analysis of \textit{Fermi} data indicates an excess in that region \citep{Zhuetal2025, SalcesPerez2026}. 

This framework of precise time-series measurements to measure accelerations directly is distinct from the traditional characterization of the Galaxy using phase-space measurements \citep{BlandHawthorn2016} that relies on modeling static snapshots of the positions and velocities of stars to estimate accelerations (which typically assume equilibrium and symmetry).  A variety of observations now show that our Galaxy has had a dynamic history, leading to large perturbations in the stars \citep{Quillenetal2009,Purcell2011,ChangChakrabarti2011,Helmi2018,Craigetal2021,CraigChakrabartietal2021} and in the gas \citep{LevineBlitzHeiles2006,ChakrabartiBlitz2009,ChakrabartiBlitz2011,Chakrabartietal2019,Craigetal2021,CraigChakrabartietal2021,Craigetal2025}.  Such a time-dependent potential may lead to an inaccurate characterization of Galactic parameters using kinematic analysis \citep{Haines2019}.  Recent work \citep{Donlon2026Nacc} also demonstrates the efficacy of direct acceleration measurements - they show that a single acceleration measurement typically has as much constraining power for the local dark matter density, $\rho_{\rm DM}$, as $\sim$ 100,000 phase-space measurements.

Motivated by the goal of precisely characterizing the gravitational potential of the Galaxy using acceleration measurements, we have been obtaining spectra with the Échelle SPectrograph for Rocky Exoplanets and Stable Spectroscopic Observations (ESPRESSO) \citep{2021A&A...645A..96P} at the Very Large Telescope (VLT) beginning in 2021, to carry out the first Galactic scale EPRV survey.  We began by obtaining reconnaissance spectra of sub-giant stars we selected from \textit{Gaia} DR-2 that were not known to be binaries, by cross-matching to APOGEE \citep{Apogee2020} and ensuring that these stars did not have large RUWE values \citep{CastroGinardetal2024,Stassun_Torres_2021}.  Recent work \citep{Luhnetal2020a,Luhnetal2020b} provided a means of identifying quiet subgiant stars, i.e., with relatively low stellar jitter.  In our formulation here, we have chosen subgiant stars as a compromise between the dwarf stars that live at the jitter minimum but are not intrinsically bright, and the brighter, giant stars that can trace the potential at larger distances but have much larger stellar jitter.  

The paper is organized as follows.  In \S \ref{sec:sample}, we describe the design and motivations of our initial sample. In \S \ref{sec:observations}, we give an overview of our observing framework, in \S \ref{sec:data_reduction}, we describe our data reduction process, and outline how we select our quiet  stars in \S \ref{sec:quietstarsample}.  \S \ref{sec:stellarparameters} describes the derivation of photospheric and fundamental stellar parameters for our sample.   \S \ref{sec:stellarcomps} describes our initial constraints on stellar companions using RV fits to our data, and reasons for removal of some sources from our survey.  We discuss future work (in particular potential improvements to our sample and achievable RV precision) and conclude in \S \ref{sec:discussion}.  The Appendix gives our observing log, discusses the differences in our initial observing set-up and current observing framework, and presents the full RV time-series for the quiet stars in our sample (that we refer to as our "gold" sample).



\section{Initial sample}
\label{sec:sample}

\begin{figure}[h]
    \centering
    {\includegraphics[width=0.5\textwidth]{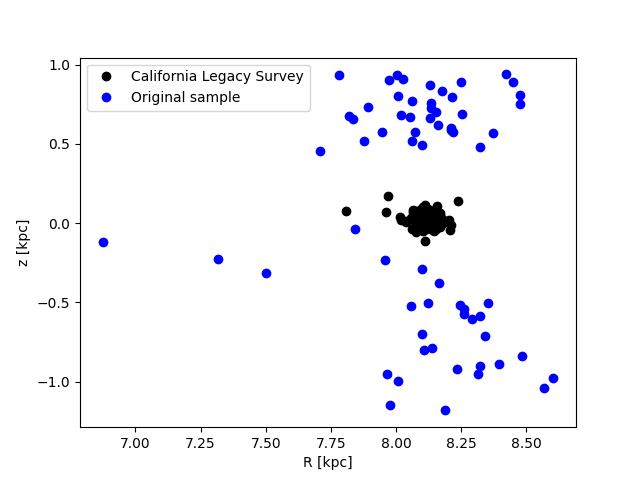}}%
    \caption{Galactocentric R and z coordinates of our original sample compared to the California Legacy Survey.}
\label{fig:CLScomp}    
\end{figure}

There are two important considerations for our survey here. Not only do we operate at high RV precision, as we describe in \S \ref{sec:observations}, but our sources are chosen to sample the Galactic potential, in a complementary manner to current direct acceleration measurements (that primarily come from pulsar timing accelerations thus far).  As shown in Figure \ref{fig:CLScomp}, which compares the Galactocentric R and z coordinates of our original sample of 66 stars to the California Legacy Survey (henceforth CLS), our original sample spans somewhat more than 2 kpc in Galactic height and more than 1 kpc in radial distance.  This sample was selected from \textit{Gaia} DR-2, and we used stellar parameters (effective temperature, luminosity, stellar radius) when available from \textit{Gaia} DR-2, selecting sources with comparable stellar parameters to sub-giant stars samples \citep{Bergeretal2020} (that have average values of effective temperature and luminosity of $\sim$ 5500 K and $\sim$ 3 $L_{\odot}$ respectively).  We cross-matched the sample with APOGEE and removed known binaries \citep{PriceWhelan2020,Mazzolaetal2020}.  We then began our ESPRESSO observations by obtaining reconnaissance spectra of this initial sample.   Having obtained reconnaissance spectra of all the stars in our original sample, we proceeded to cull the sample to find the quietest stars  as follows.  We first calculated the stellar activity indicators \citep{Wright2005,Luhn2020} and visually examined the spectra and cross-correlation functions.  At this stage, we removed stars from the sample if they either had asymmetric cross-correlation functions or multiple peaks in the cross-correlation function that indicated they were binaries, or if their stellar activity indicators from the Calcium H \& K lines suggested the star would not be quiet enough for us to measure the Galactic acceleration.  For stars that survived this initial vetting, we continued RV monitoring, and finally formulated what we refer to as our ``gold sample" on the basis of the scatter in the RV observations as a function of epoch.  We discuss this selection more quantitatively in \S \ref{sec:quietstarsample}.

\begin{figure}[h]
    \centering
    {\includegraphics[width=0.45\textwidth]{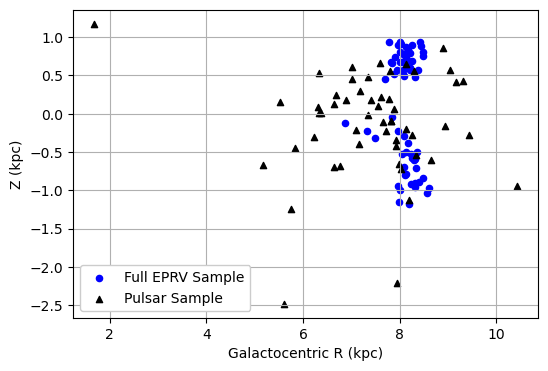}}%
    \caption{Galactocentric R and z coordinates of our original sample compared to the concurrent pulsar sample from Donlon et al (2025).}
\label{fig:Pulsarcomp}    
\end{figure}

In the reconnaissance phase of our survey, we set exposure times such that the instrumental and photon noise contribution to the RV error budget was $\sigma_{\rm inst}$ = 70 cm/s; we later proceeded to increase our RV precision to $\sim$ 50 cm/s once we began the monitoring phase of our survey (discussed in \S \ref{sec:observations}).  The CLS survey includes observations from a variety of telescopes and observed a much larger number of stars (about 719), and achieved typical RV precision, including stellar jitter, of several m/s. Given the advent of extreme-precision spectrographs and our goal to measure the Galactic acceleration, we sample a larger Galactic volume, and operate at an ambitious $\sigma_{\rm inst}$ level far below the jitter levels of our targets, in anticipation of jitter-mitigation methods \citep{Fordetal2024} that will allow us to realize this precision in our final RV time series.  That is, we do not claim that we can at the moment measure these stars' RVs to the precision of $\sigma_{\rm inst}$; we make sure to expose long enough that these data will, in the decades hence, have sufficient Doppler information content that such precision is not precluded.  Figure \ref{fig:Pulsarcomp} compares our original sample to the largest currently available pulsar timing acceleration sample \citep{Donlonetal2025}, which shows that our EPRV sample is designed to more densely sample the vertical region of the Galaxy near the Sun, in the hopes of searching for dark matter sub-structure (for which there is some tentative evidence from pulsar timing - \cite{Chakrabartietal2026}), or possibly a dark matter disk that pulsar timing accelerations cannot yet characterize \citep{Donlonetalmeanmassdensity}. 

\begin{figure}[h]
    \centering
    {\includegraphics[width=0.45\textwidth]{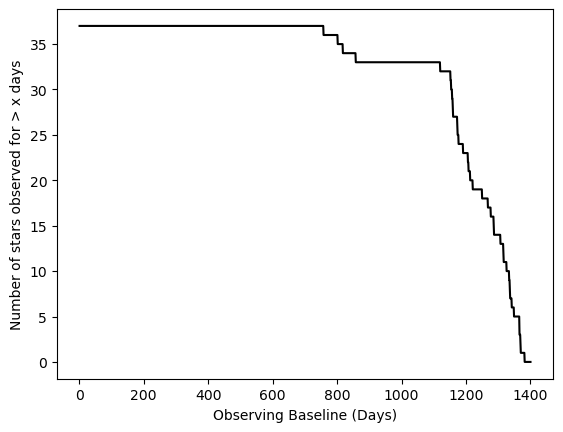}}%
    \caption{Cumulative distribution of observing time baselines.}
    \label{fig:CDF}
\end{figure}

Figure \ref{fig:CDF} shows a cumulative distribution of the observation time baselines for all the stars in our sample.  About 80 \% of our sample has been observed for at least 1200 days, and 50 \% for at least 1300 days.   The longest time baselines for the stars in our sample are 1400 days (3.83 years).

\section{Observations}
\label{sec:observations}


Observations of the sample were obtained with the \'{E}chelle SPectrograph for Rocky Exoplanets and Stable Spectroscopic Observations (ESPRESSO; \citealp{2021A&A...645A..96P}) at the Very Large Telescope (VLT), between 2021 October 04 and 2025 September 01 (ESO observing periods 108 to 115). The program consisted of two phases; a reconnaissance phase to identify the stars with the lowest stellar activity indicators, followed by a monitoring phase targeting those stars. During the first phase, the detector integration times (DITs) were set to achieve an $S/N\sim 50$ per pixel, sufficient for measuring activity indicators and identifying high-amplitude binaries. Longer DITs were used in the second phase to achieve an $S/N\sim 100$ in an effort to maximize the precision of the radial velocity measurements for the monitoring observations. This $S/N$ requirement set the limit of $V\lesssim 13$ mentioned in the previous section to avoid overly long DITs.  We carried out these observations with a typical spectral resolution of $\sim$ 140,000.

The instrument configuration was set by the requirement for extreme radial velocity precision and the faintness of the sample. Throughout the program, we used ESPRESSO in the single-UT mode with the high-resolution ($R\sim$140,000) fiber and the \texttt{2x1} detector binning mode. At the start of the program we had the Fabry-P\'{e}rot (FP) calibration source illuminate the second fiber such that an instantaneous measurement of the radial velocity drift relative to the daytime calibrations could be made. However, the bright lines from the FP calibration source caused an over-subtraction of the detector background during the data reduction process. This significantly biased measurements of the emission (or lack thereof) from the cores of the calcium lines in the bluest part of the spectrum where the overall throughput is lowest (see Appendix \ref{sec:FPbias}). As such, we modified the observing strategy to have the sky illuminate the second fiber during the science observation, followed immediately by an attached calibration with the FP calibration source illuminating both fibers without moving the telescope. A summary of all the observations including instrument configuration and observing conditions is given in the Appendix Table \ref{tab:observing_log}. 


\section{Data Reduction}
\label{sec:data_reduction}

The raw data and associated calibrations were collected from the ESO Science Archive Facility\footnote{https://archive.eso.org/} which automatically associates certified calibrations to each requested dataset. The data were organized and reduced using the ESPRESSO pipeline (v3.1.0) and ESPRESSO Data Analysis Software (DAS; v1.3.8) within the \texttt{EsoReflex} environment \citep{2013A&A...559A..96F}.

The data reduction followed the standard ESPRESSO workflow, including measurement and/or correction of: detector bias level and dark current, detector flat field, order definition, fiber flat field, inter-fiber contamination, fiber relative efficiency, wavelength calibration and spectral extraction for both fibers, and flux calibration.


The radial velocities (RVs) and chromospheric activity metric log $R'_{\rm HK}$ were extracted utilizing the preconstructed ESPRESSO-DRS and ESPRESSO-DAS pipelines available via ESO, executed with the EsoReflex GUI \cite{2013A&A...559A..96F} utilizing the G2 stellar spectral template (closest to the spectral type of our stars, that is publicly available for use in EsoRex).  Using other templates (such as F9) leads to minimal difference for the relative RVs and stellar activity indicators (the differences are at the level of 0.1 \% or lower).  This is because we use a binary mask for the CCF, so the primary dependence is on the line centers, rather than on the line shape (the latter may depend on the stellar template but the former does not).

\section{Sample of quiet stars}
\label{sec:quietstarsample}

The advent of extreme-precision spectrographs along with improvements in our understanding of stellar activity cycles indicated that the time was right to measure the Galactic acceleration via time-series EPRV observations \citep{Chakrabarti2020}.  The acceleration signal that we aim to measure ultimately with our EPRV survey is very small ($\sim$ 10 cm/s/decade) and therefore it is important that we carefully consider sources of contamination.  There are two primary contaminants here - external contaminants, that include binaries and planets, and internal contaminants, i.e., ``stellar jitter" that arises from processes internal to the star, including magnetic fields, oscillations, and granulation, leading to RV scatter that can confound our measurement \citep{Wright2005}.

\subsection{The ``gold sample"}

\begin{figure}[h]
    \centering
    {\includegraphics[width=0.5\textwidth]{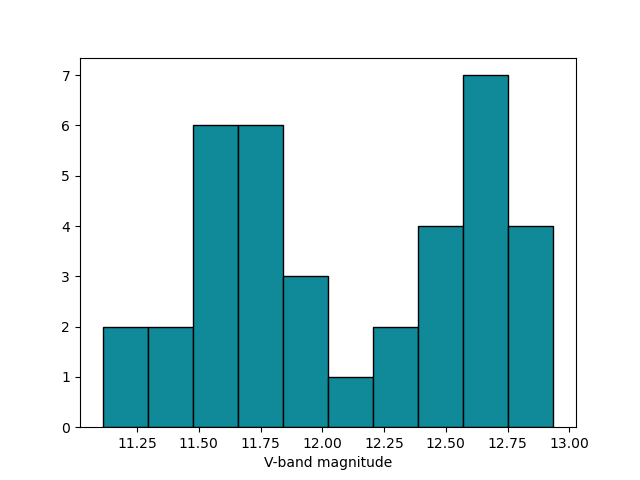}}
    \caption{Histogram of V-band magnitudes for our gold sample.}
\label{fig:vmaggold}
\end{figure}

\begin{figure}[h]
    \centering
    {\includegraphics[width=0.5\textwidth]{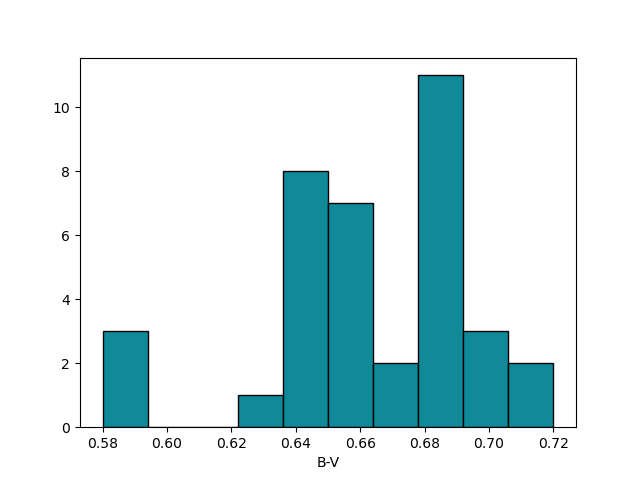}}%
    \caption{Histogram of B-V colors for our gold sample.}
    \label{fig:BVgold}
\end{figure}

\begin{figure}[h]
    \centering
    {\includegraphics[width=0.4\textwidth]{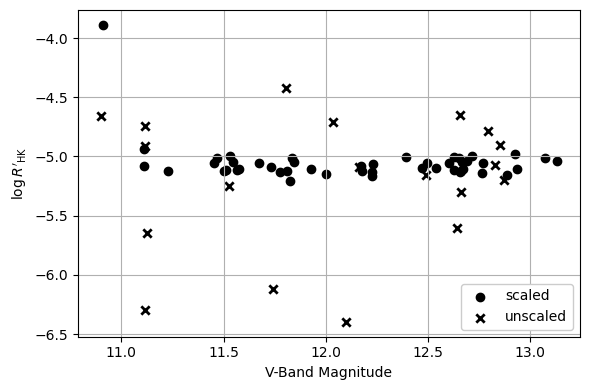}}%
    \hspace{8pt}
    {\includegraphics[width=0.4\textwidth]{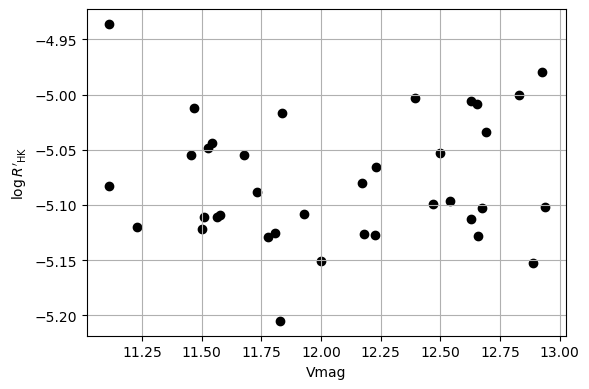}}
    \caption{V-band magnitudes and log $R'_{\rm HK}$ for entire catalog of observed stars (top) and trimmed to the gold sample (bottom). As described in Section \ref{sec:observations}, there was a disparity between observations taken with the Fabry-Perot calibration source or the sky in fiber B. We have scaled the Fabry-Perot observations that have corresponding sky measurements, but are unable to do the same for the FP-exclusive observations.}
\label{fig:RHKgold}    
\end{figure}

\begin{figure}[h]
    \centering
	{\includegraphics[width=0.4\textwidth]{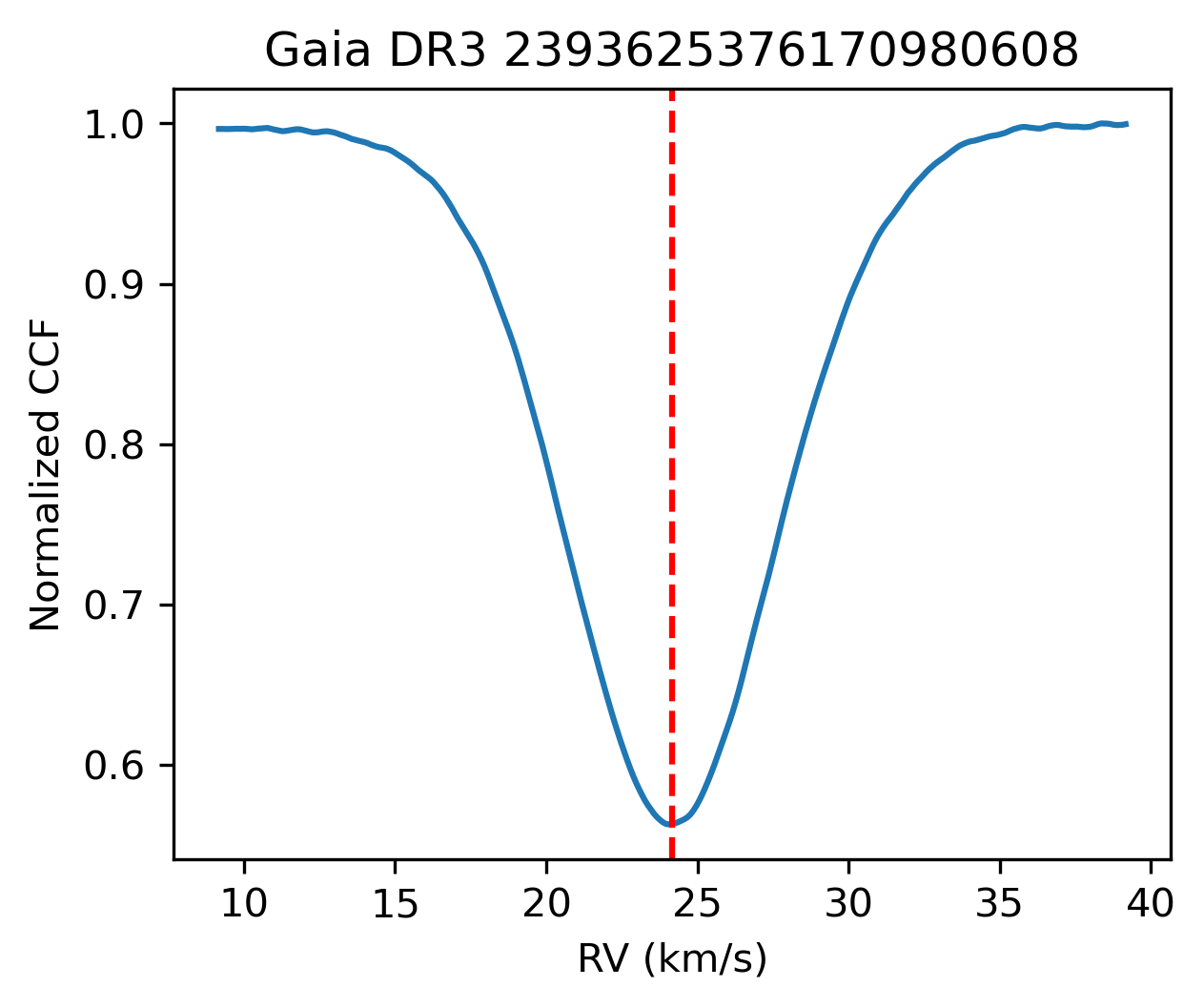}}%
    \\{\includegraphics[width=0.4\textwidth]{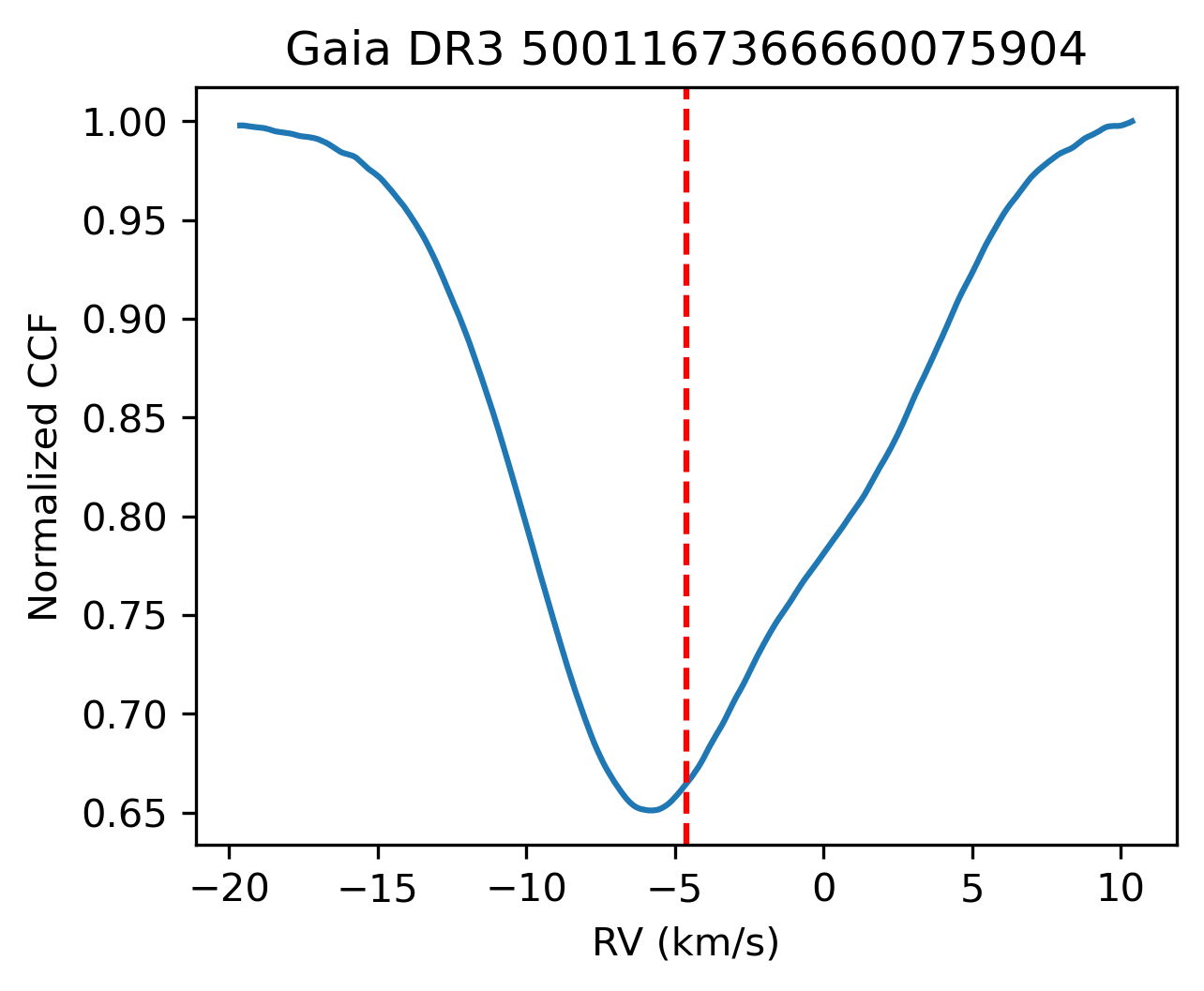}}
    \\{\includegraphics[width=0.4\textwidth]{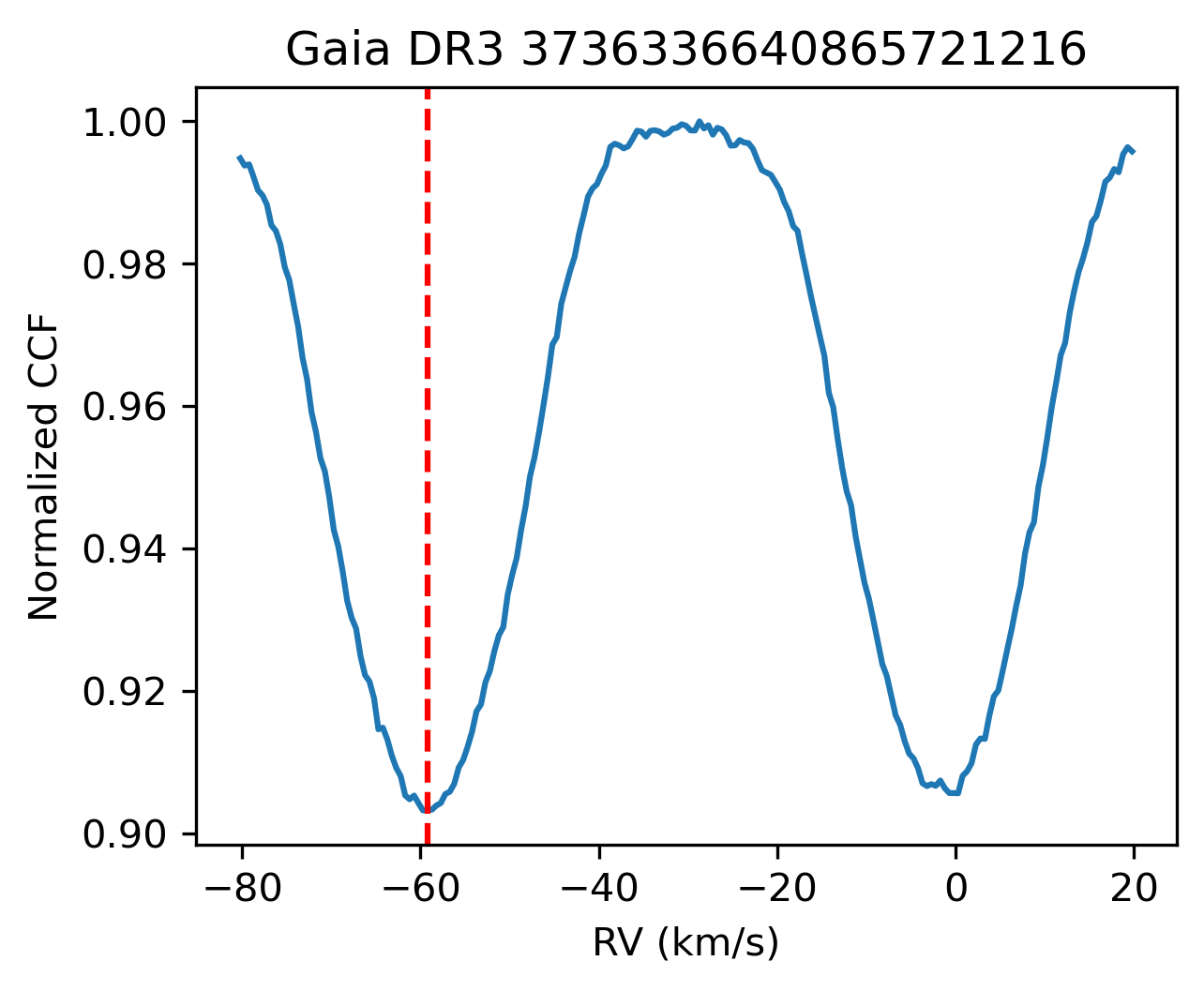}}
    \caption{Examples of a single-lined CCF (top), an asymmetric CCF (middle), and a double-lined CCF (bottom). }
\label{fig:CCFs}    
\end{figure}

\cite{Chakrabarti2020} primarily examined the effects of external contaminants to the Galactic signal, including binaries (that have large $\Delta$RV and are easily culled from the sample) and planets, given the demographics of planets and stellar binaries.  Although low-mass long-period planets are a contaminant, \cite{Chakrabarti2020} found that one should be able to statistically recover the Galactic signal to high confidence with a sample size comparable to what we select here.  In formulating our original sample, we selected sub-giants as a compromise between the dwarf stars that live at the ``jitter" minimum and the brighter, giant stars that are noisier but can trace the Galactic potential at larger distances.  As discussed earlier, in formulating our final ``gold sample", we first examined the reconnaissance spectra.  At this stage, we removed stellar binaries as well as sources that had asymmetric cross-correlation functions (CCFs) as well as stars with stellar activity indicators that were unlikely to be quiet enough for us to measure the Galactic acceleration.  We then began RV monitoring those stars that survived this initial screening.  Our ``gold sample" was then selected to minimize the scatter in the RV measurements as a function of epoch.

\cite{Luhn2020}'s analysis indicated that F- and G-type subgiants would be ideal candidates to measure Galactic accelerations using EPRV observations, as these stars are bright enough to be measured at significant distances from the Galactic midplane, while also having reduced magnetic activity to have relatively low stellar jitter.  Stellar jitter is typically characterized by chromospheric activity as measured by the Ca II H and K lines \citep{Luhn2020}, and the covection-driven component for subgiants is traced by the star's luminosity $L$.  Typical Mount Wilson survey chromospheric activity indicators include $\rm S_{\rm HK}$ and $\rm log R_{HK}'$, where the latter subtracts out the photospheric component of the emission peak and accounts for the star's color. \cite{Luhn2020} sought to identify a low-jitter metric for F-type stars using a combination of their measured $\rm log R_{HK}'$ and the luminosity.  They showed that the quantity:

\begin{equation}
j = 2(\rm log R_{HK}' + 5.4) + log L   
\end{equation}

\noindent is traced well by the stellar jitter.  They adopted a threshold value of $j < 1.53d$ to select for low-jitter ($<$ 10 m/s) F-stars.  Once we obtained reconnaissance spectra of our original sample, we calculated the $j$ values and required that $j < 1.53$.  We also visually examined the spectra and cross-correlations functions to ensure that the stars are not binaries (i.e., binaries manifest two peaks in the cross-correlation function), or have asymmetric cross-correlation functions that indicate either high stellar activity or a stellar companion that is partially resolved in the spectra.  Plots of the CCFs are shown in Figure \ref{fig:CCFs} for a single-lined, asymmetric, and double-lined CCF. We then continued RV monitoring for the set of stars that had sufficiently low ``j" values, as the ``j" metric is useful for characterizing stellar variability but it does not guarantee that the star does not have stellar or planetary companions.  Thus, the ``j" metric is a necessary but insufficient criterion for a star entering our ``gold sample".  

As noted earlier, our requirement for obtaining high $S/N$ spectra at \emph{galactic} scale distances within reasonable integration times effectively sets an upper limit on the V-band magnitudes of our final ``gold" sample.  A histogram of V-band magnitudes is shown in Figure \ref{fig:vmaggold}.  The B-V color for the gold sample is shown in Figure \ref{fig:BVgold}.  The two panels in Figure \ref{fig:RHKgold} show the $\rm log R_{HK}'$ for our original sample (top panel) as a function of the V-band magnitude, and for our gold sample (bottom panel).  As is clear, the ``gold sample's" $\rm log R_{HK}'$ values span a smaller range than our original sample.  The gold sample's $\rm log R_{HK}'$ has a mean and standard deviation of -5.079 and 0.056 respectively, typical of inactive sub-giant stars \citep{Isaacson_Fischer2010}, as compared to the original sample which has a mean and standard deviation of -5.113 and 0.383 respectively.  As described in \S \ref{sec:observations}, there are differences between our initial observing strategy and our final observing strategy, such that our initial use of the FP calibration source led to an over-subtraction of the detector background, which biased the measurements of the cores of the calcium lines and hence the $\rm log R_{HK}'$ values.  However, in most cases, those same stars were observed later with our modified observing strategy where the sky illuminated the second fiber during the science observation, which was followed by the calibration with the FP calibration source illuminating both fibers.  The latter observing strategy led to well defined emission from the cores of the Calcium lines.  Since most of the stars were observed with our final observing strategy, here in Figure \ref{fig:RHKgold}, we have scaled the $\rm log R_{HK}'$ for the FP observations to the on-sky observations (see Appendix~\ref{sec:FPbias}).

Table \ref{tab:excludedsources} gives a list of the sources that we removed from this original list based on either having a stellar companion or tentative evidence for an exoplanet or brown dwarf (discussed in \S \ref{sec:stellarcomps}) or excessively large stellar activity as determined from the analysis of the reconnaissance spectra (discussed in \S \ref{sec:quietstarsample}).  In \S \ref{sec:stellarcomps}, we study several sources that may have companions.  We have carried out fits to our RV data and give estimates of the companion mass and binary parameters.   These estimates can be refined further with the anticipated release of \textit{Gaia} DR-4 data.  Since we are primarily interested in single stars that trace the Galactic potential, we remove these stars from our ``gold sample" and will not continue to monitor them.

\begin{figure}[h]
    \centering
    {\includegraphics[width=0.5\textwidth]{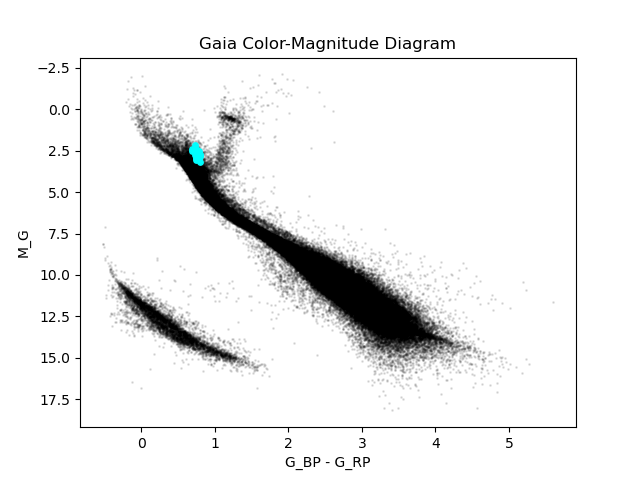}}%
    \caption{Our gold sample (in cyan) overplotted with the Gaia main-sequence (black)}
    \label{fig:gaiagoldandms}
\end{figure}

Our gold sample currently constitutes 37 stars.  Figure \ref{fig:gaiagoldandms} shows our ``gold sample" in cyan overplotted with the Gaia main-sequence in black, on the Gaia color-magnitude diagram.  As is clear, our sub-giant stars are slightly evolved relative to main-sequence stars.  Properties of the ``gold sample" are given in Table \ref{tab:goldsampleproperties}, including the mean $\rm log R_{HK}'$ value, V-band magnitude, $V_{\rm mag}$, $B-V$ color, absolute vertical height in parsec, the luminosity, and calculated $j$ value.  $V_{\rm mag}$, $B-V$, vertical height, and stellar luminosity are from \textit{Gaia} DR-3, and $j$
and the mean value $\rm log R_{HK}'$ are from our calculations here.  

\begin{figure}[h]
    \centering
    \includegraphics[width=0.3\textwidth]{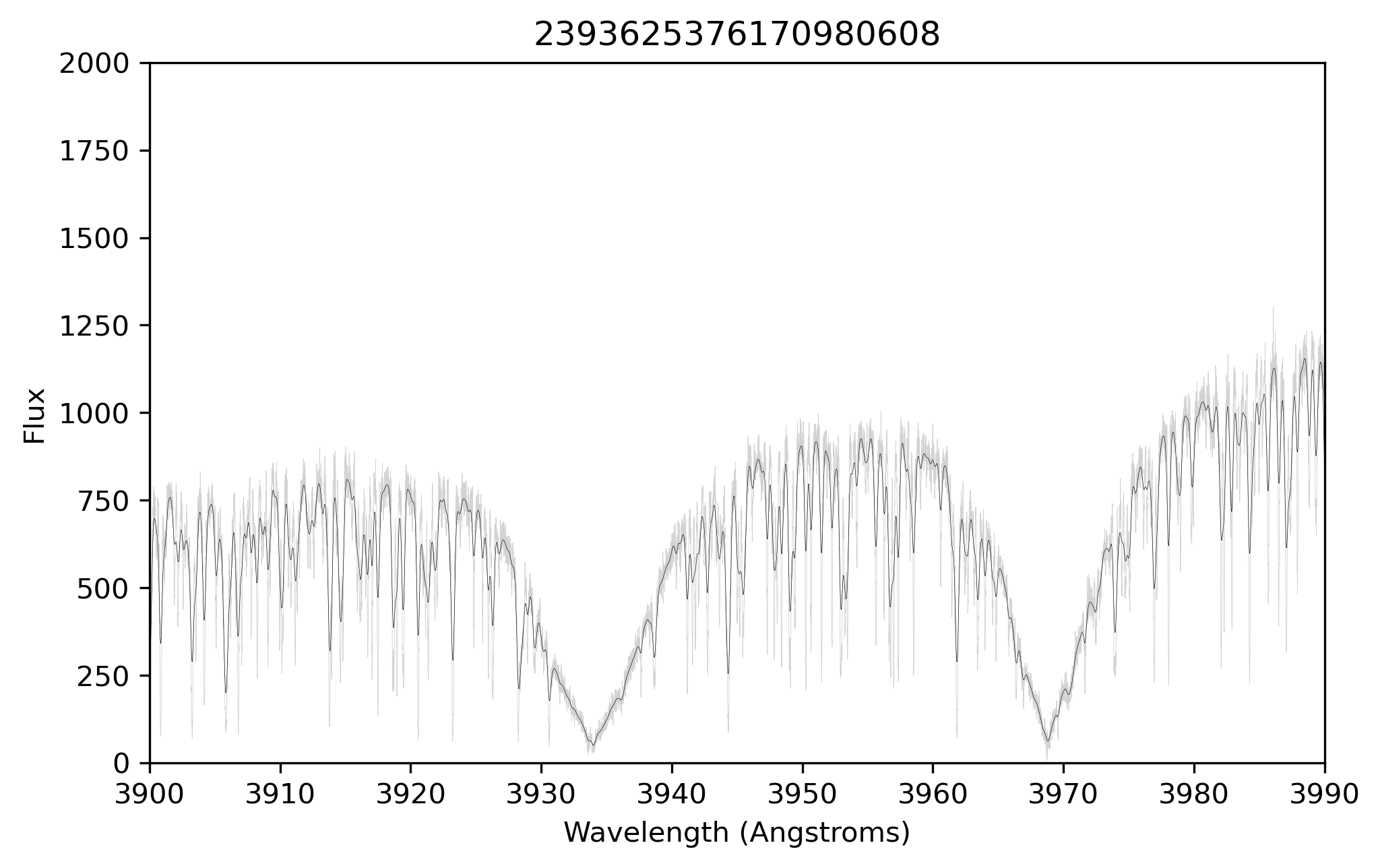}
    \includegraphics[width=0.3\textwidth]{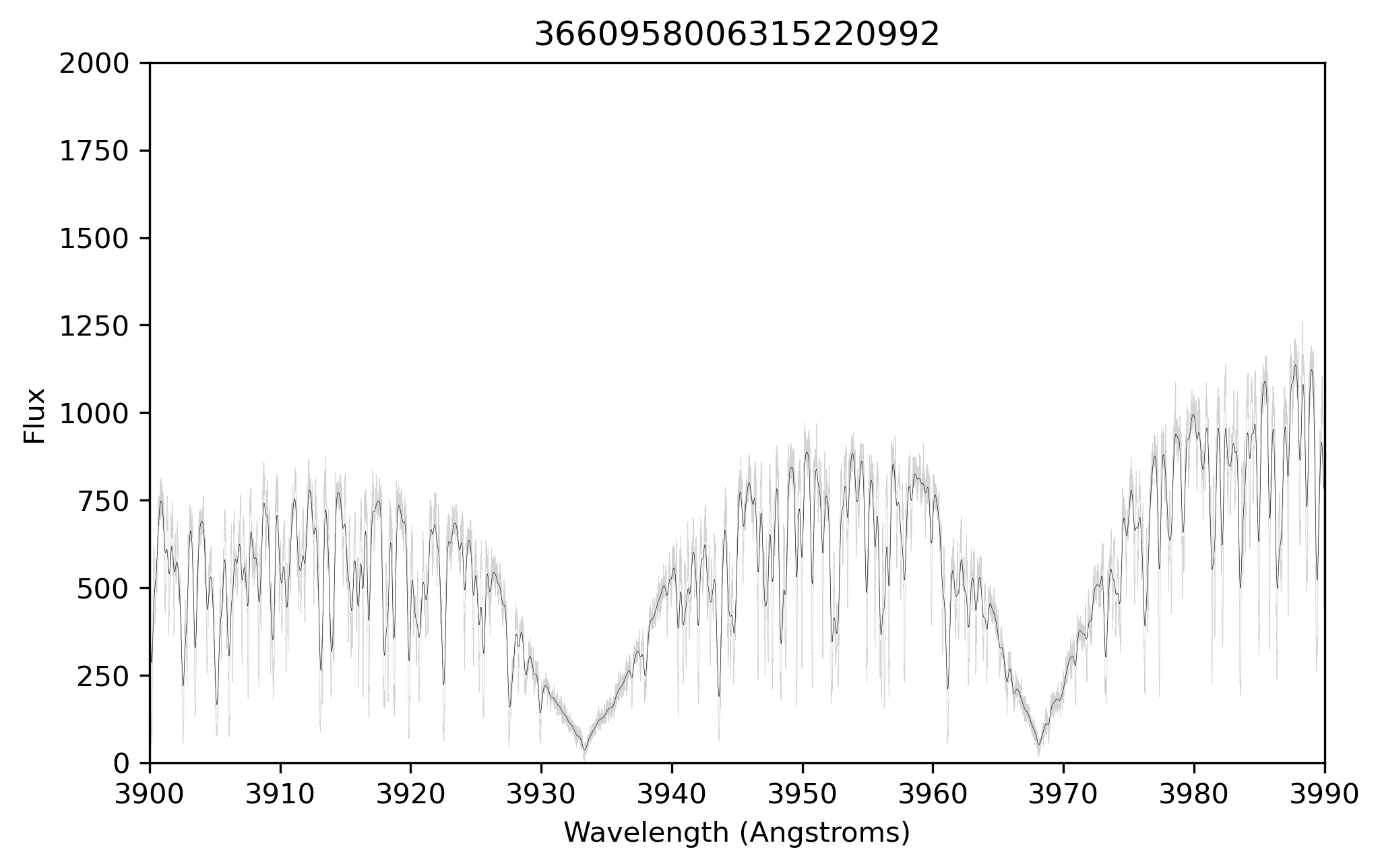}
    \includegraphics[width=0.3\textwidth]{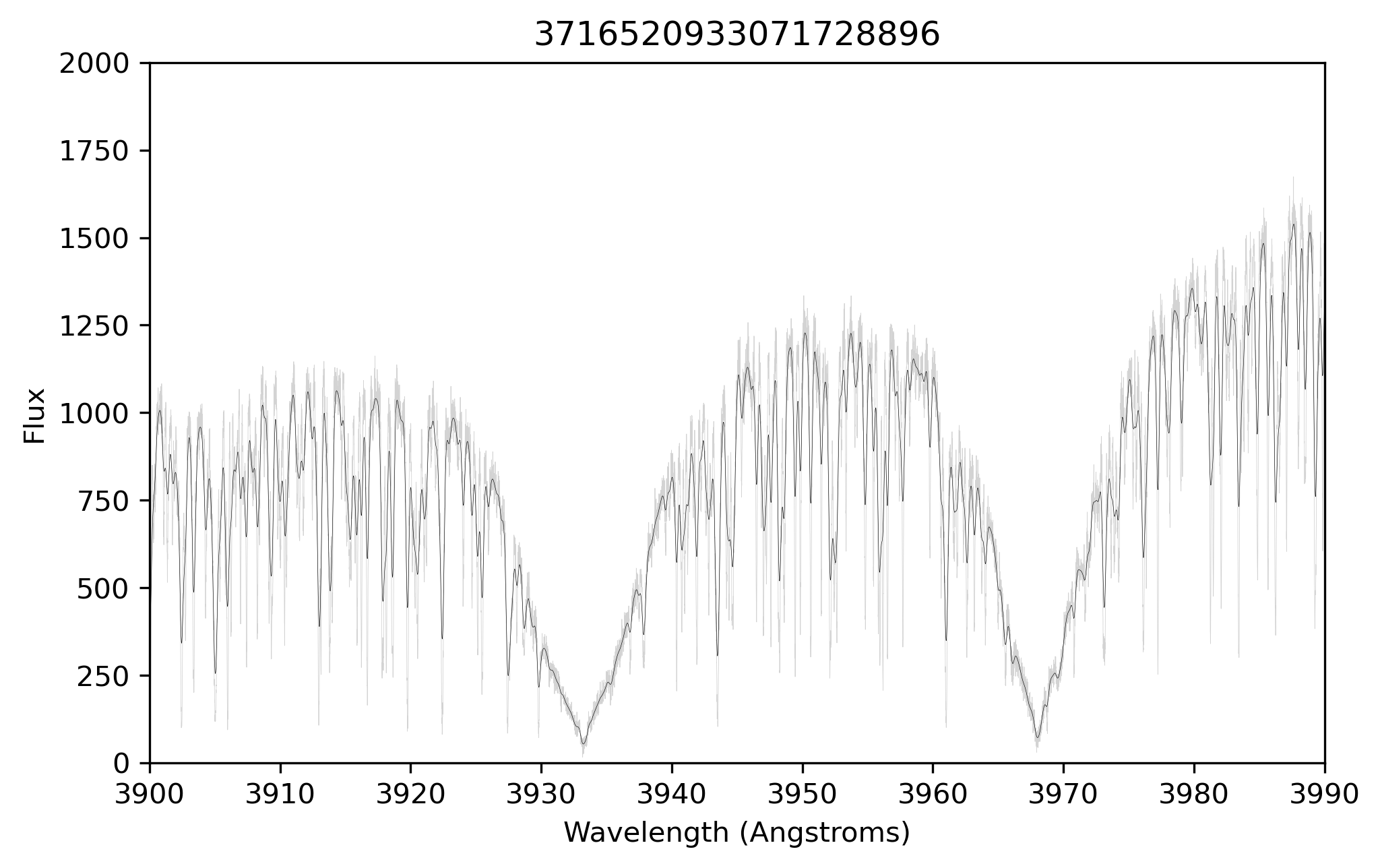}
    \medskip
    \includegraphics[width=0.3\textwidth]{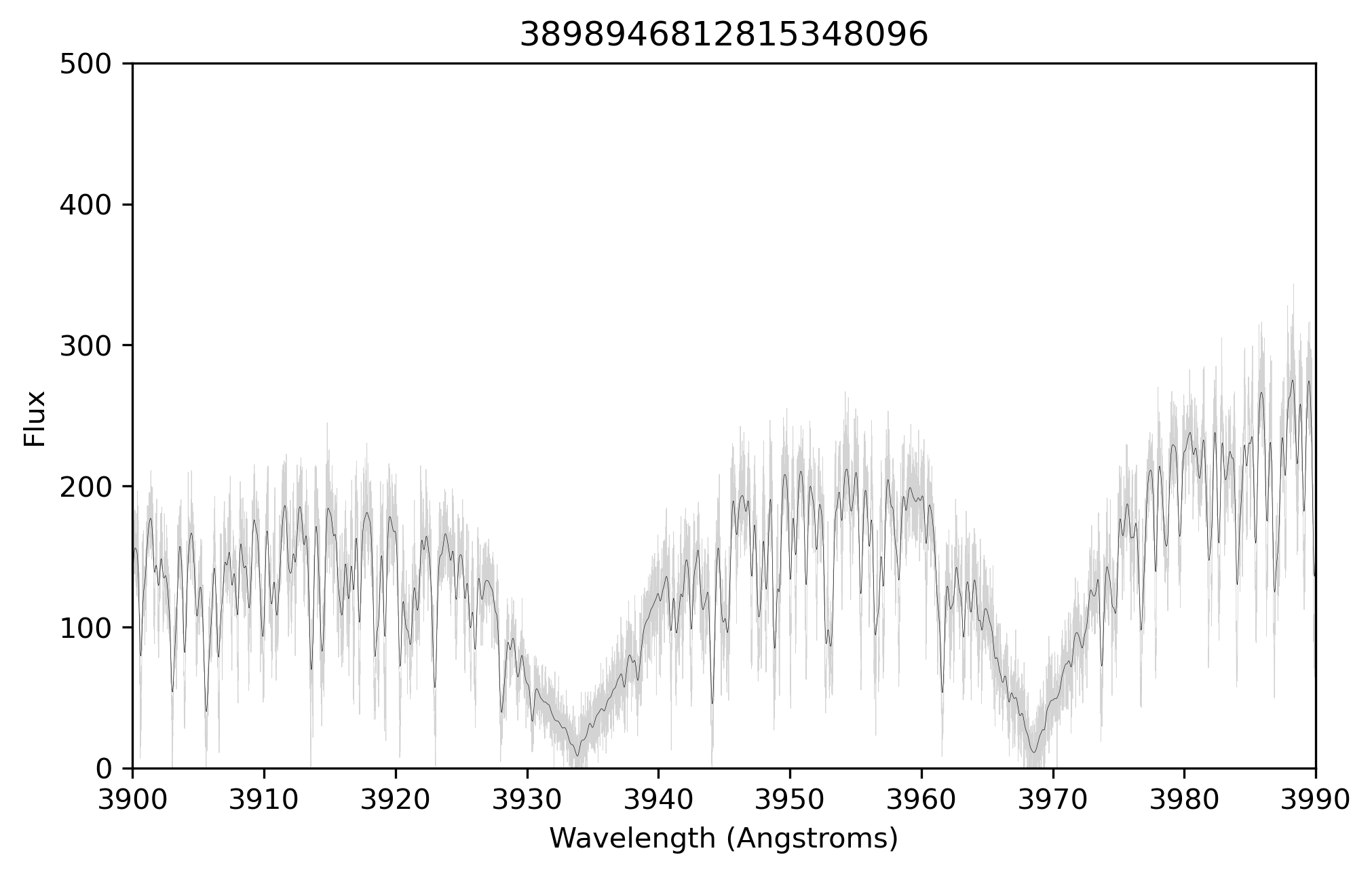}
    \includegraphics[width=0.3\textwidth]{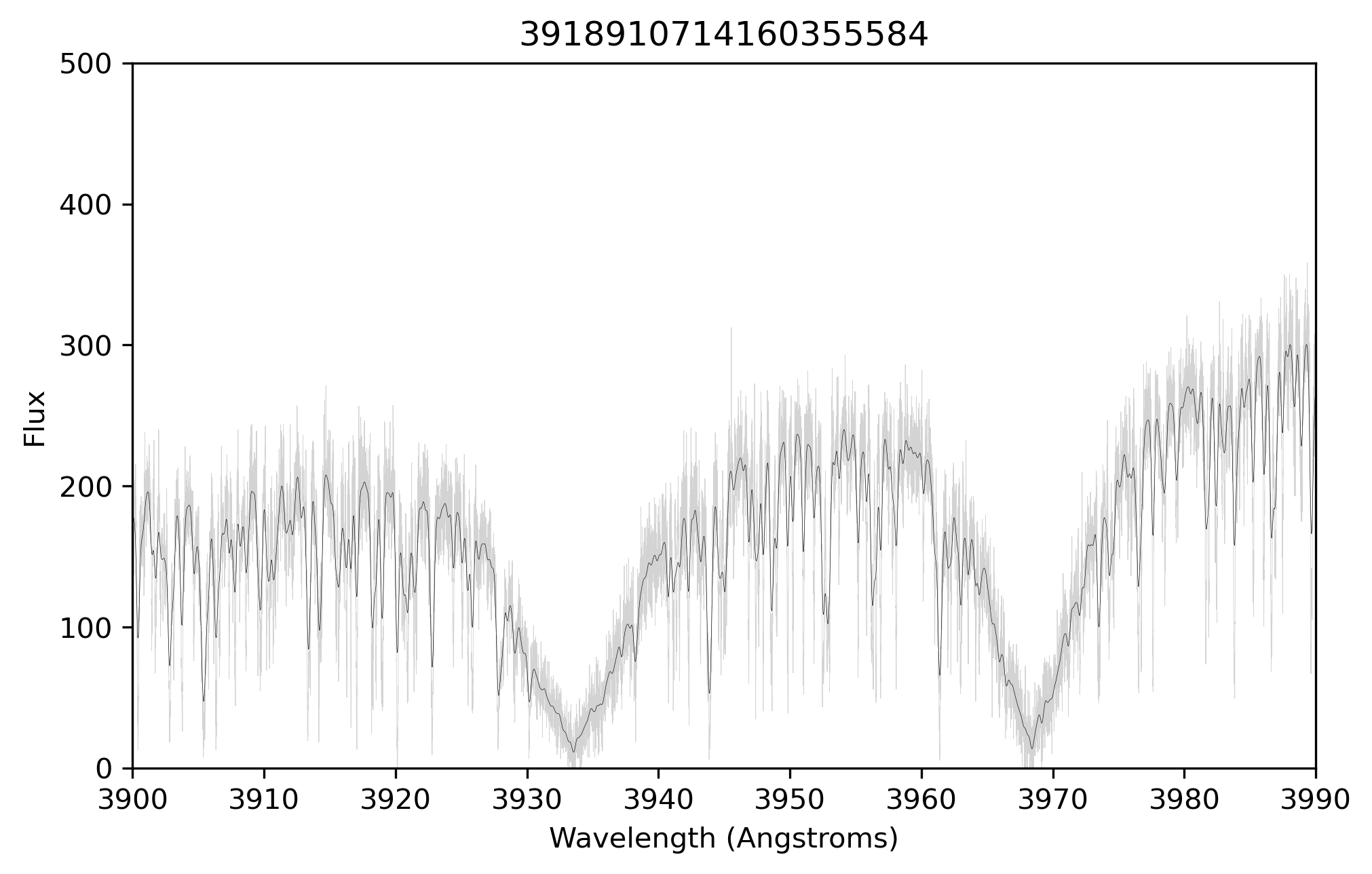}
    \includegraphics[width=0.3\textwidth]{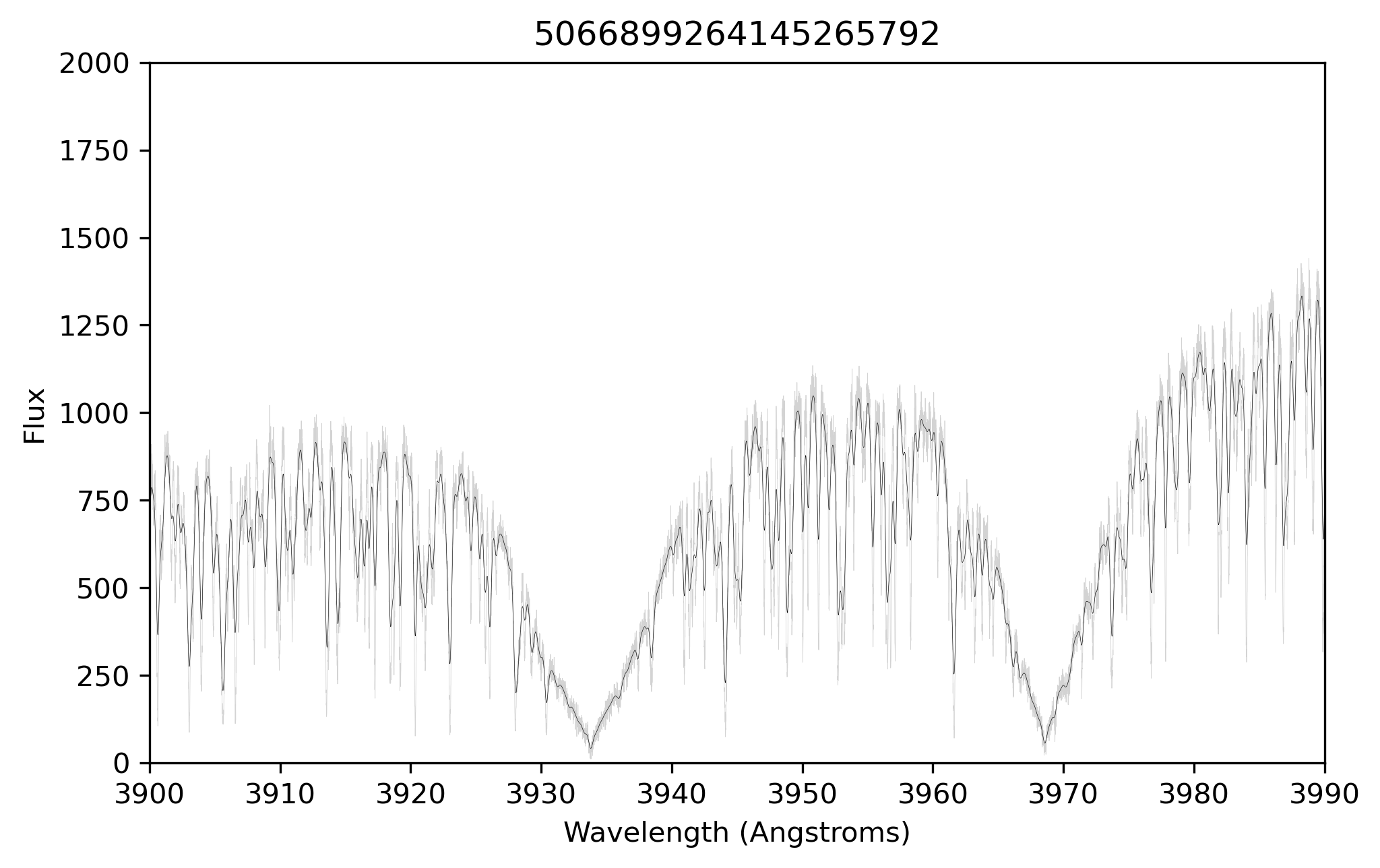}
    \caption{ESPRESSO Spectra near the Ca II H\&K lines for the six quietest stars in our ``gold sample"}
\label{fig:goldspectra}
\end{figure}

\begin{figure}[p]
    \centering
    \includegraphics[width=0.25\textwidth]{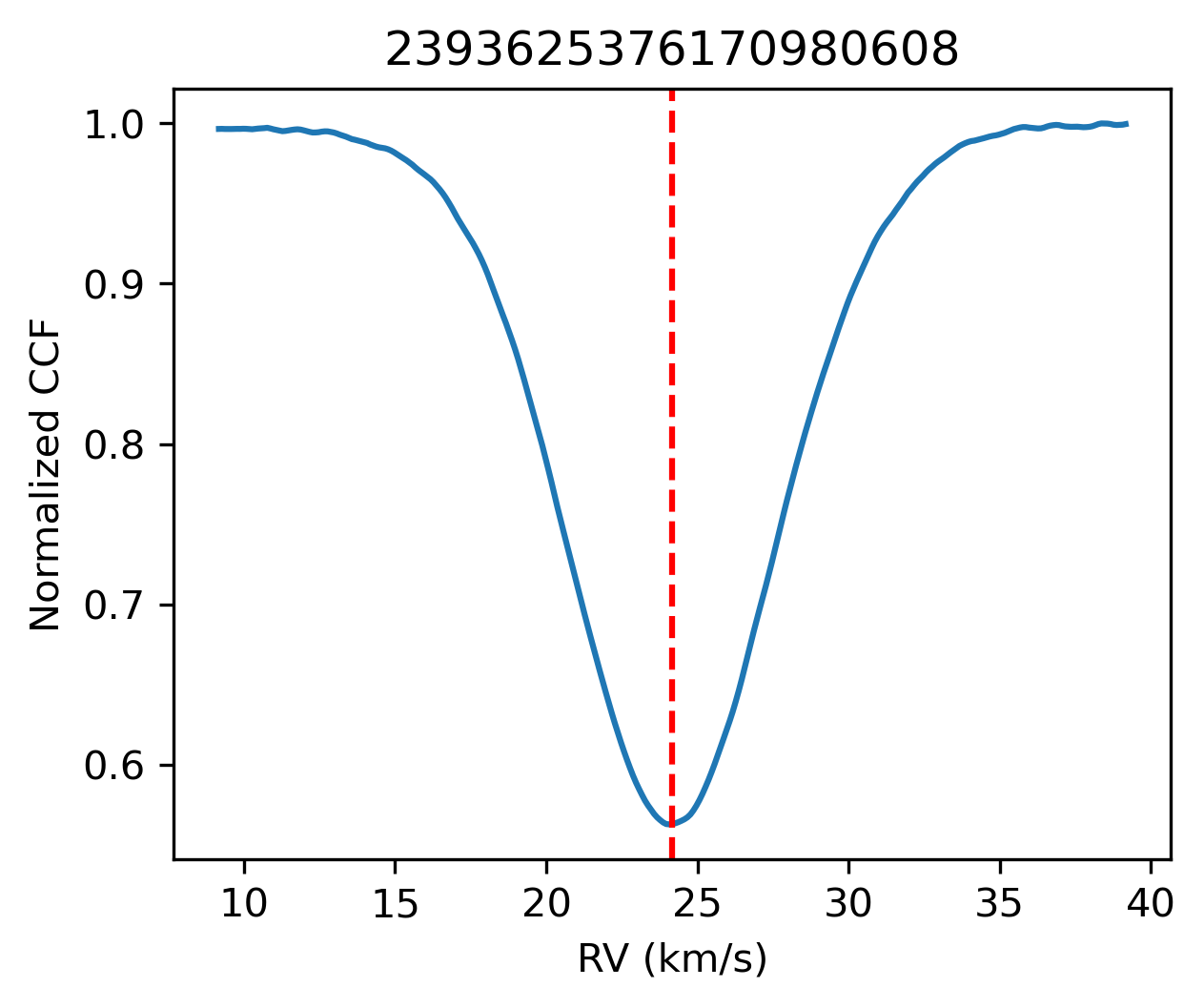}
    \includegraphics[width=0.25\textwidth]{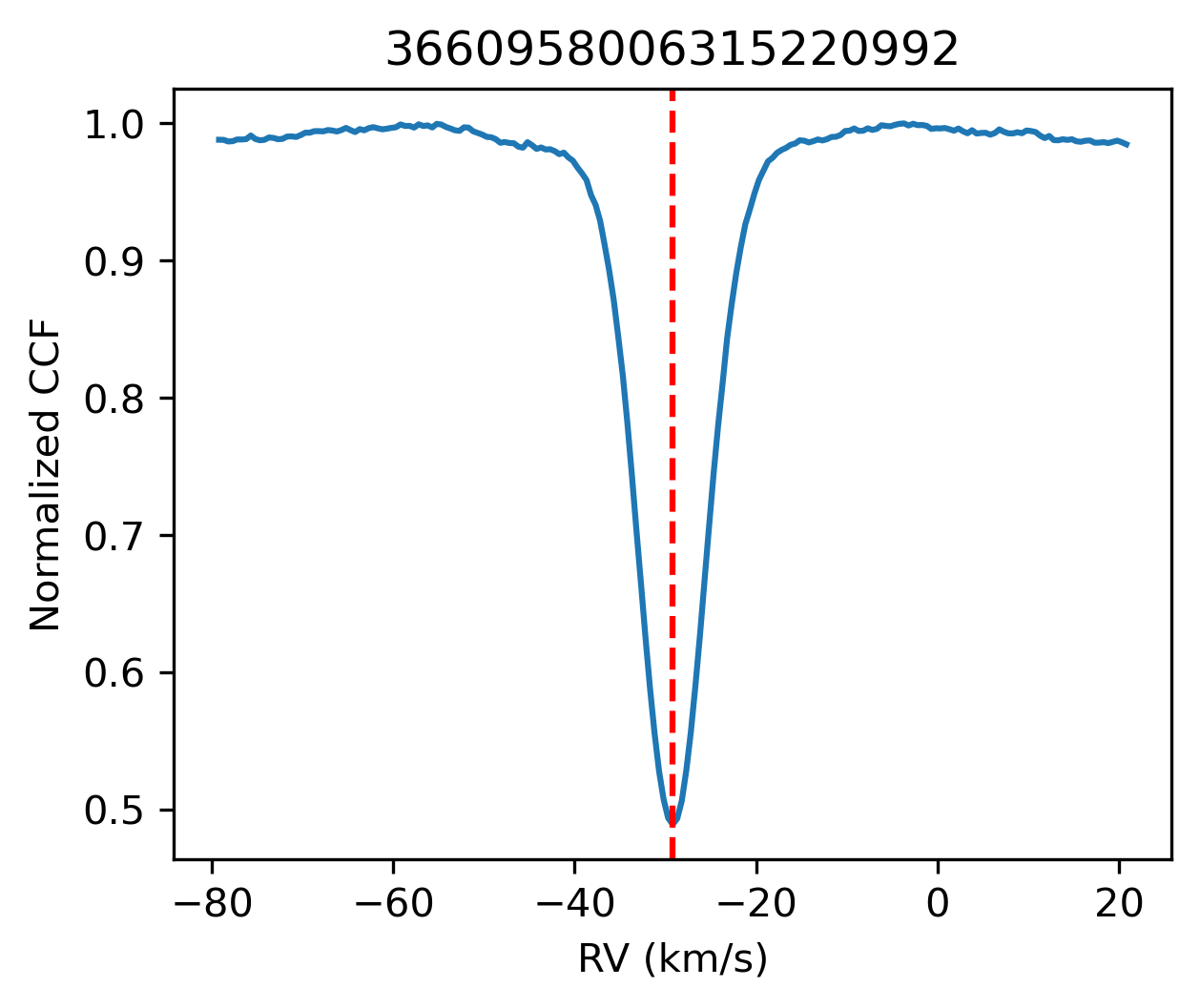}
    \includegraphics[width=0.25\textwidth]{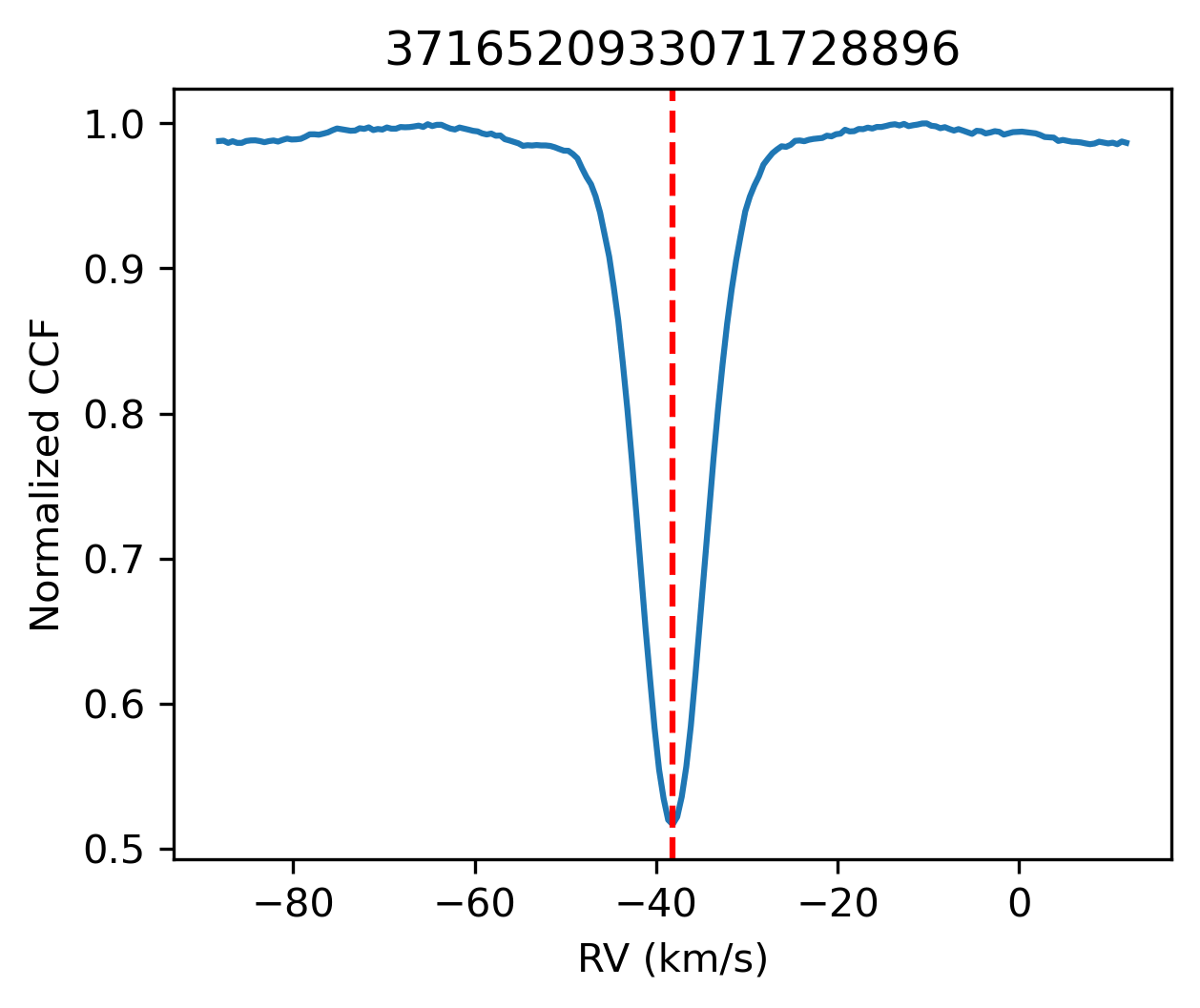}
    \medskip
    \includegraphics[width=0.25\textwidth]{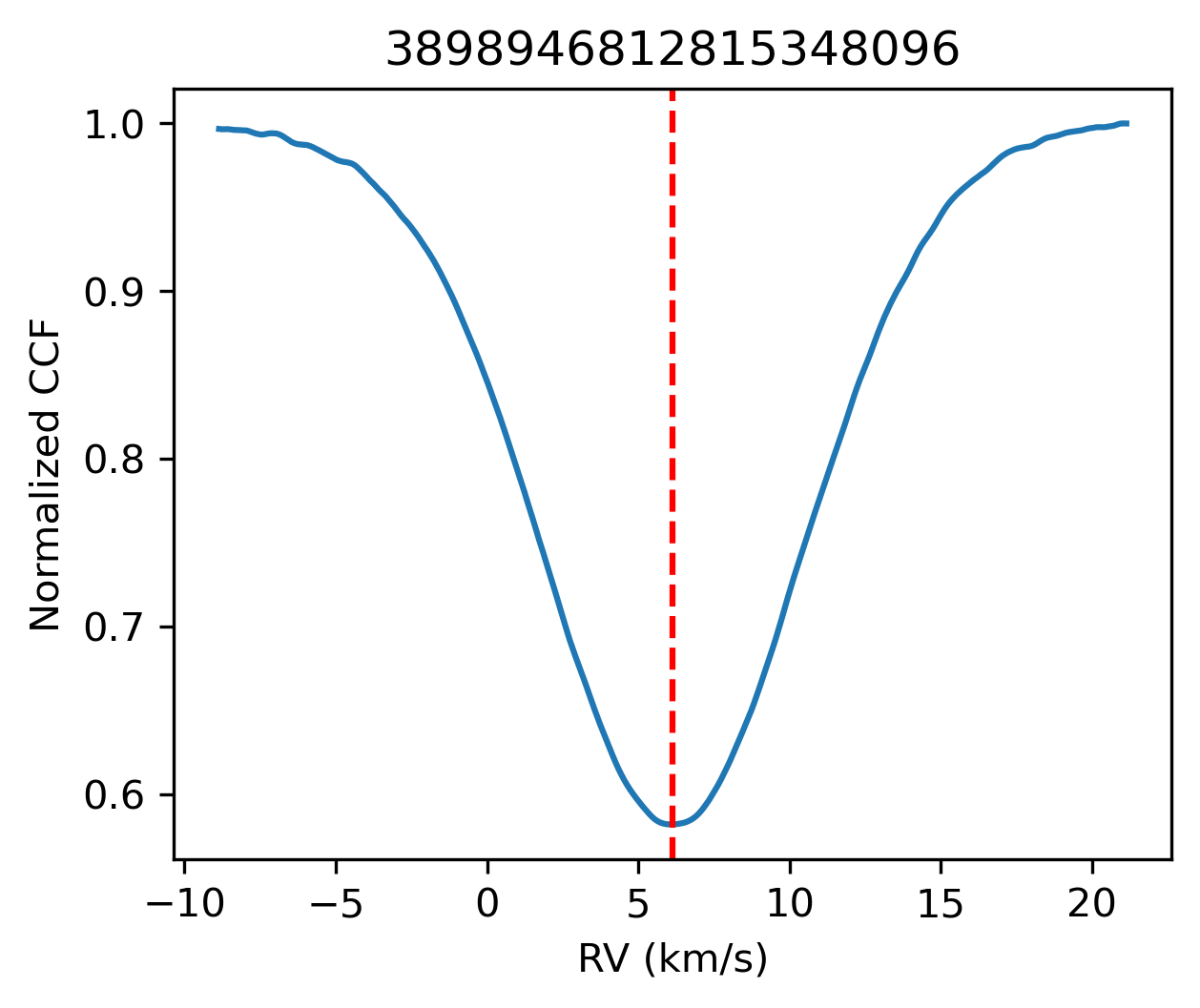}
    \includegraphics[width=0.25\textwidth]{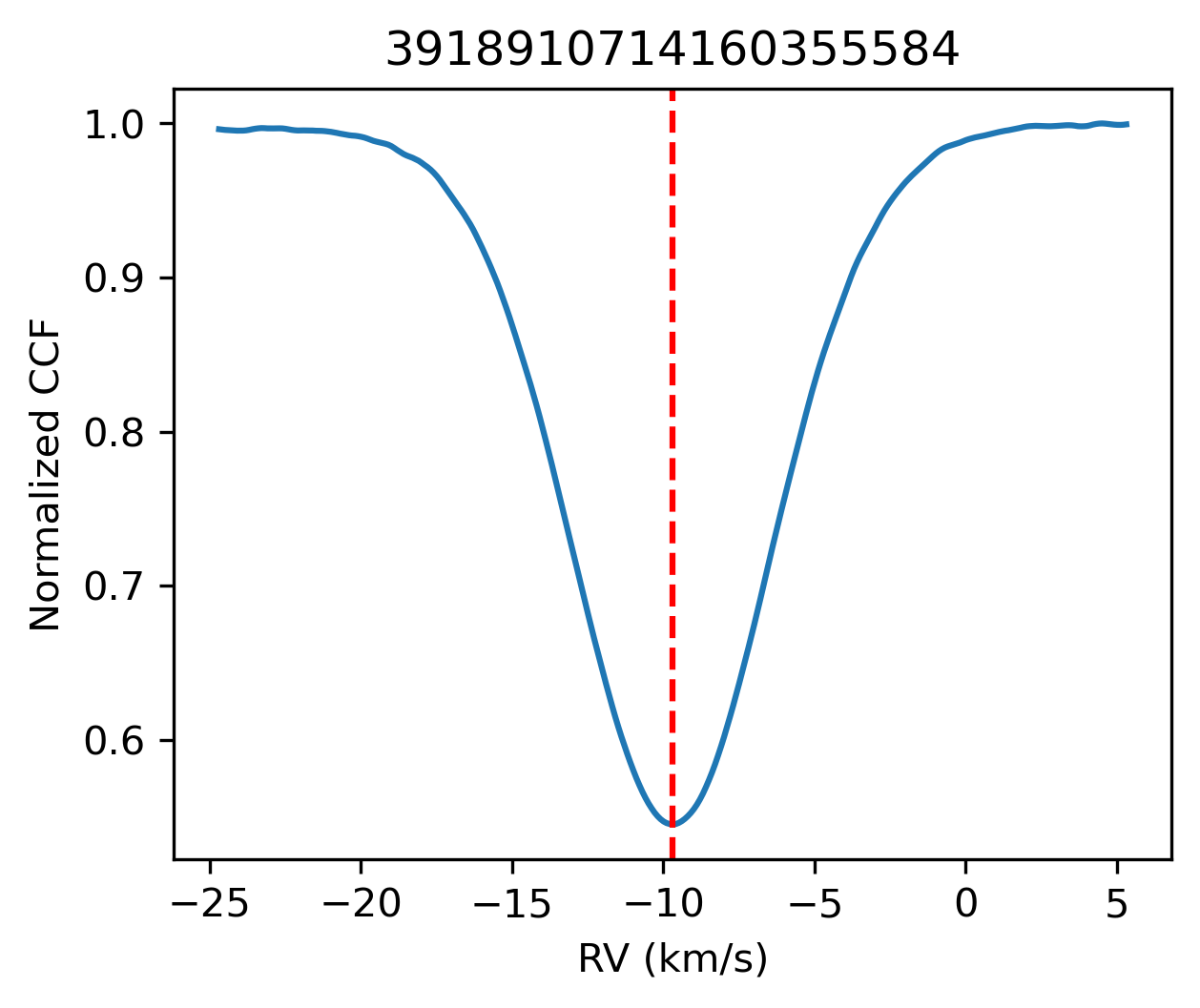}
    \includegraphics[width=0.25\textwidth]{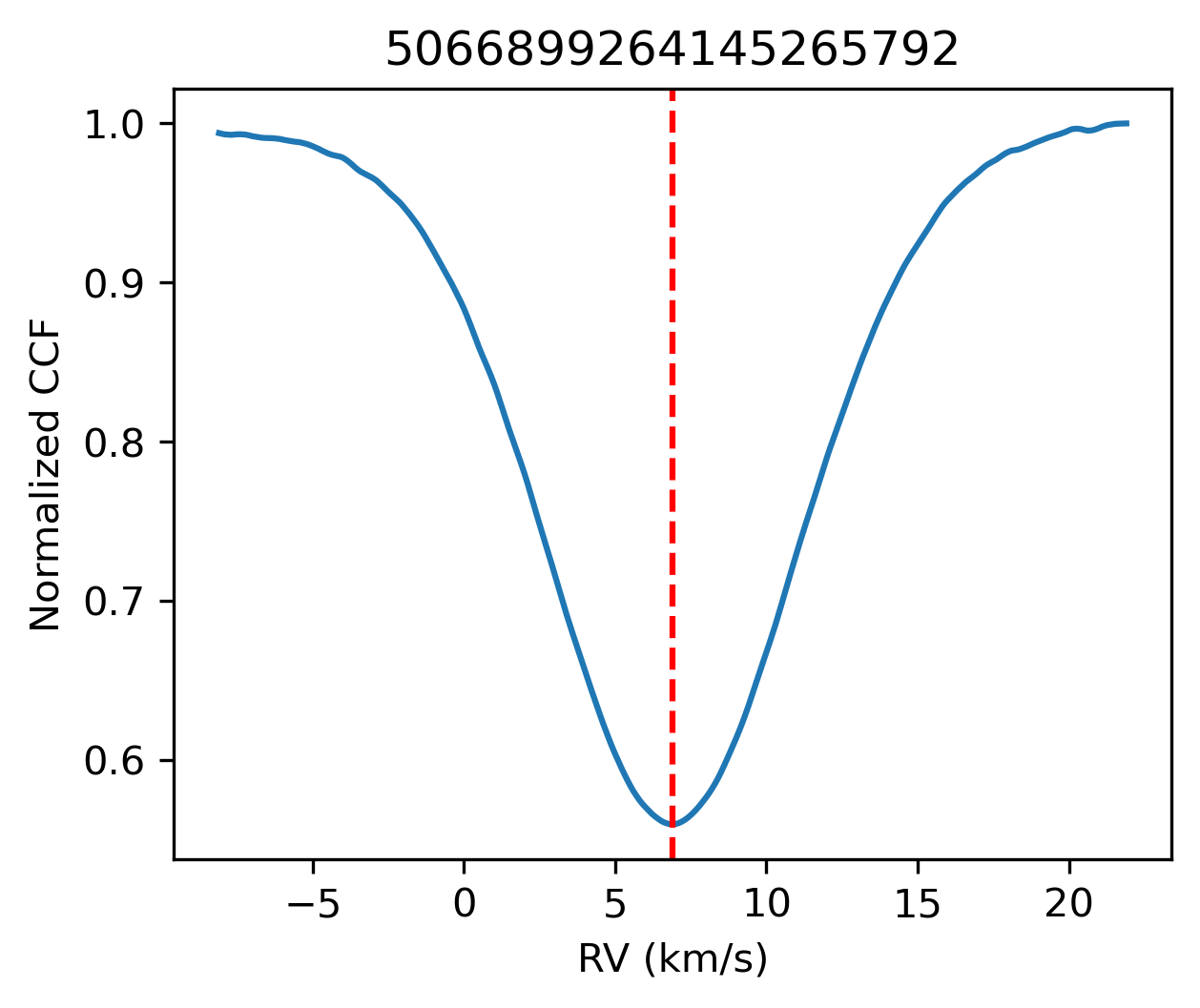}
    \caption{CCFs of the six quietest stars in our ``gold sample"}
\label{fig:goldCCF}    
\end{figure}

\begin{figure}[h]
    \centering
    {\includegraphics[width=0.55\textwidth]{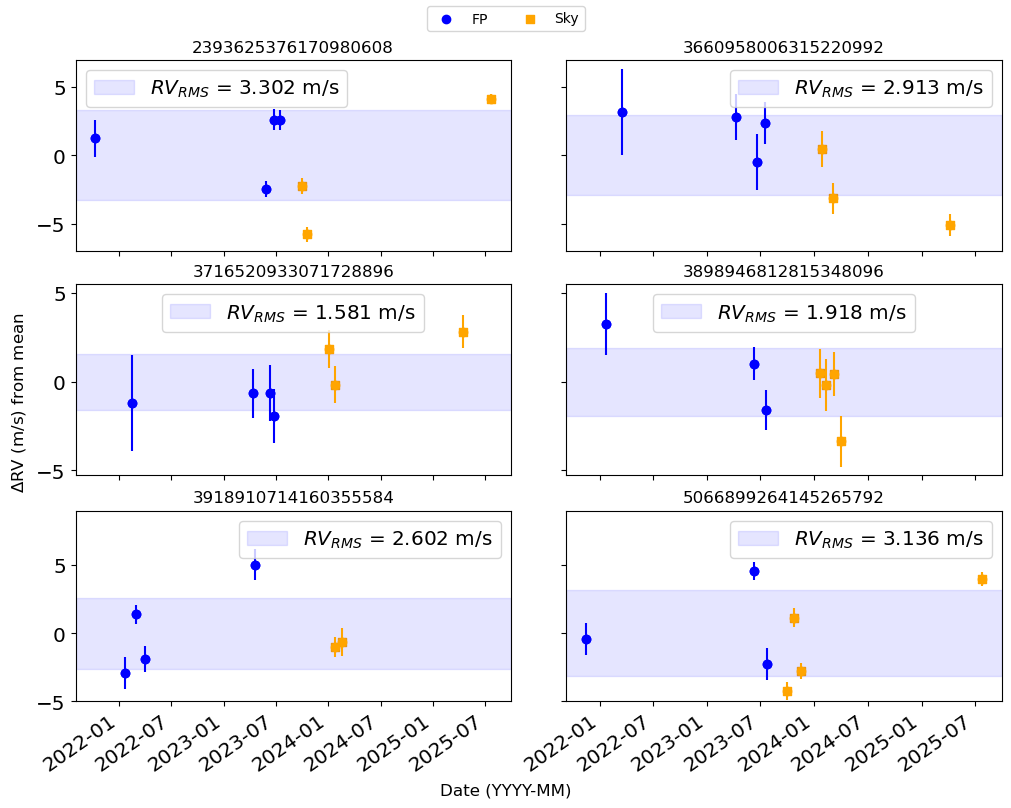}}%
    \caption{Mean-subtracted RV time-series for six of the ``quietest" stars in our ``gold sample".}
\label{fig:goldRVtimeseries}    
\end{figure}

\begin{figure}[h]
    \centering
    {\includegraphics[width=0.4\textwidth]{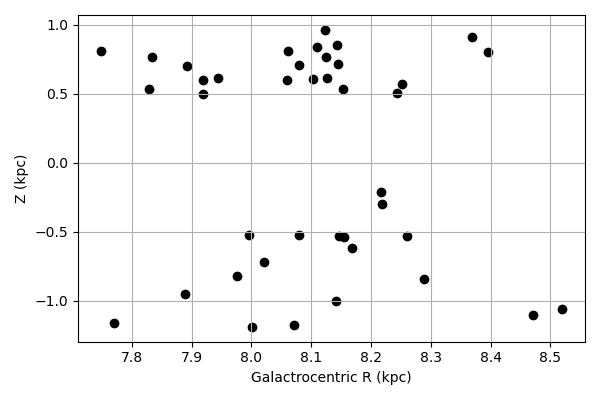}}
       \caption{Galactocentric map of the gold sample.}
\label{fig:goldRz}    
\end{figure}

If the measured stellar jitter of a star meets our expected low-jitter threshold (Eqn 1; $j < 1.53$), show symmetric cross-correlation functions, and low-enough RV scatter over a large number of epochs, we include it in our final gold sample, which we will continue to monitor over the next several years to measure a Galactic acceleration.   If we have more than 10 observations, we require RV scatter less  than 10 m/s to include the star in the gold sample.  There are a few stars that have a RV scatter larger than 10 m/s but these have a small number of epochs thus far.  We require a minimum of 5 epochs to determine if the star will remain in the gold sample, and some of these stars are still currently under investigation and may be removed in future cycles.  We also retain a few stars like Gaia DR3 3995373707693411072 that are at higher z values (about 1 kpc in this case) to maintain as large a dynamic range as we can.  Table \ref{tab:RVparamsgold} summarizes the RV measurements of the gold sample, including the number of observations, the mean RV, the $RV_{\rm RMS}$, the FWHM of the CCF, and its associated error.  Figures \ref{fig:goldspectra}, \ref{fig:goldCCF} and \ref{fig:goldRVtimeseries} show the spectra, CCFs and RV time-series (subtracting out the mean to show the RV scatter) for six of the quietest stars in our gold sample.  For this set of the six quietest stars in our gold sample, the mean $RV_{\rm RMS}$ (determined from epoch-level measurements) is 2.7 m/s.  In comparison, the quietest dwarf stars have a lower jitter floor, estimated at $\sim$ 1.7 m/s for quiet late-F/early-G dwarfs \citep{Isaacson_Fischer2010}.  The RV time-series for the entirety of the gold sample are shown in the Appendix.  Figure \ref{fig:NGstyle} shows the stars in our gold sample with the epochs of observations marked per year, and Figure \ref{fig:goldRz} shows the distribution of the gold sample in Galactocentric R and z coordinates.



\begin{figure}[h]
    \centering
    {\includegraphics[width=0.45\textwidth]{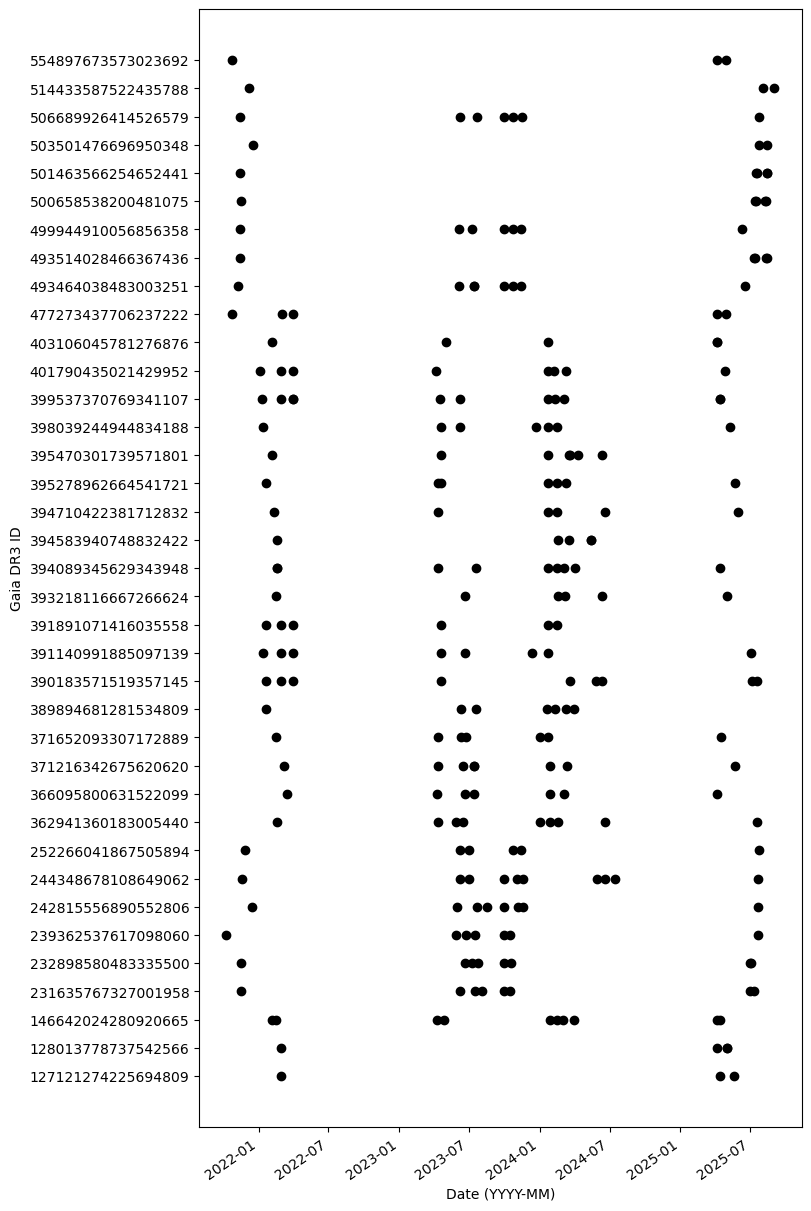}}%
    \caption{Timestamps of our gold sample observations, indicating the number of epochs per star.} 
    \label{fig:NGstyle}
\end{figure}

\begin{table*}[h]
      \caption{Properties of the ``Gold Sample"}
   \begin{tabular}{@{} ccccccccc @{}} 
      \toprule
      Gaia DR3 ID & $\overline{\log R'_{\rm HK}}$ & $V_{\rm mag}$ & $B-V$ & $\abs{z}$  (pc) & \(L_\odot\) & \textit{j} \\
      \midrule
      1271212742256948096 & $-5.152$ & 12.573   &  0.64  & 808.2 & $5.99^{+0.09}_{-0.15}$ & 1.27 \\ 
      1280137787375425664 & $-5.185$ & 12.167   &  0.64  & 703.3 & $5.88^{+0.15}_{-0.87}$ & 1.21 \\ 
      1466420242809206656 & $-5.161$ & 12.609   &  0.68  & 964.6 & $5.70^{+0.62}_{-0.59}$ & 1.23 \\
      2316357673270019584 & $-5.074$ & 12.692   &  0.72  & 819.7 & $4.41^{+0.24}_{-0.24}$ & 1.30 \\
      2328985804833355008 & $-5.182$ & 12.887   &  0.68  & 1157.9 & $5.46^{+0.20}_{-0.18}$ & 1.17 \\
      2393625376170980608 & $-5.082$ & 11.454   &  0.64  & 519.9 & $6.23^{+0.14}_{-0.08}$ & 1.43 \\
      2428155568905528064 & $-5.108$ & 11.675   &  0.64  & 534.0 & $5.57^{+0.15}_{-0.17}$ & 1.33 \\
      2443486781086490624 & $-5.133$ & 11.730   &  0.71  & 530.7 & $7.16^{+0.27}_{-0.27}$ & 1.27 \\
      2522660418675058944 & $-5.155$ & 12.541   &  0.69  & 839.7 & $7.06^{+0.07}_{-0.29}$ & 1.38 \\
      3629413601830054400 & $-5.173$ & 11.230   &  0.64  & 503.0 & $7.86^{+0.19}_{-0.19}$ & 1.35 \\
      3660958006315220992 & $-5.104$ & 11.575   &  0.58  & 539.3 & $6.91^{+0.15}_{-0.14}$ & 1.43 \\
      3712163426756206208 & $-5.183$ & 12.000   &  0.65  & 764.9 & $5.67^{+0.16}_{-0.13}$ & 1.19 \\
      3716520933071728896 & $-5.158$ & 11.500   &  0.63  & 600.7 & $8.42^{+0.40}_{-0.40}$ & 1.41 \\
      3898946812815348096 & $-5.177$ & 11.930   &  0.64  & 765.2 & $4.23^{+0.12}_{-0.05}$ & 1.06 \\
      3901835715193571456 & $-5.041$ & 12.652   &  0.65  & 812.7 & $4.82^{+0.15}_{-0.13}$ & 1.40 \\
      3911409918850971392 & $-5.087$ & 11.545   &  0.65  & 536.5 & $6.38^{+0.24}_{-0.24}$ & 1.43 \\
      3918910714160355584 & $-5.086$ & 12.497   &  0.70  & 858.2 & $5.34^{+0.14}_{-0.16}$ & 1.36 \\
      3932181166672666240 & $-5.177$ & 12.226   &  0.69  & 707.2 & $5.59^{+0.12}_{-0.08}$ & 1.19 \\
      3940893456293439488 & $-5.247$ & 11.827   &  0.68  & 602.9 & $6.36^{+0.23}_{-0.22}$ & 1.11 \\
      3945839407488324224 & $-5.145$ & 12.468   &  0.67  & 839.4 & $6.00^{+0.36}_{-0.40}$ & 1.29 \\
      3947104223817128320 & $-5.202$ & 11.565   &  0.67  & 612.3 & $7.54^{+0.02}_{-0.05}$ & 1.27 \\
      3952789626645417216 & $-5.117$ & 12.230   &  0.66  & 713.7 & $5.05^{+0.09}_{-0.08}$ & 1.27 \\
      3954703017395718016 & $-5.151$ & 11.809   &  0.68  & 611.1 & $5.08^{+0.09}_{-0.08}$ & 1.19 \\
      3980392449448341888 & $-5.068$ & 11.468   &  0.70  & 507.7 & $5.39^{+0.11}_{-0.10}$ & 1.24 \\
      3995373707693411072 & $-5.031$ & 12.627   &  0.66  & 806.4 & $4.41^{+0.11}_{-0.10}$ & 1.38 \\
      4017904350214299520 & $-5.072$ & 11.527   &  0.64  & 512.8 & $7.08^{+0.40}_{-0.40}$ & 1.51 \\
      4031060457812768768 & $-5.083$ & 12.828   &  0.70  & 914.7 & $5.34^{+0.28}_{-0.28}$ & 1.33 \\ 
      4772734377062372224 & $-5.114$ & 11.844   &  0.68  & 298.0 & $7.73^{+0.09}_{-0.09}$ & 1.46 \\ 
      4934640384830032512 & $-5.180$ & 11.778   &  0.65  & 521.7 & $4.73^{+0.07}_{-0.07}$ & 1.11\\
      4935140284663674368 & $-5.130$ & 12.936   &  0.64  & 1188.7 & $7.92^{+0.19}_{-0.23}$ & 1.44 \\ 
      4999449100568563584 & $-5.056$ & 12.393   &  0.69  & 718.1 & $4.54^{+0.07}_{-0.10}$ & 1.35 \\
      5006585382004810752 & $-5.142$ & 12.672   &  0.68  & 1176.6 & $9.44^{+0.37}_{-0.37}$ & 1.49 \\ 
      5014635662546524416 & $-5.168$ & 12.656   &  0.68  & 1000.4 & $7.66^{+0.44}_{-0.44}$ & 1.35 \\ 
      5035014766969503488 & $-5.065$ & 11.836   &  0.68  & 615.3 & $5.72^{+0.14}_{-0.13}$ & 1.43 \\ 
      5066899264145265792 & $-5.149$ & 11.511   &  0.65  & 533.5 & $7.84^{+0.21}_{-0.21}$ & 1.40 \\
      5144335875224357888 & $-4.998$ & 12.927   &  0.58  & 1060.3 & $4.45^{+0.01}_{-0.01}$ & 1.47 \\ 
      5548976735730236928 & $-4.928$ & 11.110   &  0.58  & 210.1 & $5.57^{+0.15}_{-0.17}$ & 1.82 \\ 
       \bottomrule
   \end{tabular}
\label{tab:goldsampleproperties}
\end{table*}

\begin{table*}[h]
      \caption{Radial velocity parameters of the ``Gold sample" stars}
   \begin{tabular}{@{} ccccccccc @{}} 
      \toprule
      Gaia DR3 ID & N$_{obs}$ & $\overline{RV}$& $\overline{\sigma_{RV}}$& $RV_{\rm rms}$ & CCF FWHM & FWHM Error \\
      & & (km.s$^{-1}$) & (m.s$^{-1}$) & (m.s$^{-1}$) & (km.s$^{-1}$) & (m.s$^{-1}$) \\
      \midrule
      1271212742256948096 & 3 & 11.482 & 2.243 & 32.757 & 10.198 & 4.485  \\ 
      1280137787375425664 & 4 & -3.554 & 1.150 & 3.858  & 8.891 & 2.141 \\ 
      1466420242809206656 & 10 & 3.600 & 0.808 & 8.704 & 9.265 &  1.616 \\
      2316357673270019584 & 8 & 12.188 & 0.637 & 10.881 & 7.610 & 1.275  \\
      2328985804833355008 & 8 & $-8.288$ & 0.892 & 14.617 & 10.688 & 1.785 \\
      2393625376170980608 & 7 & 24.166 & 0.713 & 3.566 & 8.165 & 1.426 \\
      2428155568905528064 & 8 & 1.776 & 0.814 & 6.729  & 10.204 & 1.628 \\
      2443486781086490624 & 10 & 5.547 & 0.572 & 6.232 & 9.000 & 1.145 \\
      2522660418675058944 & 6 & 14.097 & 0.917 & 4.920 & 10.213 & 1.834  \\
      3629413601830054400 & 9 & $-8.709$ & 2.473 & 3.861 & 13.248 & 4.947 \\
      3660958006315220992 & 7 & $-29.197$ & 1.672 & 3.146 & 9.151 & 3.343  \\
      3712163426756206208 & 8 & 23.194 & 1.682 & 12.598 & 8.886 & 3.364  \\
      3716520933071728896 & 7 & $-38.175$ & 1.458 & 1.707 & 8.841 & 2.917  \\
      3898946812815348096 & 7 & 6.142 & 1.330 & 2.071  & 10.217 & 2.661 \\
      3901835715193571456 & 9 & 31.081 & 0.897 & 15.511 & 8.935 & 1.794  \\
      3911409918850971392 & 8 & 31.392 & 0.812 & 3.918 & 8.367 & 1.626 \\
      3918910714160355584 & 6 & $-9.696$ & 0.948 & 2.851 & 8.177 & 1.896 \\
      3932181166672666240 & 6 & 12.489 & 1.600 & 4.698 & 8.417 & 3.200  \\
      3940893456293439488 & 9 & 6.998 & 1.709 & 11.119 & 9.271 & 3.267  \\
      3945839407488324224 & 5 & 20.658 & 1.630 & 27.293 & 8.644 & 3.260  \\
      3947104223817128320 & 6 & $-16.971$ & 2.354 & 6.463 & 9.236 & 4.708  \\
      3952789626645417216 & 7 & $-5.589$ & 0.832 & 5.433 & 8.090 & 1.665  \\
      3954703017395718016 & 7 & $-18.710$ & 0.701 & 2.999 & 8.618 & 1.402  \\
      3980392449448341888 & 7 & $-2.040$ & 0.984 & 8.092 & 9.461 & 1.788 \\
      3995373707693411072 & 11 & 21.014 & 0.798 & 13.325 & 8.447 & 1.595  \\
      4017904350214299520 & 8 & $-17.645$ & 0.780 & 12.189 & 8.147 & 1.560 \\
      4031060457812768768 & 5 & 11.648 & 1.089 & 10.744 & 9.197 & 2.177  \\ 
      4772734377062372224 & 5 & 23.436 & 1.079 & 5.607 & 11.234 & 2.158  \\ 
      4934640384830032512 & 8 & 15.497 & 0.921 & 8.247 & 10.258 & 1.843 \\
      4935140284663674368 & 5 & $-35.232$ & 0.825 & 4.196 & 10.473 & 1.649 \\ 
      4999449100568563584 & 7 & 14.393 & 0.674 & 5.882 & 7.938 & 1.349 \\
      5006585382004810752 & 5 & 17.564 & 0.862 & 36.307 & 8.944 & 1.723  \\ 
      5014635662546524416 & 5 & 17.982 & 0.730 & 6.434 & 9.756 & 1.459 \\ 
      5035014766969503488 & 3 & $-22.660$ & 0.611 & 7.431 & 7.903 & 1.223 \\ 
      5066899264145265792 & 7 & 6.904 & 0.780 & 3.388 & 9.886 & 1.559 \\
      5144335875224357888 & 3 & 50.837 & 0.882 & 32.801 & 8.237 & 1.765 \\ 
      5548976735730236928 & 3 & 59.726 & 2.437 & 41.334 & 20.752 & 4.875  \\ 
       \bottomrule
   \end{tabular}
\label{tab:RVparamsgold}   
\end{table*}

\section{Stellar Parameters}
\label{sec:stellarparameters}

\begin{deluxetable*}{lcccccc}
\tablecaption{Photospheric and Fundamental Stellar Parameters. The Table is also available, in its entirety, online}
\label{table:stellar_params}
\tablewidth{0pt}
\tablehead{
\colhead{Star} & \colhead{T$_{\rm eff}$} & \colhead{$\log g$} & \colhead{[Fe/H]} & \colhead{V$_{T}$} & \colhead{Mass}          & \colhead{Age ($\tau$)} \\ 
\colhead{}     & \colhead{(K)}       & \colhead{}       & (dex)            & (km.s$^{-1}$)     & \colhead{(M$_{\odot}$)} & \colhead{(Gyr)} } 
\startdata
Gaia DR3 2393625376170980608 & $5966^{+51}_{-32}$ & $3.76^{+0.02}_{-0.01}$ & $-0.51\pm{0.040}$ & $1.26\pm{0.05}$ & $1.21^{+0.03}_{-0.02}$ & $4.10^{+0.16}_{-0.19}$ \\ 
Gaia DR3 5066899264145265792 & $5950^{+51}_{-28}$ & $3.73^{+0.01}_{-0.01}$ & $-0.20\pm{0.041}$ & $1.27\pm{0.05}$ & $1.39^{+0.02}_{-0.02}$ & $3.02^{+0.09}_{-0.09}$ \\ 
Gaia DR3 4935140284663674368 & $5976^{+53}_{-29}$ & $3.73^{+0.01}_{-0.01}$ & $-0.16\pm{0.043}$ & $1.27\pm{0.05}$ & $1.40^{+0.02}_{-0.02}$ & $2.99^{+0.08}_{-0.09}$ \\ 
Gaia DR3 2522660418675058944 & $5873^{+50}_{-31}$ & $3.74^{+0.02}_{-0.01}$ & $-0.26\pm{0.043}$ & $1.27\pm{0.05}$ & $1.35^{+0.02}_{-0.02}$ & $3.30^{+0.12}_{-0.12}$ \\ 
Gaia DR3 3712163426756206208 & $5904^{+50}_{-33}$ & $3.83^{+0.02}_{-0.02}$ & $-0.16\pm{0.042}$ & $1.25\pm{0.05}$ & $1.29^{+0.06}_{-0.06}$ & $4.09^{+0.25}_{-0.22}$ \\ 
Gaia DR3 3629413601830054400 & $5977^{+53}_{-30}$ & $3.73^{+0.01}_{-0.01}$ & $-0.12\pm{0.044}$ & $1.27\pm{0.05}$ & $1.39^{+0.02}_{-0.02}$ & $2.99^{+0.09}_{-0.09}$ \\ 
Gaia DR3 2443486781086490624 & $5791^{+72}_{-47}$ & $3.73^{+0.02}_{-0.02}$ & $-0.32\pm{0.059}$ & $1.27\pm{0.05}$ & $1.27^{+0.03}_{-0.02}$ & $3.77^{+0.17}_{-0.17}$ \\ 
Gaia DR3 1280137787375425664 & $5994^{+44}_{-33}$ & $3.83^{+0.03}_{-0.02}$ & $0.01\pm{0.041}$ & $1.25\pm{0.05}$ & $1.31^{+0.08}_{-0.06}$ & $3.81^{+0.25}_{-0.25}$ \\ 
Gaia DR3 4772734377062372224 & $6156^{+40}_{-29}$ & $3.77^{+0.02}_{-0.02}$ & $-0.14\pm{0.034}$ & $1.25\pm{0.05}$ & $1.32^{+0.05}_{-0.05}$ & $3.24^{+0.16}_{-0.12}$ \\ 
Gaia DR3 3911409918850971392 & $5890^{+66}_{-40}$ & $3.75^{+0.02}_{-0.02}$ & $-0.44\pm{0.052}$ & $1.01\pm{0.05}$ & $1.25^{+0.05}_{-0.05}$ & $3.89^{+0.28}_{-0.21}$ \\ 
Gaia DR3 5142451621532291840 & $5850^{+64}_{-36}$ & $3.80^{+0.02}_{-0.02}$ & $-0.17\pm{0.054}$ & $0.97\pm{0.08}$ & $1.30^{+0.03}_{-0.03}$ & $3.94^{+0.22}_{-0.20}$ \\ 
Gaia DR3 5014635662546524416 & $5917^{+46}_{-29}$ & $3.81^{+0.02}_{-0.02}$ & $-0.12\pm{0.046}$ & $1.01\pm{0.10}$ & $1.33^{+0.05}_{-0.05}$ & $3.73^{+0.17}_{-0.16}$ \\ 
Gaia DR3 5144335875224357888 & $5926^{+61}_{-40}$ & $3.78^{+0.02}_{-0.02}$ & $-0.36\pm{0.051}$ & $0.79\pm{0.09}$ & $1.24^{+0.03}_{-0.03}$ & $4.10^{+0.24}_{-0.24}$ \\ 
Gaia DR3 4999449100568563584 & $5869^{+57}_{-37}$ & $3.88^{+0.02}_{-0.02}$ & $-0.22\pm{0.051}$ & $0.92\pm{0.09}$ & $1.20^{+0.06}_{-0.06}$ & $5.11^{+0.44}_{-0.35}$ \\ 
Gaia DR3 4934640384830032512 & $5997^{+35}_{-30}$ & $3.96^{+0.02}_{-0.04}$ & $0.07\pm{0.048}$ & $0.99\pm{0.12}$ & $1.40^{+0.03}_{-0.10}$ & $3.39^{+1.01}_{-0.35}$ \\ 
Gaia DR3 3660958006315220992 & $5973^{+53}_{-33}$ & $3.78^{+0.02}_{-0.02}$ & $-0.07\pm{0.047}$ & $1.21\pm{0.08}$ & $1.37^{+0.05}_{-0.04}$ & $3.38^{+0.11}_{-0.13}$ \\ 
Gaia DR3 3901835715193571456 & $5932^{+52}_{-30}$ & $3.85^{+0.02}_{-0.02}$ & $-0.34\pm{0.045}$ & $0.93\pm{0.09}$ & $1.16^{+0.03}_{-0.03}$ & $5.18^{+0.30}_{-0.33}$ \\ 
Gaia DR3 2328985804833355008 & $5908^{+49}_{-26}$ & $3.85^{+0.02}_{-0.02}$ & $0.09\pm{0.052}$ & $1.08\pm{0.12}$ & $1.35^{+0.02}_{-0.02}$ & $3.90^{+0.14}_{-0.17}$ \\ 
Gaia DR3 2428155568905528064 & $6017^{+82}_{-36}$ & $3.87^{+0.02}_{-0.01}$ & $-0.02\pm{0.062}$ & $1.08\pm{0.09}$ & $1.31^{+0.04}_{-0.02}$ & $3.97^{+0.13}_{-0.28}$ \\ 
Gaia DR3 3898946812815348096 & $5977^{+56}_{-28}$ & $3.94^{+0.02}_{-0.02}$ & $-0.06\pm{0.055}$ & $0.76\pm{0.15}$ & $1.22^{+0.02}_{-0.02}$ & $4.94^{+0.21}_{-0.29}$ \\ 
Gaia DR3 3980392449448341888 & $5805^{+56}_{-36}$ & $3.81^{+0.02}_{-0.02}$ & $0.01\pm{0.063}$ & $0.79\pm{0.16}$ & $1.30^{+0.05}_{-0.06}$ & $4.10^{+0.24}_{-0.17}$ \\ 
Gaia DR3 3940893456293439488 & $5840^{+80}_{-46}$ & $3.79^{+0.02}_{-0.02}$ & $0.01\pm{0.069}$ & $1.11\pm{0.10}$ & $1.33^{+0.06}_{-0.06}$ & $3.74^{+0.25}_{-0.21}$ \\ 
Gaia DR3 1466420242809206656 & $5983^{+42}_{-32}$ & $3.85^{+0.05}_{-0.03}$ & $0.13\pm{0.056}$ & $0.90\pm{0.15}$ & $1.33^{+0.18}_{-0.07}$ & $3.83^{+0.24}_{-1.09}$ \\ 
Gaia DR3 1271212742256948096 & $5963^{+149}_{-94}$ & $3.81^{+0.03}_{-0.03}$ & $-0.20\pm{0.112}$ & $0.98\pm{0.11}$ & $1.28^{+0.07}_{-0.07}$ & $3.91^{+0.35}_{-0.37}$ \\ 
Gaia DR3 4031060457812768768 & $5911^{+139}_{-91}$ & $3.88^{+0.03}_{-0.03}$ & $-0.12\pm{0.113}$ & $0.66\pm{0.18}$ & $1.19^{+0.06}_{-0.07}$ & $5.11^{+0.61}_{-0.63}$ \\ 
Gaia DR3 3995373707693411072 & $6053^{+123}_{-89}$ & $3.91^{+0.03}_{-0.03}$ & $-0.21\pm{0.092}$ & $1.02\pm{0.07}$ & $1.20^{+0.07}_{-0.08}$ & $4.83^{+0.69}_{-0.67}$ \\ 
Gaia DR3 5035014766969503488 & $5931^{+145}_{-93}$ & $3.75^{+0.03}_{-0.03}$ & $-0.50\pm{0.105}$ & $0.84\pm{0.10}$ & $1.14^{+0.06}_{-0.06}$ & $4.72^{+0.54}_{-0.50}$ \\ 
Gaia DR3 3932181166672666240 & $5739^{+102}_{-65}$ & $3.79^{+0.03}_{-0.03}$ & $-0.36\pm{0.079}$ & $0.85\pm{0.06}$ & $1.22^{+0.06}_{-0.07}$ & $4.60^{+0.43}_{-0.31}$ \\ 
Gaia DR3 3918910714160355584 & $5993^{+150}_{-137}$ & $3.78^{+0.03}_{-0.03}$ & $-0.49\pm{0.116}$ & $0.99\pm{0.07}$ & $1.17^{+0.06}_{-0.06}$ & $4.50^{+0.62}_{-0.55}$ \\ 
Gaia DR3 4017904350214299520 & $6049^{+132}_{-92}$ & $3.70^{+0.03}_{-0.03}$ & $-0.47\pm{0.094}$ & $1.00\pm{0.09}$ & $1.25^{+0.06}_{-0.07}$ & $3.43^{+0.39}_{-0.30}$ \\ 
Gaia DR3 3947104223817128320 & $6022^{+148}_{-103}$ & $3.72^{+0.03}_{-0.03}$ & $-0.21\pm{0.108}$ & $1.01\pm{0.08}$ & $1.31^{+0.07}_{-0.07}$ & $3.25^{+0.31}_{-0.25}$ \\ 
Gaia DR3 2316357673270019584 & $5853^{+133}_{-93}$ & $3.86^{+0.04}_{-0.03}$ & $-0.31\pm{0.104}$ & $0.95\pm{0.05}$ & $1.09^{+0.08}_{-0.07}$ & $6.22^{+0.75}_{-0.86}$ \\ 
Gaia DR3 3716520933071728896 & $6253^{+149}_{-148}$ & $3.74^{+0.05}_{-0.03}$ & $-0.19\pm{0.115}$ & $1.37\pm{0.05}$ & $1.37^{+0.14}_{-0.07}$ & $2.80^{+0.24}_{-0.44}$ \\ 
Gaia DR3 3952789626645417216 & $6003^{+126}_{-91}$ & $3.82^{+0.03}_{-0.03}$ & $-0.34\pm{0.093}$ & $1.05\pm{0.05}$ & $1.16^{+0.07}_{-0.07}$ & $4.76^{+0.59}_{-0.55}$ \\ 
Gaia DR3 5548976735730236928 & $6316^{+161}_{-144}$ & $3.82^{+0.06}_{-0.02}$ & $-0.12\pm{0.121}$ & $1.55\pm{0.05}$ & $1.35^{+0.15}_{-0.04}$ & $3.07^{+0.29}_{-0.92}$ \\ 
Gaia DR3 3945839407488324224 & $5962^{+131}_{-106}$ & $3.76^{+0.03}_{-0.03}$ & $-0.21\pm{0.106}$ & $1.06\pm{0.05}$ & $1.27^{+0.06}_{-0.07}$ & $3.73^{+0.36}_{-0.29}$ \\ 
Gaia DR3 3954703017395718016 & $5978^{+145}_{-108}$ & $3.83^{+0.03}_{-0.03}$ & $-0.12\pm{0.113}$ & $1.05\pm{0.05}$ & $1.25^{+0.07}_{-0.07}$ & $4.24^{+0.46}_{-0.46}$ \\ 
Gaia DR3 5006585382004810752 & $6108^{+151}_{-95}$ & $3.67^{+0.04}_{-0.03}$ & $-0.13\pm{0.106}$ & $1.09\pm{0.05}$ & $1.44^{+0.09}_{-0.07}$ & $2.48^{+0.16}_{-0.19}$ \\ 
Gaia DR3 2314056743326275840 & $6115^{+104}_{-74}$ & $3.76^{+0.03}_{-0.03}$ & $-0.48\pm{0.075}$ & $0.96\pm{0.05}$ & $1.20^{+0.06}_{-0.06}$ & $3.97^{+0.42}_{-0.35}$ \\ 
Gaia DR3 1282810871941219968 & $5955^{+128}_{-85}$ & $3.77^{+0.03}_{-0.03}$ & $-0.28\pm{0.097}$ & $1.06\pm{0.05}$ & $1.23^{+0.07}_{-0.07}$ & $4.08^{+0.44}_{-0.37}$ 
\enddata

\end{deluxetable*}

We derive photospheric and fundamental stellar parameters for our program stars using the algorithm described in \citet{reggiani2021,reggiani2022} that makes use of both an spectroscopic approach\footnote{The classical spectroscopic approach simultaneously minimizes line-by-line iron abundance inference difference between \ion{Fe}{1} \& \ion{Fe}{2}-based abundances as well as their dependencies on excitation potential and reduced equivalent widths to infer T$_{\text{eff}}$, log$g$, and [Fe/H].} and isochrones to infer accurate, precise, and self-consistent photospheric (T$_{\text{eff}}$, log$g$, and [Fe/H]) and fundamental (mass, luminosity, and radius) stellar parameters.  

For our isochrone fitting we use high-quality multiwavelength photometry: Gaia DR3 G \citep{gaia2016,gaia2018,arenou2018,evans2018,hambly2018,riello2018, gaia2021,fabricius2021,lindegren2021a,lindegren2021b,torra2021}, J, H, and Ks bands from the Two Micron All Sky Survey (2MASS) All-Sky Point Source Catalog \citep[PSC,][]{skrutskie2006}, and W1 and W2 bands from the Wide-field Infrared Survey Explorer (WISE) AllWISE mid-infrared data \citep{wright2010,mainzer2011}, and SkyMapper DR4  \textit{u, v, g, r, i, and z} magnitudes \citep{Onken:2024PASA...41...61O}. The included data is based on availability and on data flag-based cuts, as detailed in \cite{Nataf:2024ApJ...976...87N}. We also include the Gaia DR3-based \textit{geophotometric} distances from \citep{bailer-jones2021} of our targets in our priors.  Finally, we include extinction $A_V$ inferences based on the \textit{Bayestar} extinction \cite{Green:2019ApJ...887...93G}.

For the spectroscopic-based inferences ([Fe/H], and microturbulent velocities, $V_{\rm T}$) we use the equivalent widths (EWs) of \ion{Fe}{1} and \ion{Fe}{2} atomic absorption lines. The EWs were measured from our ESPRESSO spectra using Gaussian profiles with the \texttt{REvIEW} semi-automated code \cite{McKenzie:2022MNRAS.516.3515M}. All EWs were individually reviewed, and bad fits were removed from the analysis. Abundances were calculated using the 1D local themodynamic equilibrium radiative transfer code MOOG \citep{Sneden:1973PhDT.......180S}, using plane-parallel MARCS model atmospheres \citep{Gustafsson:2008A&A...486..951G}, and adopting \cite{asplund2021} solar abundances.

As described in detail in \citet{reggiani2022}, we use the \texttt{isochrones} package\footnote{\url{https://github.com/timothydmorton/isochrones}} \citep{morton2015} to fit the MESA Isochrones and Stellar Tracks \cite[MIST;][]{dotter2016,choi2016,paxton2011,paxton2013,paxton2015,paxton2018,paxton2019} library to our photospheric stellar parameters as well as our input multiwavelength photometry, parallax, and extinction data using \texttt{MultiNest}\footnote{\url{https://ccpforge.cse.rl.ac.uk/gf/project/multinest/}} \citep{feroz2008,feroz2009,feroz2019} via \texttt{PyMultinest} \citep{buchner2014}.

Our adopted stellar parameters ($\rm{T_{eff}}$ and surface gravity from the isochrone analysis, and [Fe/H] and $\xi$ inferred from the atomic \ion{Fe}{1} and \ion{Fe}{2} lines) are in Table \ref{table:stellar_params}. All of the uncertainties quoted in Table \ref{table:stellar_params} include random uncertainties only.  That is, they are uncertainties derived under the unlikely assumption that the MIST isochrone grid we use in our analyses perfectly reproduces all stellar properties.  The derived stellar parameters of the gold sample are typical of sub-giant stars \citep{Bergeretal2020}. The average $T_{\rm eff}$ and $\rm logg$ values for the gold sample are 5966K and 3.8 respectively.  The average mass of these stars is 1.3 $M_{\odot}$, with an average age of 3.9 Gyr.  To roughly estimate where our gold sample stars fall on theoretical model grids, we have used these measured values for stellar parameters ($M_{\star}, T_{\rm eff}, \rm logg$, Fe/H and age) and interpolated these parameters within the \cite{VanSaders2013}'s subgiant evolutionary/rotation grids to derive an approximate model rotation period.  If our metallicities fall outside the model grids, the nearest model metallicity is used.  This indicates that our sample has a median inferred model period of approximately \(12.6\) days; 13 stars fell below 10 days, and 24 between 10 and 40 days.

\section{Stellar Companion Candidates}
\label{sec:stellarcomps}

\begin{table*}[h]
      \caption{Excluded Sources}
   \begin{tabular}{@{} ccccccccccc @{}} 
      \toprule
      Gaia DR3 & Reason for Removal & $\log R'_{\rm HK}$ & $V_{\rm mag}$ & $B-V$ & $\abs{z}$  (pc) \\
      \midrule
      1225731203253635584 & $RV_{RMS}$ $>$ 16 m.s$^{-1}$ - Exoplanet Candidate & -5.124 & 13.131 & 0.73 & 1027.4 \\ 
      1250840234901215744 & Binary (via CCF) & -4.904 & 12.852 & 0.74 & 1038.4 \\
      1257728537810125184 & Asymmetric CCF & -5.602 & 12.640 & 0.67 & 822.7 \\ 
      1282810871941219968 & $RV_{RMS}$ $>$ 100 m.s$^{-1}$ - Exoplanet Candidate & -5.076 & 11.844 & 0.68 & 614.5 \\
      1282813070964475392 & Spectroscopic Binary (via Gaia) & -5.375 & 11.532 & 0.66 & 512.8 \\
      1448931857534138112 & $RV_{RMS}$ $>$ 180 m.s$^{-1}$ - Exoplanet Candidate & -5.311 & 12.603 & 0.68 & 947.1 \\
      1455717081228013696 & CCF FWHM $>$ 12 km.s$^{-1}$ & -4.711 & 12.036 & 0.63 & 726.2 \\
      1469142530521510016 & CCF indicated stellar companion & -6.118 & 11.740 & 0.68 & 618.8 \\
      2314056743326275840 & $RV_{RMS}$ $>$ 1 km$^{-1}$ - Exoplanet Candidate & -5.311 & 12.603 & 0.68 & 947.1 \\
      2370995468366361344 & j $>$ 1.53 cutoff & -4.783 & 12.795 & 0.64 & 938.1 \\
      2735962096754761088 & CCF FWHM $>$ 30 km.s$^{-1}$ & -6.294 & 11.116 & 0.58 & 293.1 \\
      2810741192525443968 & Asymmetric CCF & -5.649 & 11.124 & 0.61 & 334.5 \\
      2903113569558464256 & CCF FWHM $>$ 25 km.s$^{-1}$ & -4.912 & 11.115 & 0.59 & 207.1 \\
      3698120056225966976 & Binary (via CCF) & -6.396 & 12.100 & 0.64 & 743.8 \\
      3736336640865721216 & Binary (via CCF) & -4.654 & 12.656 & 0.65 & 928.3 \\
      3903785286749087744 & $RV_{RMS}$ $>$ 800 m.s$^{-1}$ - Exoplanet Candidate & -5.301 & 12.765 & 0.68 & 966.0 \\
      3937811525200355328 & $RV_{RMS}$ $>$ 1.5 km.s$^{-1}$ - Exoplanet Candidate & -5.396 & 12.224 & 0.65 & 724.4 \\
      3946715718255175168 & $RV_{RMS}$ $>$ 7.5 km.s$^{-1}$ -Exoplanet Candidate & -5.408 & 12.661 & 0.68 & 894.5 \\
      4007351482424776576 & Binary (via CCF) & -6.531 & 12.713 & 0.65 & 990.1 \\
      4029722313506309760 & Binary (via CCF) & -5.965 & 12.665 & 0.66 & 1119.2 \\
      4264320784544595328 & High Stellar Activity (A- or B-type star) & -3.939 & 10.910 & 0.61 & 21.1 \\
      5001167366660075904 & Irregular CCF & -5.155 & 12.490 & 0.68 & 849.1 \\
      5015490635916742912 & CCF FWHM $>$ 120 km.s$^{-1}$ & -4.422 & 11.806 & 0.68 & 601.8 \\
      5040226172911337344 & Binary (via CCF) & -5.087 & 12.160 & 0.67 & 702.8 \\
      5040747242639330432 & Binary (via CCF) & -5.201 & 12.870 & 0.66 & 1011.8 \\
      5117683953885263360 & Binary (via CCF) & -4.660 & 10.900 & 0.61 & 500.6 \\
      5142271095466719744 & Binary (via CCF) & -5.290 & 12.659 & 0.69 & 829.9 \\
      5142451621532291840 & $RV_{RMS}$ $>$ 100 m.s$^{-1}$ - Exoplanet Candidate & -5.055 & 13.072 & 0.65 & 1100.8 \\
      5295916025702362368 & CCF indicated active spectral type & -4.741 & 11.116 & 0.61 & 93.7 \\
       \bottomrule
   \end{tabular}
\label{tab:excludedsources}   
\end{table*}

After eliminating binaries and sources with unusual CCFs from the initial sample, eight sources exhibited radial velocity variations with scatter $>$ 15 m$^{-1}$ over a minimum of 5 epochs, which indicated potential companions.  Information about these sources and other sources removed from the gold sample are available in Table \ref{tab:excludedsources}. Although these stars did not display double-peaked CCF diagrams during the reconnaissance survey period, the observed changes in the measured RV values indicate massive companions. Of these eight sources, we derive tentative constraints on the orbital parameters and mass of the companion for four sources, as described in  Section \ref{sec:companionswithorbit} and the accompanying Table \ref{orbit_parameters_table}. For the remaining four sources, our observations do not capture enough of the orbit to make well-informed predictions of orbital parameters and companion mass, but we provide values for acceleration and curvature terms in Table \ref{incomplete_orbit_table}.   

Characterizing exoplanets via analysis of RV measurements has a storied history dating back to the discovery of the first exoplanet orbiting a Sun-like star in \cite{1995Natur.378..355M}. In the decades since, the RV precision of spectrographs has significantly increased, leading to this paper's main aim of measuring the Galactic acceleration of low-jitter stars with ESPRESSO.  As we discuss below, these observations have also enabled an initial characterization of eight stellar companion candidates.  Our observations have measured a full orbital cycle for four sources for which we provide some constraints on their orbital parameters.  For the other four sources (likely longer-period) for which our observations did not track a full orbital cycle, we present a linear acceleration term and higher order curvature term.  We characterize stellar companions using the binary mass function $f(M)$:

\begin{equation}
f(M)= \frac {(M \sin i)^{3}}{(M + M_\star)^{2}} =  K^3 \frac{P_{orb}}{2\pi G}(1-e^{2})^{3/2}   
\label{binary-mass-function}
\end{equation}

where $K$ refers to the RV semi-amplitude, $P_{\rm orb}$ is the orbital period of the companion, and $e$ is the eccentricity of the orbit. $M$ refers to the mass of the companion while $M_\star$ refers to the observed star. 
Because RV observations alone do not constrain the orbital inclination \(i\), the true companion mass cannot be uniquely determined. Instead, the RV orbit constrains the mass function, from which we derive the minimum companion mass (corresponding to \(i=90^\circ\)). The determination of the companion mass also depends on the mass of the primary star. Our high-resolution ESPRESSO spectra enable precise measurements of the photospheric and fundamental stellar parameters of the targets, including stellar masses with typical uncertainties of order \(0.1\,M_\odot\) or better (Table \ref{table:stellar_params}).
The uncertainty in the companion mass comes mainly from uncertainties in the derived orbits. Further observations could refine these orbital predictions.


\subsection{Companions with Derived Orbits}
\label{sec:companionswithorbit}

\begin{deluxetable*}{lccccc}
\tablecaption{Derived Orbital Parameters for Potential Companions}
\label{orbit_parameters_table}
\tablewidth{0pt}
\tablehead{
\colhead{Gaia DR3 ID} & \colhead{Period ($P$)} & \colhead{Eccentricity ($e$)} &  \colhead{Semi-amplitude ($K$)} & \colhead{Systemic velocity ($v_{0}$)} & \colhead{$M\sin i$} \\   
\colhead{}     & \colhead{(days)}              &  \colhead{}  &\colhead{(km.s$^{-1}$)} & \colhead{(km.s$^{-1}$)} &   \colhead{(M$_{J}$)}}
\startdata
1225731203253635584* & 133$^{+10.54}_{-2.94}$ & 0.49$^{+0.186}_{-0.156}$&  0.031$^{+0.017}_{-0.008}$& $-3.373^{+0.006}_{-0.006}$ & 0.58$^{+0.26}_{-0.15}$ \\ 
3903785286749087744 & 1238$^{+106.5}_{-62.8}$ & 0.432$^{+0.032}_{-0.012}$&  4.203$^{+0.157}_{-0.037}$& $-21.62^{+0.228}_{-0.202}$ &  267.9$^{+14.6}_{-4.06}$ \\ 
3946715718255175168 & 306.1$^{+0.432}_{-0.406}$ & 0.33$^{+0.003}_{-0.003}$&  3.733$^{+0.016}_{-0.014}$& $-9.25^{+0.024}_{-0.023}$ &  132.75$^{+0.77}_{-0.67}$ \\ 
5142451621532291840 & 262.2$^{+101.6}_{-74.89}$ & 0.39$^{+0.10}_{-0.22}$&  0.375$^{+0.114}_{-0.117}$& $-39.58^{+0.062}_{-0.228}$ &  13.27$^{+3.47}_{-5.47}$ \\ 
\enddata
\tablecomments{*Refer to subsection \ref{sec:stellarcomps}.1.1 for a more detailed discussion of this source's ambiguity.}
\end{deluxetable*}

\begin{figure}[!h]
    \centering
    \includegraphics[width=0.33\textwidth]{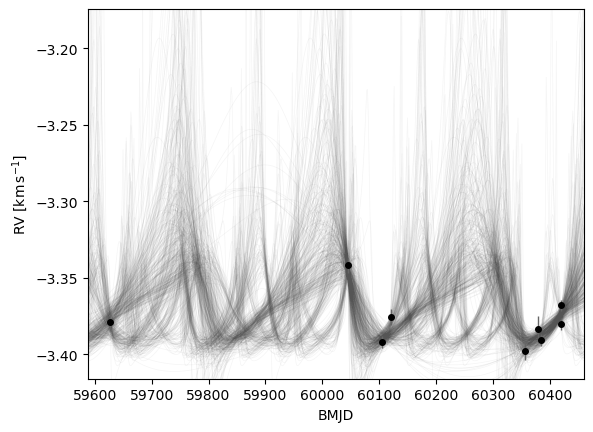}%
    \\

    \includegraphics[width=0.33\textwidth]{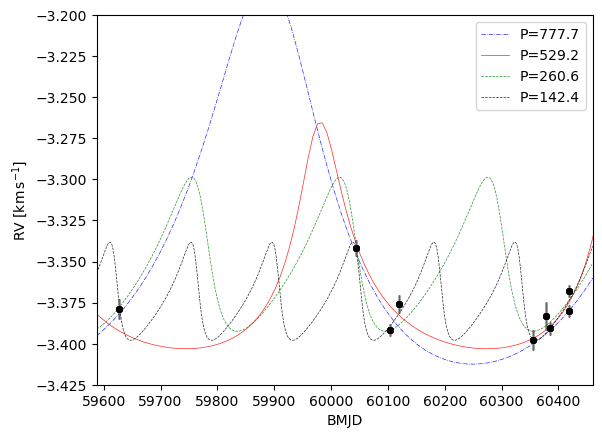}%
    \\ \includegraphics[width=0.33\textwidth]{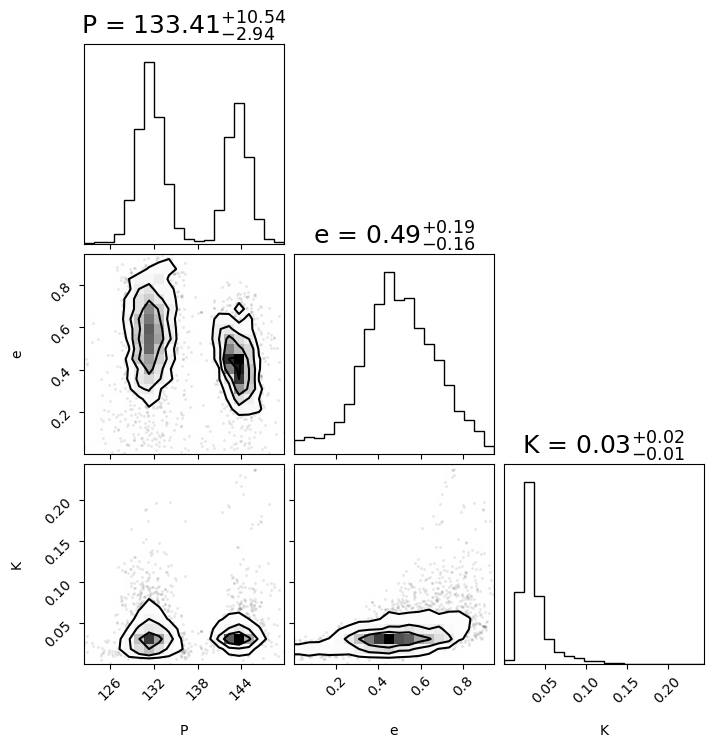}
    \\ \includegraphics[width=0.30\textwidth]{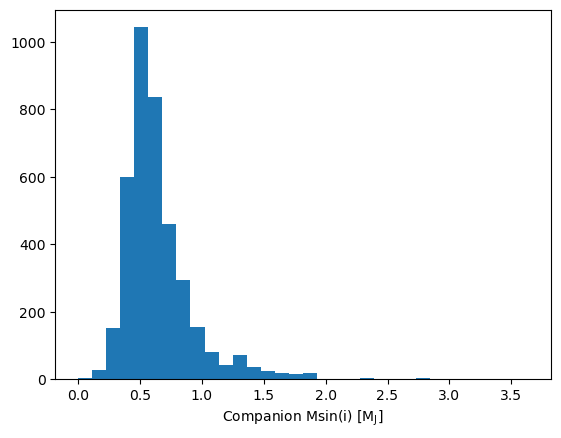}
    \caption{Analysis of Gaia DR3 1225731203253635584.  RV curves from \textit{TheJoker}'s rejection sampler, which indicates a multitude of prominent orbital solutions (top). The formal errors on the RVs are plotted, but too small to be seen. A clear look at the four period aliases (upper middle). Corner plots showing the posterior distributions for the orbital period, eccentricity, and RV semi-amplitude (lower middle). Posterior distribution of $Msin(i)$ (bottom).}
    \label{fig:122OrbitPlots}
\end{figure}

\begin{figure}[!h]
    \centering
    \includegraphics[width=0.35\textwidth]{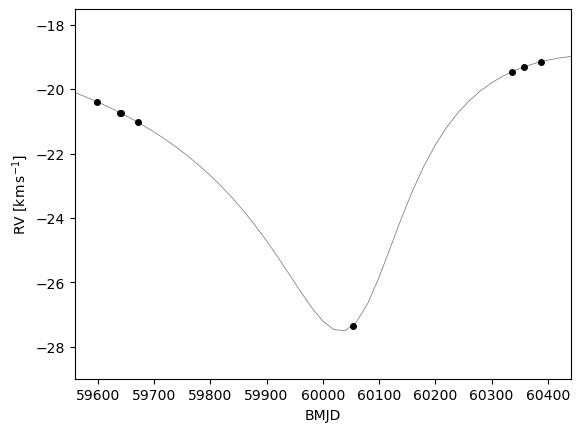}%
    \\ 

    \includegraphics[width=0.45\textwidth]{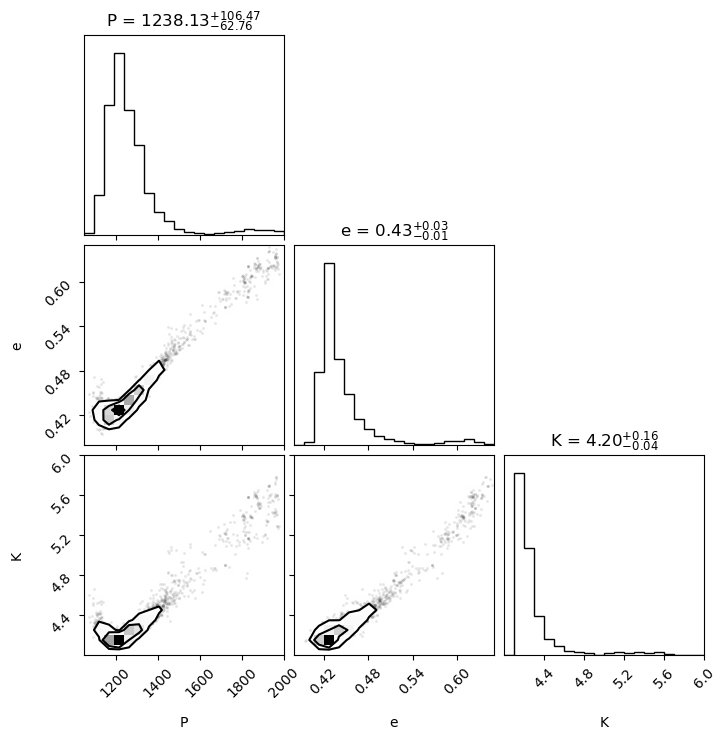}%
    \\ \includegraphics[width=0.35\textwidth]{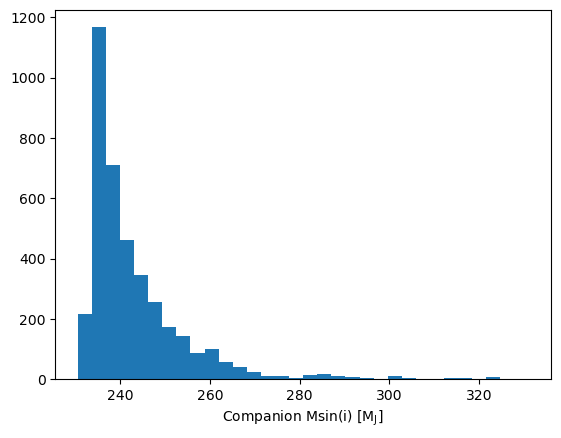}
    \caption{Analysis of Gaia DR3 3903785286749087744. RV curves from \textit{TheJoker}'s rejection sampler (top), which indicates a single initial orbital solution. The formal errors on the RVs are plotted, but too small to be seen. We show the corner plot from \textit{TheJoker}'s parameter exploration (middle) as well as a histogram with the $Msin(i)$ distribution from the posterior results (bottom).
        } 
    \label{fig:39037OrbitPlots}
\end{figure}

For four sources, our observations captured a full orbital cycle. For these sources, orbital solutions and their uncertainties are presented in Table \ref{orbit_parameters_table}, with the median of the posterior reported as the primary value and the 16th and 84th percentiles of the posterior as the lower and upper uncertainties. These orbital fits were determined using \textit{TheJoker} \citep{2017ApJ...837...20P}, a custom MCMC sampler that solves Kepler's equations given RV observations. Here, we analyze the RV data assuming a single companion.  For our use case, 10 million prior samples were generated for each source with periods ranging up to 2000 days.  Our sparse observations do not allow us to properly explore short period ($P <50$ days) orbits. We also incorporate a lognormal distribution for an additional error term (referred to as ``s") added in quadrature; RV standard stars traditionally have additional error terms on the scale of 2-3 m s$^{-1}$ for high-precision instruments, but as these stars have not been historically well-observed and are at distances larger than in typical EPRV surveys, we allowed this error parameter to explore higher magnitudes, up to 10-15 m s$^{-1}$. This is also due to the reported errors of each ESPRESSO measurement being extremely small, usually on the order of meters per second. The default prior is described in \cite{2017ApJ...837...20P}. After initializing the prior samples, \textit{TheJoker} enacts a rejection sampling step using a marginalized likelihood that will reject a number of the generated samples that fail to reach a likelihood threshold. Ideally, this process returns a smaller number of samples, but in the case that multiple possible orbits survive, the remaining orbits are intended to represent a true sampling of the posterior. Of the four sources we investigated using this software, only one of them returned multiple orbits after this rejection sampling step (Gaia DR3 1225731203253635584). Post-rejection sampling, the surviving orbit is used as an initializing basis for Markov Chain Monte Carlo (MCMC) parameter exploration.  Figures \ref{fig:122OrbitPlots}-\ref{fig:514OrbitPlots} show the results from our \textit{TheJoker} analysis, and below we describe the sources and orbital analysis in detail. 

\subsubsection{Gaia DR3 1225731203253635584}

Gaia DR3 1225731203253635584 (as shown in Figure \ref{fig:122OrbitPlots}) is the only star for which the rejection sampler returned more than one orbital solution (approximately 500), likely due to the induced RV shift of the companion being significantly smaller in amplitude than the other companion candidates. The rejection sampler results are in the top plot of Figure \ref{fig:122OrbitPlots}. The observed RV semi-amplitude K is only on the order of 50 m s$^{-1}$, on the lower threshold of sources we investigated for companions. The cadence of observations for this star coincide with phases of the companion's orbit such that the eccentricity and period are not well determined, even with a higher number of observations than the other sources with potential companions. Attempts to determine orbital parameters via TheJoker's implemented MCMC functions returned four distinct possible orbits of approximately equal likelihood with multiple period peaks appear to coincide with multiples of 130, with corresponding increases in RV semi-amplitude as the period becomes longer. The shortest period of $P = 142.4$ days had the least residual error compared to the other periods and required less assumptions of RV amplitude than the other periods. We therefore report the orbital parameters associated with this period in a localized MCMC search in Table \ref{orbit_parameters_table}, but more information is required to make a conclusive determination. Based on the results of the RV analysis as presented in Table \ref{orbit_parameters_table}, the companion for this star is most likely a Saturn- or Jupiter-mass exoplanet.  


\subsubsection{Gaia DR3 3903785286749087744}

Gaia DR3 3903785286749087744 is a more fortuitous example; our sparse observations have been effective at capturing local extrema for the companion's orbit. The eight observations shown in Figure \ref{fig:39037OrbitPlots} span three epochs, but neither the first nor the third batch of observations provide much evidence for a companion. It is the second epoch that displays a drastic RV shift from the other observations and informed the decision to further investigate for a companion. Only one orbit was returned from \textit{TheJoker}'s rejection sampler, which served as the basis for continued posterior exploration. The resulting parameter distributions are unimodal, albeit with a tail that stretches toward greater $P$, $e$, and $K$ values, which also informs the shape of the companion $M \sin i$ distribution. Notably, the mass is more consistent with a stellar body than a planet: a mass of 240 $M_{J}$ or 0.23 $M_{\odot}$ suggests the companion is likely a dim dwarf star, not an exoplanet. In that case, Gaia DR3 3903785286749087744 would be a binary star system.

\begin{figure}[!h]
    \centering
    \includegraphics[width=0.35\textwidth]{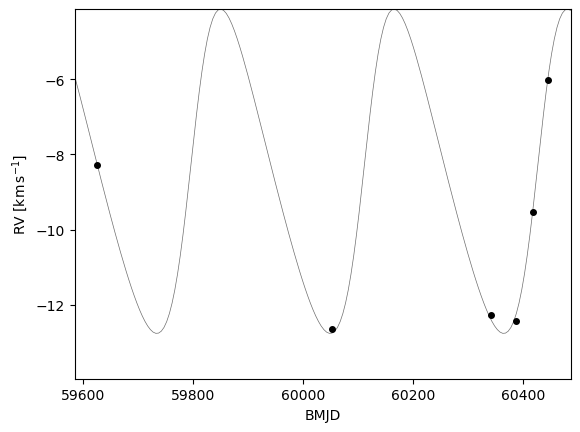}%
    \\

    \includegraphics[width=0.45\textwidth]{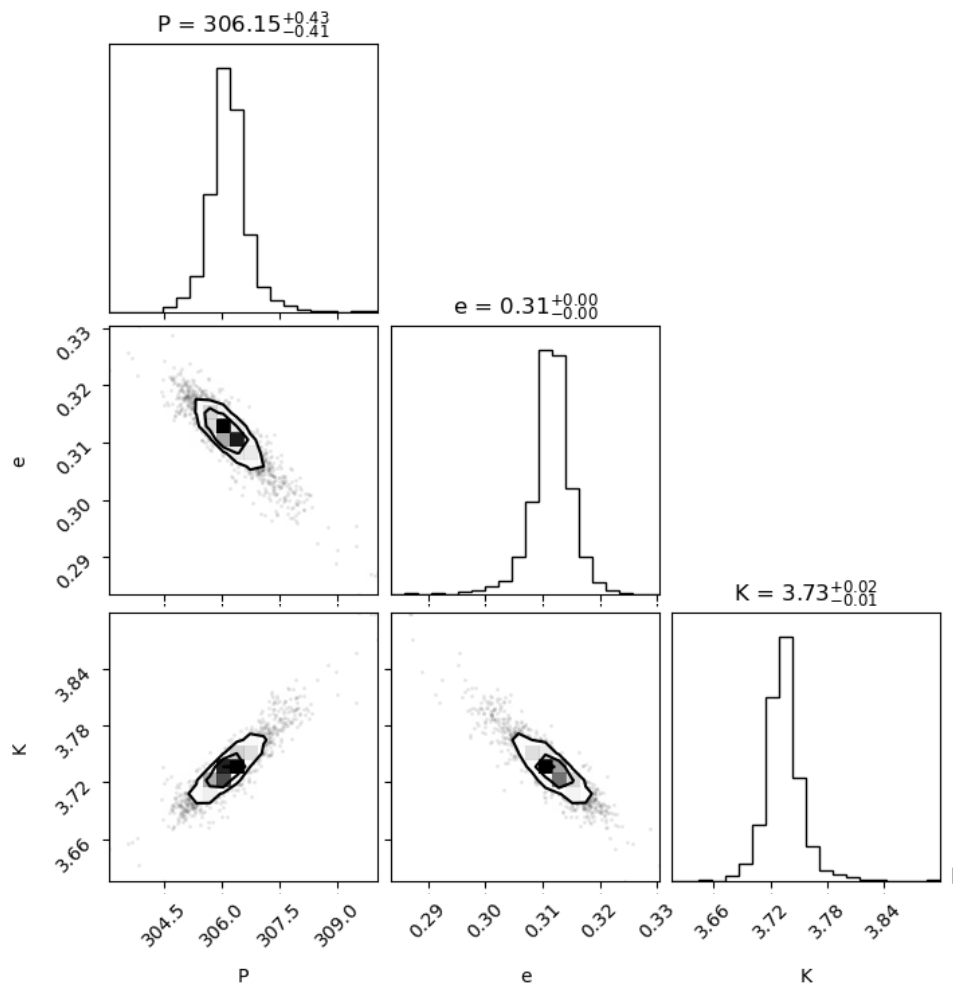}

    \includegraphics[width=0.35\textwidth]{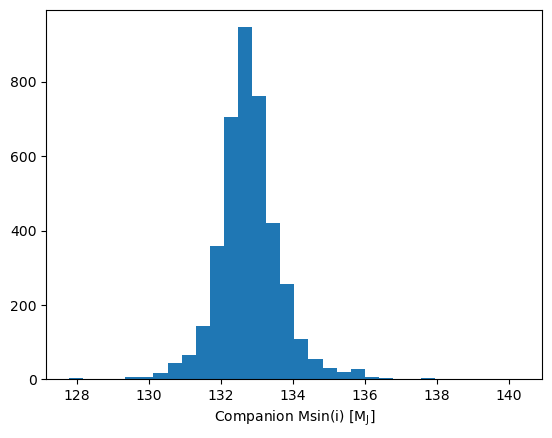}
    
    \caption{Analysis of Gaia DR3 3946715718255175168, beginning with \textit{TheJoker}'s rejection sampler, which indicates a single initial orbital solution (top). The formal errors on the RVs are plotted, but too small to be seen. We show the corner plots from \textit{TheJoker}'s refined search about the most likely period (middle) and the subsequent $Msin(i)$ distribution (bottom).}
    
    \label{fig:39467OrbitPlots}
\end{figure}

\subsubsection{Gaia DR3 3946715718255175168}

Gaia DR3 3946715718255175168, like the previously discussed Gaia DR3 3903785286749087744, is also likely to have a dim dwarf star companion as opposed to an exoplanet. Though the rejection sampler returned only one orbit, with a period of 310 days, an initial survey of the posterior pdf returned multimodal solutions of varying parameters except semi-amplitude, which also caused a multimodal distribution for companion mass. As the differing period peaks corresponded to different probabilities, with the highest likelihood belonging to the period corresponding to the initial rejection sampler, we refined the MCMC search around the most likely orbital configuration. This refined corner plot is shown in Figure \ref{fig:39467OrbitPlots} with the resulting orbital parameters included in Table \ref{orbit_parameters_table}. This indicates that this source has a dim stellar companion with a mass of 133 $M{_J}$ or 0.127 $M{_\odot}$. 

\begin{figure}[!h]
    \centering
    \includegraphics[width=0.35\textwidth]{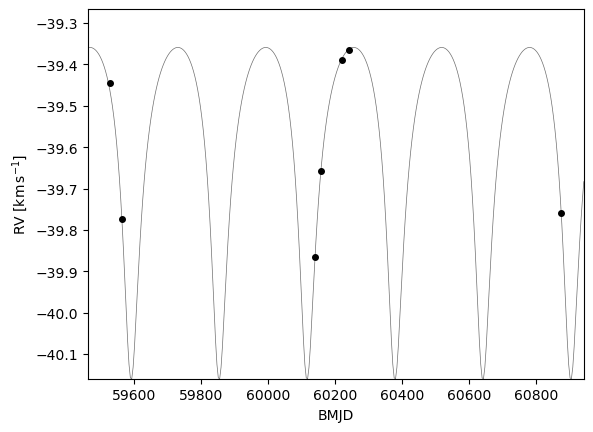}%

    \includegraphics[width=0.45\textwidth]{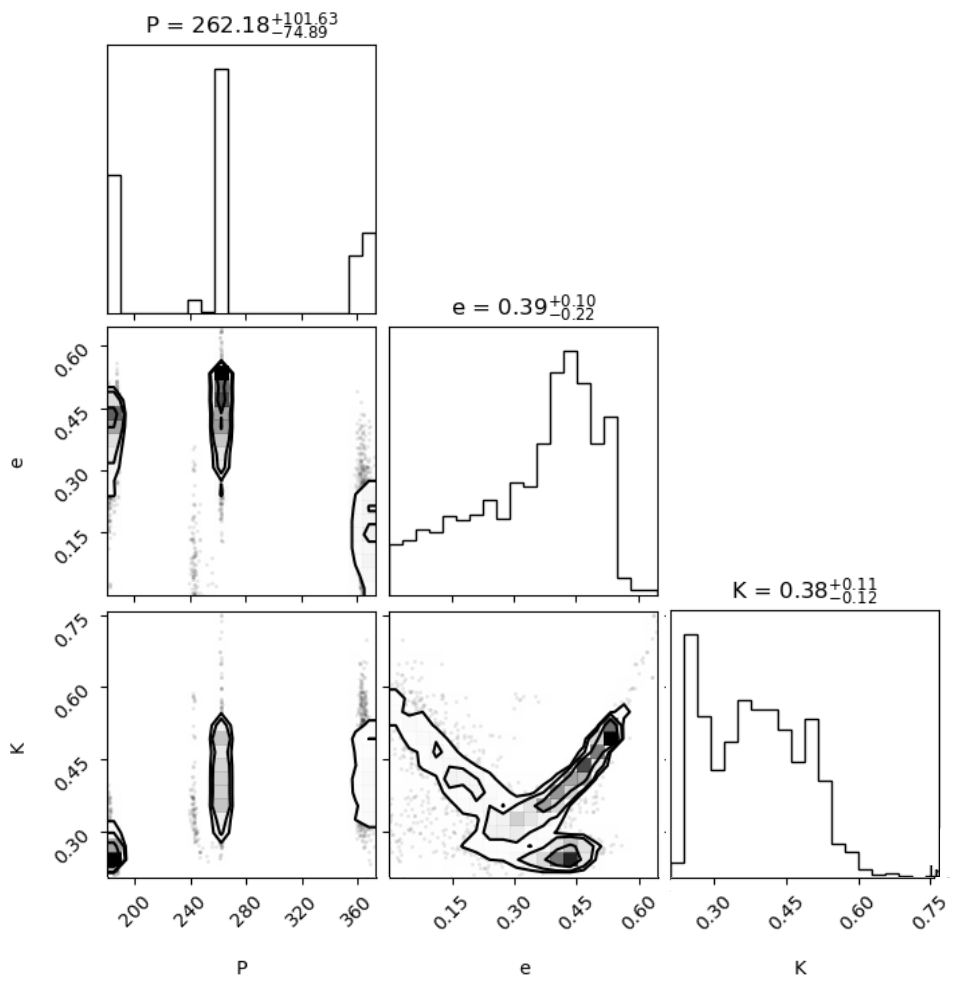}%
    \hfill \includegraphics[width=0.35\textwidth]{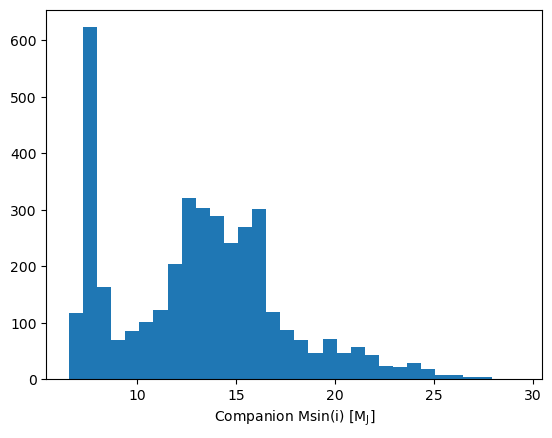}

    \caption{Analysis of Gaia DR3 5142451621532291840, beginning with \textit{TheJoker}'s rejection sampler, which indicates a single initial orbital solution (top). The formal errors on the RVs are plotted, but too small to be seen. We show the corner plot from \textit{TheJoker}'s parameter exploration (middle) as well as a histogram with the $Msin(i)$ distribution from the posterior results (bottom).  }
    \label{fig:514OrbitPlots}
\end{figure}

\subsubsection{Gaia DR3 5142451621532291840}

Gaia DR3 5142451621532291840 is projected to have a massive companion, potentially a giant exoplanet or brown dwarf. In analyzing the star with \textit{TheJoker}, only one sample was returned via the rejection sampler, however the MCMC-determined posterior distribution was significantly more varied, with a bimodal distribution in \textit{Msin(i)} centered around 4 Jupiter masses and 14 Jupiter masses. The MCMC chains tested four different period options without any sense of convergence, with the shortest period corresponding to the first peak in \textit{K}. More observations may help refine these orbital characterizations to resolve the degeneracies in period and the shape of the \textit{K} distribution. Figures \ref{fig:514OrbitPlots} showcase these results.  As with Gaia DR3 1225731203253635584, the final companion mass distribution is mostly defined by the distribution of \textit{K}. Though this companion is less well-characterized than the others, the observed RV behavior is still indicative of a massive companion.


\begin{figure}[h!]
    \centering
    \includegraphics[width=0.40\textwidth]{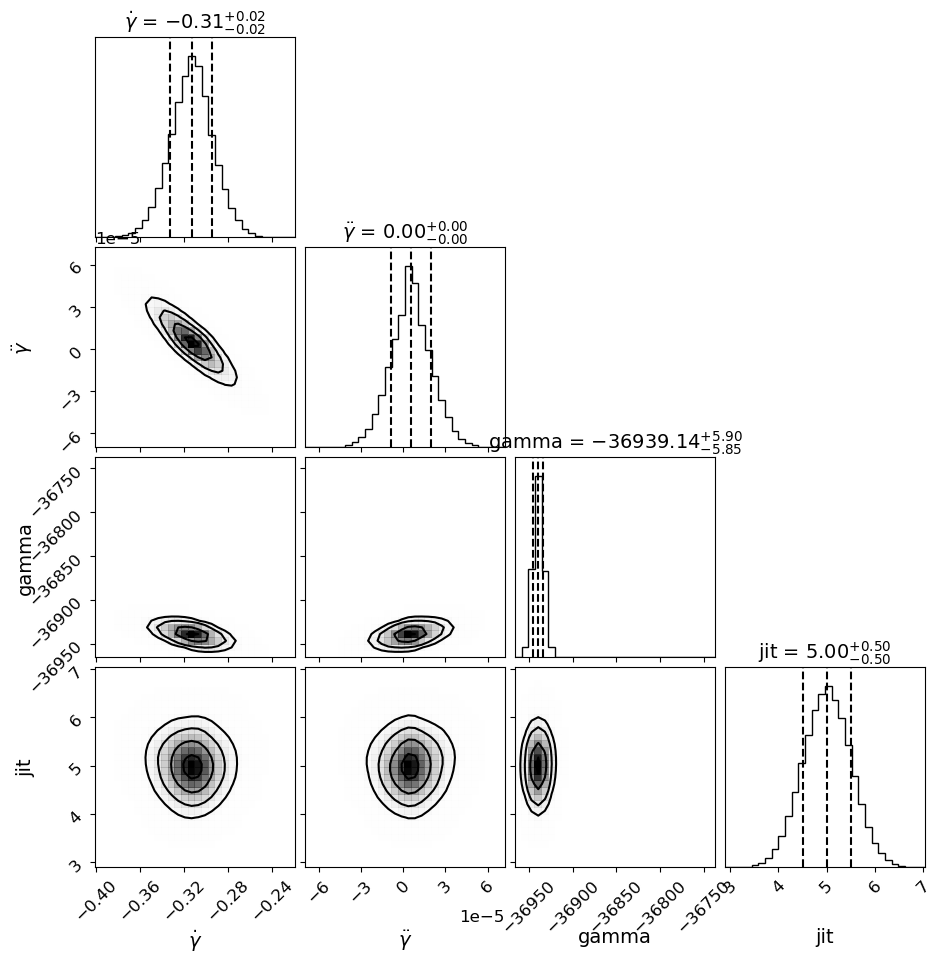}
    \\
    \includegraphics[width=0.40\textwidth]{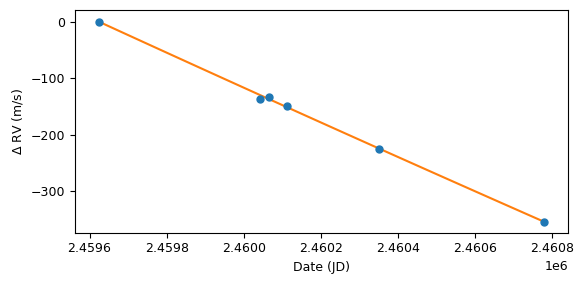}
    \caption{Corner plot of posterior distribution of Gaia DR3 1282810871941219968 (top) and time series with ``best-fit" line using RadVel's linear regression features (bottom). The formal errors on the RVs are plotted, but too small to be seen. The y-axis of the time-series plot is the RV change relative to the first observation.}
    \label{fig:128Radvel}
\end{figure}

\begin{figure}[h]
    \centering
    \includegraphics[width=0.40\textwidth]{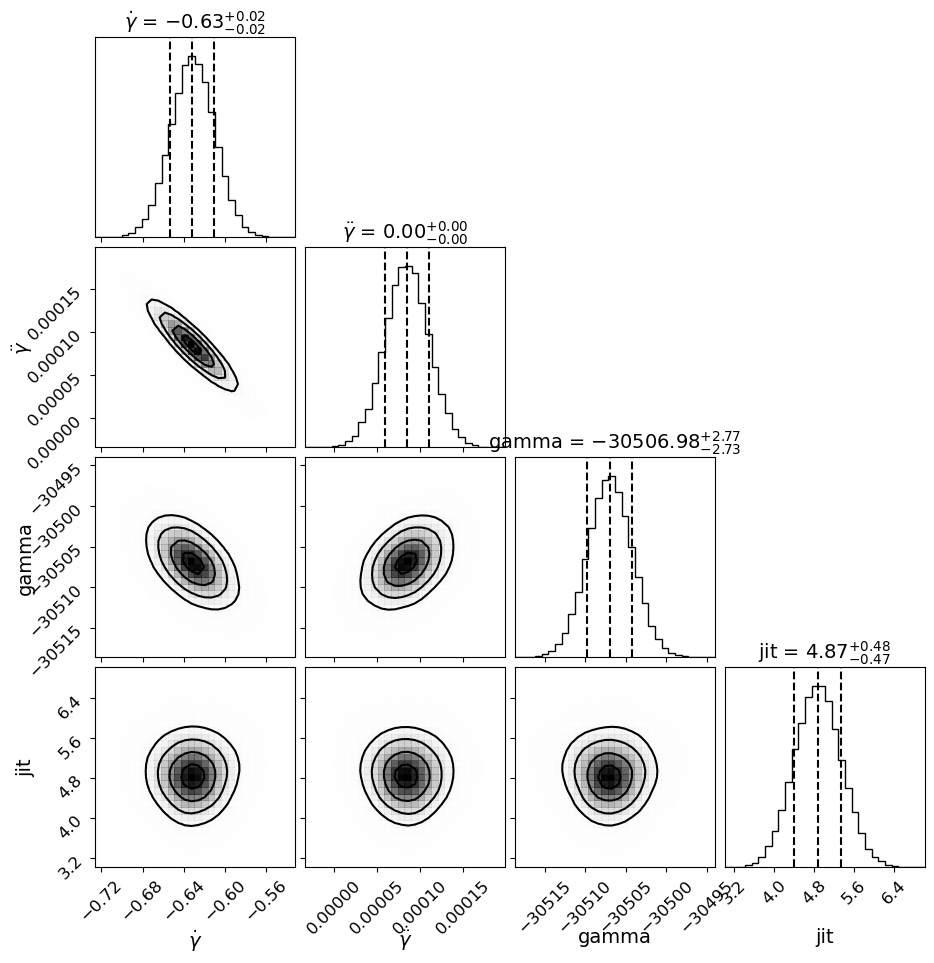}%
    \\
    \includegraphics[width=0.40\textwidth]{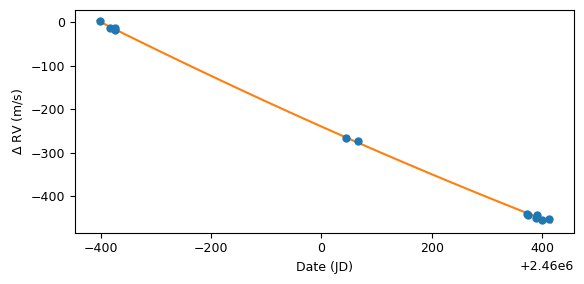}
    \caption{Corner plot of posterior distribution of Gaia DR3 1448931857534138112 (top) and time series with ``best-fit" line (bottom) using RadVel's linear regression features. The formal errors on the RVs are plotted, but too small to be seen.}
    \label{fig:144Radvel}
\end{figure}
\begin{figure}[h]
    \centering
    \includegraphics[width=0.40\textwidth]{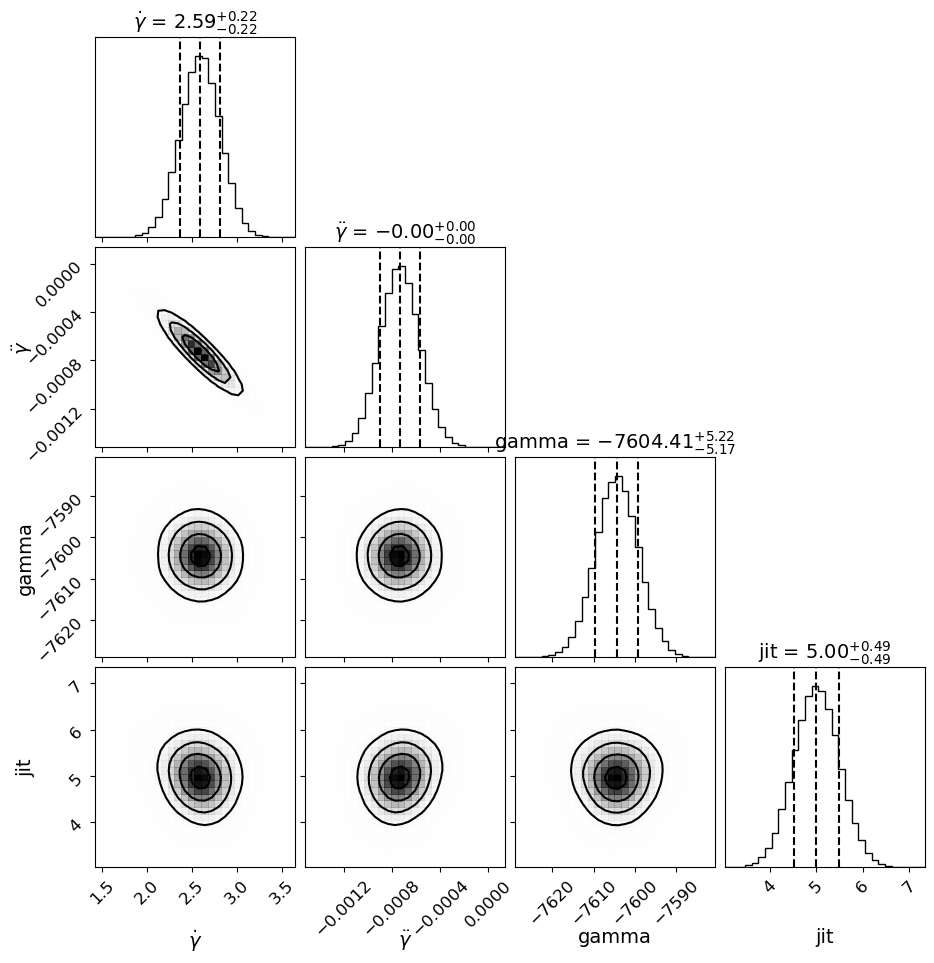}%
    \\
    \includegraphics[width=0.40\textwidth]{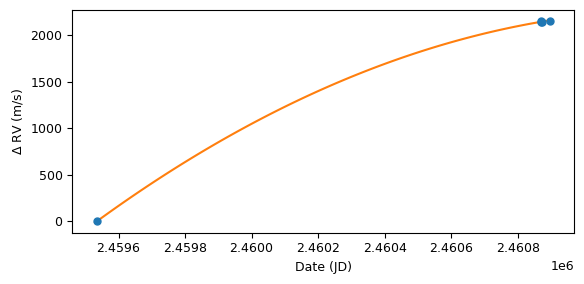}
    \caption{Corner plot of posterior distribution of Gaia DR3 2314056743326275840 (top) and time series with ``best-fit" line (bottom) using RadVel's linear regression features. The formal errors on the RVs are plotted, but too small to be seen.}
    \label{fig:231Radvel}
\end{figure}
\begin{figure}[h]
    \centering
    \includegraphics[width=0.40\textwidth]{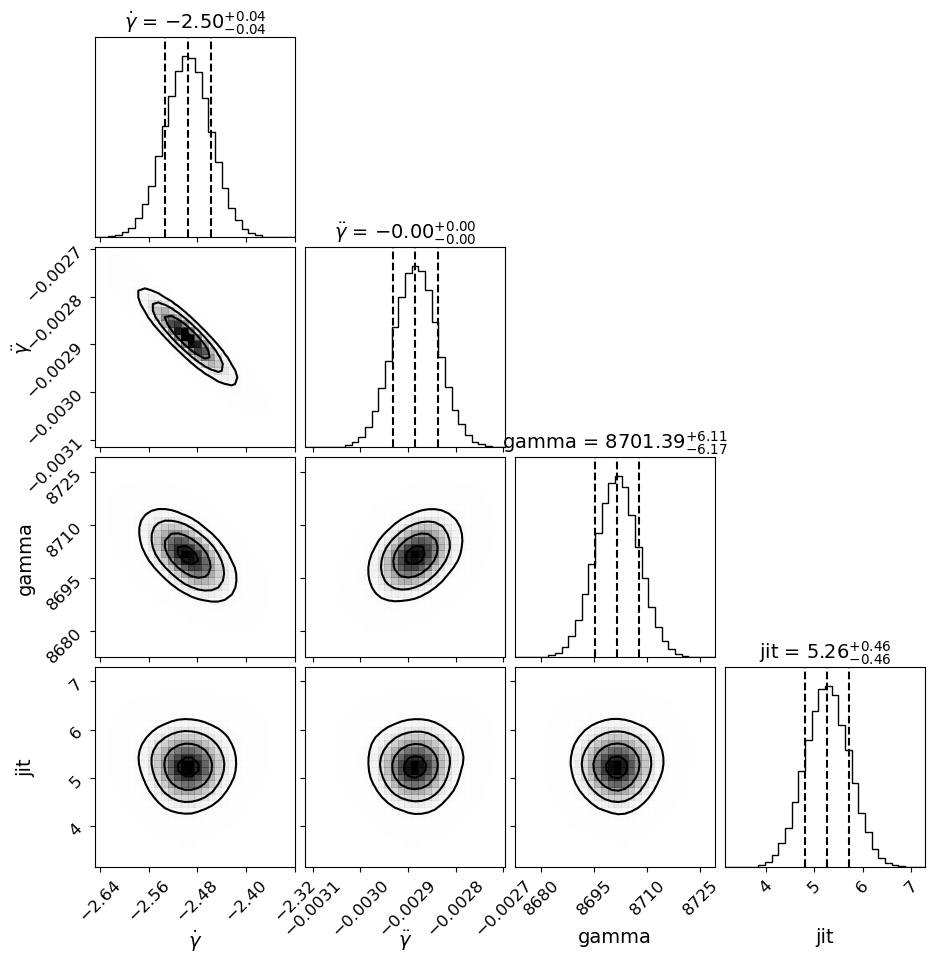}%
    \\
    \includegraphics[width=0.40\textwidth]{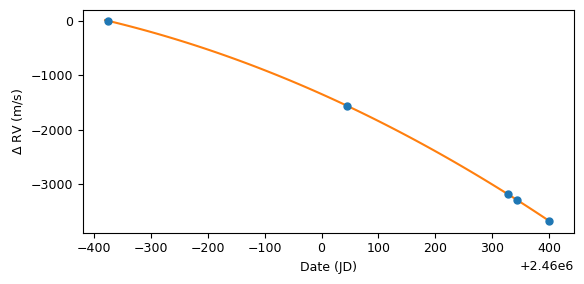}
    \caption{Corner plot of posterior distribution of Gaia DR3 3937811525200355328 (top) and time series with ``best-fit" line (bottom) using RadVel's linear regression features. The formal errors on the RVs are plotted, but too small to be seen.}
    \label{fig:393Radvel}
\end{figure}

\subsection{Long-Period Companion Candidates}
In addition to the four sources that we discussed in  in \S \ref{sec:companionswithorbit}, there are additionally four long-period companion candidates.  For these sources, our observations do not provide a complete orbital solution, only a small linear fragment of the orbit. Therefore, for these four sources, we present a linear acceleration term and a higher order curvature term in Table \ref{incomplete_orbit_table}, from a \textit{RadVel} linear regression analysis as described in (\cite{2018PASP..130d4504F}.  The RV time series and posterior distributions for the acceleration
and curvature terms of these four long-period companion candidates
are shown in Figures \ref{fig:128Radvel}--\ref{fig:393Radvel}.  These analyses were performed with simple Gaussian priors centered around zero for the acceleration and curvature terms, as well as a Gaussian prior centered at 5 m s$^{-1}$ for an additional instrumental ``jitter" term. The linear acceleration terms are well constrained, as are the curvature terms with the exception of Gaia DR3 1282810871941219968. These are likely long-period companions of significant masses, based on the disparate magnitudes of observed RV shift. As we aimed to formulate here a ``gold Sample" of stars with minimal RV shifts, these stars were excised from our observing plan when it became clear they did not fit within the low-jitter regime. Future observations may enable a more accurate characterization of these systems and translate these accelerations to mass predictions; given the last observations for these sources were two years before the writing of this paper, new observations would likely reveal promising information about the nature of these companions' orbits and capture distinct phase elements. 

\begin{deluxetable*}{lcc}
\tablecaption{Derived Linear/Curvature Terms for Potential Companions} \label{incomplete_orbit_table}
\tablewidth{0pt}
\tablehead{
\colhead{Gaia DR3} & \colhead{$\dot{\gamma}$} & \colhead{$\ddot{\gamma}$} \\
\colhead{}     & \colhead{(m.s$^{-1}$.day$^{-1}$)}       & \colhead{m.s$^{-1}$.day$^{-2}$)}}   
\startdata
1282810871941219968 & $-0.313 \pm{0.019}$ & $5 \times 10^{-6} \pm{1.4 \times 10^{-5}}$ \\
1448931857534138112 & $-0.632 \pm{0.021}$ & $8.5 \times 10^{-5} \pm{2.5 \times 10^{-5}}$\\
2314056743326275840 & 2.59 $\pm{0.22}$ & $-7.4 \times 10^{-4} \pm{1.7 \times 10^{-4}}$ \\ 
3937811525200355328 & $-2.496 \pm{0.038}$  & $-0.00289  \pm 4.7 \times 10^{-5} $ \\
\enddata
\end{deluxetable*}

\section{Discussion \& Conclusion}
\label{sec:discussion}

With several years (2021 - 2025) of observations from ESPRESSO, we have established the first Galactic-scale EPRV survey designed to directly measure the gravitational acceleration of stars in the Milky Way. Beginning with an initial sample of 66 subgiant stars, we used stellar activity indicators, cross-correlation functions, and multi-epoch RV measurements to identify a sample of 37 stars with empirically low stellar activity and RV variability. We refer to these stars as our “gold sample” and will continue to monitor them over the coming years. The longest observational baselines in the current data set already approach four years, with approximately 80\% of the sample having baselines longer than 1200 days.  We have also discussed a sample of stars that we excluded from our gold sample due to large radial velocity variation, known or suspected companions, and/or high stellar activity.  We provide an initial characterization of the companions in this latter sample from our RV time-series data.  
  
  Our goal in this paper was to begin a survey for which future data reduction algorithms can be applied to generate the required precision to make the Galactic acceleration measurement, given the RV scatter of the gold sample that we have identified here ($\sim$ few m/s).  As we intend to monitor these stars for $\sim$ decade, we are hopeful that this survey will eventually yield the precision promised in \cite{Chakrabarti2020}.  A rough scaling assuming uniformly spaced, independent measurements given our typical RV scatter of the gold sample suggests that we need about 1000 RV epochs.  
However, our ability to effectively mitigate stellar jitter will ultimately determine how many RV epochs will be needed to measure the Galactic acceleration. Magnetic activity, oscillations, and granulation introduce RV variations over a wide range of timescales and can therefore mimic or obscure the small secular acceleration we are trying to measure. Our observing strategy addresses this problem first by selecting quiet sub-giant stars following recent work \citep{Luhnetal2020a,Luhnetal2020b}, and by empirically characterizing their long-term RV behavior. The resulting gold sample has a narrow distribution in chromospheric activity, with a mean \(\log R'_{\rm HK}=-5.079\) and standard deviation of 0.056, characteristic of inactive subgiant stars. The next stage of this program will require both increased observing cadence and techniques for separating stellar variability from long-term RV trends.  Recent analyses of solar spectra  demonstrate that modeling spectral variability can substantially reduce activity-induced RV scatter \citep{Fordetal2024} by about an order of magnitude, and application of their training algorithm on a smaller subset already yielded a reduction in the RMS of the solar RV to better than a factor of 5.  In future work, we will use simulated observations incorporating stellar variability and the actual sampling of our survey to determine the cadence and temporal baseline required to recover Galactic accelerations from the gold sample.  

Here, we have opted to cite the composite CCF FWHM as produced by DAS; v1.3.8 within the \texttt{EsoReflex} environment \citep{2013A&A...559A..96F}, for which we typically used a window of 15 km/s.  In future work (Wagner et al., in prep), we explore whether an optimized line-by-line analysis may improve our ultimate RV precision for subgiant stars.  The median composite CCF FWHM for our gold sample is 9 km/s.  However,  individual lines can have FWHM that are smaller than this composite value, and we note that variations in line selection and weighting can lead to some differences in the composite CCF FWHM \citep{Bourrier2021}.  In future work, we also consider whether including a mixture of very quiet dwarf stars (with jitter at $\sim$ m/s level rather than the $\sim$ 3 m/s level we find here for our quietest subgiant stars) may be the optimal choice for such a survey, given that the RV precision scales as in proportion to the S/N and the line width, $\sigma_{e}$, ($\propto (S/N) \sigma_{e}^{-3/2}$) \citep{Jackson2015} in the photon noise limit.  The question of whether we can mitigate stellar jitter effectively for this sample is beyond the scope of this paper.    

An important feature of this survey is its spatial extent. The stars in the gold sample span approximately 1 kpc in Galactocentric radius and approximately 2 kpc in vertical height, and were selected to complement existing pulsar acceleration measurements \citep{Chakrabarti2021,Donlonetal2024,Donlonetal2025}.  We selected subgiant stars in our survey as they provide a favorable balance between stellar jitter and Galactic reach: dwarfs exhibit the lowest intrinsic stellar jitter but are relatively faint, while giants can probe the Galactic potential at greater distances but have substantially larger intrinsic stellar jitter.  Measurements of secular RV trends for our stars will sample the Galactic gravitational field at physically distinct locations. This approach differs fundamentally from conventional Galactic dynamical measurements, which infer gravitational accelerations from instantaneous phase-space measurements and generally require assumptions about the distribution function, dynamical equilibrium, or symmetry. EPRV measurements instead provide a direct measurement of the line-of-sight acceleration of individual stars. 

Combining stellar EPRV accelerations with pulsar timing accelerations will provide joint constraints on the Galactic potential \citep{Craig2022} because the two techniques have different spatial sampling and observational systematics. Direct acceleration measurements can constrain the local dark matter density and the distribution of baryonic matter while simultaneously providing tests for departures from a smooth, equilibrium Galactic potential. Of particular interest is the possibility of identifying localized perturbations to the acceleration field produced by dark matter substructure. Recent pulsar timing measurements provide tentative evidence for such substructure near the Sun \citep{Chakrabartietal2026}, and analysis of \textit{Fermi} data indicates an intriguing excess in that region \citep{Zhuetal2025, SalcesPerez2026}.  The stellar acceleration measurements developed here will provide an independent probe of the gravitational field over this region of the Galaxy and may therefore provide an important test of dark substructures.


The construction of the gold sample also illustrates the importance of identifying external sources of acceleration.  During the process of formulating our gold sample, we discovered eight sources that we identify as having stellar companions, given that the semi-amplitude of the RV observations is larger than 50 m/s.  For four of these sources, our observations covered a full orbital cycle and we present initial constraints on the orbital parameters.  These sources have $M \sin i$ ranging from Jupiter mass companions to low-mass stars, with orbital periods between $\sim 100 - 1000$ days.  The other four sources that have large variations in their RV measurements do not yield constrained orbital parameters but we present a linear acceleration term and higher order curvature term.  Continued observations, together with future astrometric information, should enable improved characterization of these systems.


This first Galactic scale EPRV survey can serve as a pathfinder for future EPRV surveys that can be done on the extremely large telescopes with extreme-precision spectrographs \citep{Marconietal2024} in the next decade \citep{Chakrabarti2020,ChakrabartiSnowmassObsFacilities}.  
Although much of the discussion of long-baseline spectroscopy with ELTs has focused on measurements of cosmological redshift drift \citep{Liske2008,Cristiani2023}, Galactic accelerations can be measured using the same fundamental observable: the secular change in the measured RV.

\acknowledgements 
SC gratefully acknowledges support from NASA EPSCoR CAN AL-80NSSC24M0104 and the Margaret Burbidge fellowship. JW thankfully acknowledges financial support from the NASA Alabama Space Grant Fellowship award (NASA Training Grant NNH24ZHAA03C). This research has been conducted based on observations collected at the European Southern Observatory under ESO programmes  108.227J.001, 111.24V7.001, 112.25TQ.001, 113.26HG.001, and 115.284Y.001.

\FloatBarrier
\pagebreak
\appendix



\section{Observing log}
\small
\begin{longtable*}{lccccc}
\caption{Observing log.}\label{tab:observing_log} \\
\toprule
Gaia DR3 & Date & FP & SKY & DIT [s] & DIMM [$^{\prime\prime}$] \\
\midrule
\endfirsthead
\multicolumn{6}{c}{{\tablename\ \thetable{} -- continued}} \\
\toprule
Gaia DR3 & Date & FP & SKY & DIT [s] & DIMM [$^{\prime\prime}$] \\
\midrule
\endhead
\midrule
\multicolumn{6}{r}{Continued on next page} \\
\endfoot
\bottomrule
\endlastfoot
1225731203253635584 & 2022-02-17 06:48 & \checkmark &  & 1500 & $0.41 \pm 0.02$ \\
 & 2023-04-10 07:36 & \checkmark &  & 3417 & $0.68 \pm 0.10$ \\
 & 2023-06-09 03:51 & \checkmark &  & 3417 & $0.67 \pm 0.14$ \\
 & 2023-06-24 23:29 & \checkmark &  & 3417 & $0.90 \pm 0.12$ \\
 & 2024-02-15 08:03 &  & \checkmark & 3297 & $0.60 \pm 0.05$ \\
 & 2024-03-09 08:44 &  & \checkmark & 1032 & $0.58 \pm 0.06$ \\
 & 2024-03-15 08:11 &  & \checkmark & 3297 & $0.52 \pm 0.04$ \\
 & 2024-04-19 04:47 &  & \checkmark & 3092 & $0.45 \pm 0.10$ \\
 & 2024-04-19 05:47 &  & \checkmark & 3092 & $0.69 \pm 0.07$ \\
\addlinespace
1250840234901215744 & 2022-02-14 07:09 & \checkmark &  & 1200 & $0.45 \pm 0.06$ \\
\addlinespace
1257728537810125184 & 2022-02-17 06:26 & \checkmark &  & 970 & $0.45 \pm 0.05$ \\
\addlinespace
1271212742256948096 & 2022-02-28 07:31 & \checkmark &  & 610 & $0.46 \pm 0.00$ \\
 & 2025-04-13 06:16 &  & \checkmark & 3429 & $0.41 \pm 0.04$ \\
 & 2025-05-20 04:13 &  & \checkmark & 3429 & $0.56 \pm 0.14$ \\
\addlinespace
1280137787375425664 & 2022-02-28 07:17 & \checkmark &  & 610 & $0.47 \pm 0.07$ \\
 & 2025-04-07 05:46 &  & \checkmark & 3429 & $0.47 \pm 0.04$ \\
 & 2025-05-01 05:29 &  & \checkmark & 3429 & $0.83 \pm 0.08$ \\
 & 2025-05-03 04:46 &  & \checkmark & 3429 & $0.65 \pm 0.07$ \\
\addlinespace
1282810871941219968 & 2022-02-14 08:14 & \checkmark &  & 460 & $0.48 \pm 0.05$ \\
 & 2023-04-08 06:07 & \checkmark &  & 1032 & $0.68 \pm 0.09$ \\
 & 2023-05-02 03:44 & \checkmark &  & 1032 & $0.53 \pm 0.05$ \\
 & 2023-06-16 02:27 & \checkmark &  & 1032 & $0.52 \pm 0.06$ \\
 & 2024-02-11 08:29 &  & \checkmark & 1032 & $0.55 \pm 0.03$ \\
 & 2025-04-13 04:29 &  & \checkmark & 2746 & $0.41 \pm 0.03$ \\
\addlinespace
1282813070964475392 & 2022-02-14 08:05 & \checkmark &  & 350 & $0.59 \pm 0.06$ \\
 & 2023-04-08 07:23 & \checkmark &  & 779 & $0.57 \pm 0.05$ \\
 & 2023-05-02 04:05 & \checkmark &  & 779 & $0.64 \pm 0.03$ \\
 & 2023-06-16 02:48 & \checkmark &  & 779 & $0.53 \pm 0.05$ \\
 & 2024-02-10 08:44 &  & \checkmark & 779 & $0.35 \pm 0.01$ \\
 & 2024-03-02 07:05 &  & \checkmark & 779 & -- \\
\addlinespace
1448931857534138112 & 2022-01-19 08:28 & \checkmark &  & 970 & $0.73 \pm 0.18$ \\
 & 2022-02-05 08:40 & \checkmark &  & 970 & $0.94 \pm 0.08$ \\
 & 2022-02-14 06:48 & \checkmark &  & 970 & $0.47 \pm 0.12$ \\
 & 2022-02-14 07:46 & \checkmark &  & 970 & $0.45 \pm 0.07$ \\
 & 2023-04-10 06:15 & \checkmark &  & 2184 & $0.67 \pm 0.08$ \\
 & 2023-05-02 04:22 & \checkmark &  & 2184 & $0.50 \pm 0.07$ \\
 & 2024-03-02 07:26 &  & \checkmark & 2184 & $1.08 \pm 0.21$ \\
 & 2024-03-05 06:39 &  & \checkmark & 2184 & $0.65 \pm 0.16$ \\
 & 2024-03-20 05:48 &  & \checkmark & 2184 & $1.01 \pm 0.15$ \\
 & 2024-03-21 04:31 &  & \checkmark & 2184 & $0.64 \pm 0.09$ \\
 & 2024-03-30 05:34 &  & \checkmark & 4368 & $0.72 \pm 0.16$ \\
 & 2024-04-11 03:31 &  & \checkmark & 2184 & $0.49 \pm 0.04$ \\
\addlinespace
1455717081228013696 & 2022-02-14 07:33 & \checkmark &  & 560 & $0.45 \pm 0.07$ \\
\addlinespace
1466420242809206656 & 2022-02-05 08:17 & \checkmark &  & 910 & $0.74 \pm 0.09$ \\
 & 2022-02-14 06:30 & \checkmark &  & 910 & $0.45 \pm 0.11$ \\
 & 2023-04-10 05:16 & \checkmark &  & 2184 & $0.51 \pm 0.06$ \\
 & 2023-04-27 02:49 & \checkmark &  & 2184 & $0.51 \pm 0.04$ \\
 & 2024-01-29 07:33 &  & \checkmark & 2184 & $0.52 \pm 0.06$ \\
 & 2024-02-16 07:23 &  & \checkmark & 2184 & $0.51 \pm 0.04$ \\
 & 2024-03-02 06:21 &  & \checkmark & 2184 & $0.45 \pm 0.04$ \\
 & 2024-03-31 05:13 &  & \checkmark & 4368 & $0.65 \pm 0.10$ \\
 & 2025-04-07 04:34 &  & \checkmark & 2544 & $0.57 \pm 0.03$ \\
 & 2025-04-13 03:05 &  & \checkmark & 2544 & $0.59 \pm 0.07$ \\
\addlinespace
1469142530521510016 & 2022-02-03 08:45 & \checkmark &  & 420 & $1.08 \pm 0.16$ \\
 & 2022-02-04 08:47 & \checkmark &  & 420 & $1.03 \pm 0.09$ \\
\addlinespace
2314056743326275840 & 2021-11-16 02:12 & \checkmark &  & 1100 & $0.65 \pm 0.05$ \\
 & 2025-07-12 06:42 &  & \checkmark & 2762 & $1.58 \pm 0.00$ \\
 & 2025-07-13 08:27 &  & \checkmark & 2762 & -- \\
 & 2025-07-16 05:39 &  & \checkmark & 2762 & $0.52 \pm 0.03$ \\
 & 2025-08-08 05:28 &  & \checkmark & 2762 & $0.95 \pm 0.09$ \\
 & 2025-08-09 09:02 &  & \checkmark & 2762 & $0.66 \pm 0.06$ \\
\addlinespace
2316357673270019584 & 2021-11-16 01:52 & \checkmark &  & 1010 & $0.65 \pm 0.06$ \\
 & 2023-06-09 07:22 & \checkmark &  & 2184 & $0.67 \pm 0.05$ \\
 & 2023-07-18 09:37 & \checkmark &  & 2184 & $0.50 \pm 0.07$ \\
 & 2023-08-03 05:57 & \checkmark &  & 2184 & $0.59 \pm 0.15$ \\
 & 2023-09-30 05:13 &  & \checkmark & 2184 & $0.96 \pm 0.10$ \\
 & 2023-10-17 02:28 &  & \checkmark & 2184 & $0.79 \pm 0.12$ \\
 & 2025-07-02 06:47 &  & \checkmark & 2544 & $0.82 \pm 0.07$ \\
 & 2025-07-11 09:19 &  & \checkmark & 2544 & $0.56 \pm 0.04$ \\
\addlinespace
2328985804833355008 & 2021-11-16 01:27 & \checkmark &  & 1240 & $0.71 \pm 0.04$ \\
 & 2023-06-22 07:47 & \checkmark &  & 2635 & $0.52 \pm 0.07$ \\
 & 2023-07-09 05:14 & \checkmark &  & 2635 & $0.73 \pm 0.10$ \\
 & 2023-07-24 05:19 & \checkmark &  & 2635 & $0.67 \pm 0.07$ \\
 & 2023-09-30 03:49 &  & \checkmark & 2635 & $0.90 \pm 0.17$ \\
 & 2023-10-20 01:23 &  & \checkmark & 2635 & $0.86 \pm 0.11$ \\
 & 2025-07-02 08:38 &  & \checkmark & 3000 & $0.54 \pm 0.10$ \\
 & 2025-07-03 06:56 &  & \checkmark & 3000 & $0.49 \pm 0.11$ \\
\addlinespace
2370995468366361344 & 2021-11-18 01:25 & \checkmark &  & 1100 & $0.57 \pm 0.03$ \\
\addlinespace
2393625376170980608 & 2021-10-09 01:11 & \checkmark &  & 320 & $0.84 \pm 0.06$ \\
 & 2023-05-28 09:50 & \checkmark &  & 709 & $0.78 \pm 0.08$ \\
 & 2023-06-23 10:00 & \checkmark &  & 709 & $0.73 \pm 0.05$ \\
 & 2023-07-16 09:56 & \checkmark &  & 709 & $0.51 \pm 0.07$ \\
 & 2023-10-01 04:45 &  & \checkmark & 709 & $0.48 \pm 0.04$ \\
 & 2023-10-17 02:08 &  & \checkmark & 709 & $0.67 \pm 0.12$ \\
 & 2025-07-21 04:30 &  & \checkmark & 2187 & $0.57 \pm 0.07$ \\
\addlinespace
2428155568905528064 & 2021-12-15 03:15 & \checkmark &  & 390 & $0.64 \pm 0.10$ \\
 & 2023-05-30 09:42 & \checkmark &  & 855 & $0.73 \pm 0.06$ \\
 & 2023-07-22 07:00 & \checkmark &  & 855 & $0.62 \pm 0.04$ \\
 & 2023-08-16 08:29 & \checkmark &  & 855 & $0.73 \pm 0.07$ \\
 & 2023-10-01 05:09 &  & \checkmark & 855 & $0.46 \pm 0.03$ \\
 & 2023-11-06 00:26 &  & \checkmark & 855 & $0.59 \pm 0.08$ \\
 & 2023-11-20 00:11 &  & \checkmark & 855 & $0.79 \pm 0.10$ \\
 & 2025-07-22 05:25 &  & \checkmark & 2486 & $0.49 \pm 0.03$ \\
\addlinespace
2443486781086490624 & 2021-11-18 03:03 & \checkmark &  & 420 & $0.73 \pm 0.04$ \\
 & 2023-06-09 08:22 & \checkmark &  & 939 & $0.58 \pm 0.05$ \\
 & 2023-06-30 09:51 & \checkmark &  & 939 & $0.75 \pm 0.08$ \\
 & 2023-10-01 05:29 &  & \checkmark & 939 & $0.42 \pm 0.03$ \\
 & 2023-11-04 00:12 &  & \checkmark & 939 & $0.50 \pm 0.06$ \\
 & 2023-11-20 00:31 &  & \checkmark & 939 & $0.63 \pm 0.13$ \\
 & 2024-05-28 08:45 &  & \checkmark & 1878 & $0.91 \pm 0.12$ \\
 & 2024-06-19 09:31 &  & \checkmark & 1878 & $1.05 \pm 0.19$ \\
 & 2024-07-16 06:08 &  & \checkmark & 1878 & $0.65 \pm 0.07$ \\
 & 2025-07-21 05:35 &  & \checkmark & 2558 & $0.56 \pm 0.09$ \\
\addlinespace
2522660418675058944 & 2021-11-26 04:07 & \checkmark &  & 880 & $0.63 \pm 0.04$ \\
 & 2023-06-08 09:15 & \checkmark &  & 1989 & $0.58 \pm 0.09$ \\
 & 2023-06-30 09:15 & \checkmark &  & 1989 & $0.82 \pm 0.07$ \\
 & 2023-10-25 02:26 &  & \checkmark & 1989 & $0.59 \pm 0.05$ \\
 & 2023-11-13 04:24 &  & \checkmark & 1989 & $0.86 \pm 0.06$ \\
 & 2025-07-23 06:09 &  & \checkmark & 4692 & $0.44 \pm 0.11$ \\
\addlinespace
2735962096754761088 & 2021-10-05 04:33 & \checkmark &  & 240 & $0.63 \pm 0.01$ \\
\addlinespace
2810741192525443968 & 2021-10-06 04:40 & \checkmark &  & 240 & $0.39 \pm 0.03$ \\
\addlinespace
2903113569558464256 & 2021-10-26 07:44 & \checkmark &  & 240 & $0.41 \pm 0.01$ \\
\addlinespace
3629413601830054400 & 2022-02-16 05:51 & \checkmark &  & 270 & $0.68 \pm 0.10$ \\
 & 2023-04-11 07:26 & \checkmark &  & 588 & $0.61 \pm 0.05$ \\
 & 2023-05-28 02:15 & \checkmark &  & 588 & $1.04 \pm 0.12$ \\
 & 2023-06-17 02:40 & \checkmark &  & 588 & $0.52 \pm 0.04$ \\
 & 2024-01-01 07:51 &  & \checkmark & 588 & $0.40 \pm 0.02$ \\
 & 2024-01-27 07:10 &  & \checkmark & 588 & $0.43 \pm 0.03$ \\
 & 2024-02-18 08:59 &  & \checkmark & 588 & $0.79 \pm 0.06$ \\
 & 2024-06-18 23:38 &  & \checkmark & 1176 & -- \\
 & 2025-07-19 23:09 &  & \checkmark & 1939 & $0.77 \pm 0.04$ \\
\addlinespace
3660958006315220992 & 2022-03-15 04:00 & \checkmark &  & 350 & $0.49 \pm 0.02$ \\
 & 2023-04-10 08:37 & \checkmark &  & 779 & $0.60 \pm 0.03$ \\
 & 2023-06-21 02:28 & \checkmark &  & 779 & $0.87 \pm 0.05$ \\
 & 2023-07-15 02:22 & \checkmark &  & 779 & $0.82 \pm 0.04$ \\
 & 2024-01-28 07:20 &  & \checkmark & 779 & $0.53 \pm 0.07$ \\
 & 2024-03-05 09:07 &  & \checkmark & 779 & $0.49 \pm 0.05$ \\
 & 2025-04-07 08:30 &  & \checkmark & 2330 & $0.72 \pm 0.07$ \\
\addlinespace
3698120056225966976 & 2022-01-13 05:54 & \checkmark &  & 610 & $0.78 \pm 0.15$ \\
\addlinespace
3712163426756206208 & 2022-03-07 09:12 & \checkmark &  & 560 & $0.39 \pm 0.03$ \\
 & 2023-04-11 07:03 & \checkmark &  & 1133 & $0.66 \pm 0.04$ \\
 & 2023-06-16 03:32 & \checkmark &  & 1133 & $0.55 \pm 0.03$ \\
 & 2023-07-15 01:22 & \checkmark &  & 1133 & $0.62 \pm 0.03$ \\
 & 2023-07-15 01:58 & \checkmark &  & 1133 & $0.78 \pm 0.04$ \\
 & 2024-01-28 05:59 &  & \checkmark & 1133 & $0.76 \pm 0.13$ \\
 & 2024-03-11 09:02 &  & \checkmark & 1133 & $0.73 \pm 0.05$ \\
 & 2025-05-24 03:22 &  & \checkmark & 2953 & $0.75 \pm 0.08$ \\
\addlinespace
3716520933071728896 & 2022-02-14 09:01 & \checkmark &  & 350 & $0.77 \pm 0.12$ \\
 & 2023-04-11 06:48 & \checkmark &  & 709 & $0.61 \pm 0.05$ \\
 & 2023-06-10 03:51 & \checkmark &  & 709 & $0.46 \pm 0.02$ \\
 & 2023-06-24 23:12 & \checkmark &  & 709 & $0.83 \pm 0.03$ \\
 & 2024-01-01 08:07 &  & \checkmark & 1133 & $0.34 \pm 0.03$ \\
 & 2024-01-23 08:24 &  & \checkmark & 1133 & $0.43 \pm 0.04$ \\
 & 2025-04-16 07:21 &  & \checkmark & 2187 & $0.72 \pm 0.16$ \\
\addlinespace
3736336640865721216 & 2022-02-25 08:52 & \checkmark &  & 970 & $0.82 \pm 0.12$ \\
\addlinespace
3898946812815348096 & 2022-01-19 07:22 & \checkmark &  & 510 & $0.85 \pm 0.08$ \\
 & 2023-06-10 01:15 & \checkmark &  & 1133 & $0.69 \pm 0.14$ \\
 & 2023-07-18 23:16 & \checkmark &  & 1133 & $0.63 \pm 0.07$ \\
 & 2024-01-21 05:59 &  & \checkmark & 1133 & $0.76 \pm 0.06$ \\
 & 2024-02-09 08:36 &  & \checkmark & 1133 & $1.20 \pm 0.14$ \\
 & 2024-03-09 08:18 &  & \checkmark & 1133 & $0.57 \pm 0.08$ \\
 & 2024-03-31 02:40 &  & \checkmark & 1133 & $0.77 \pm 0.04$ \\
\addlinespace
3901835715193571456 & 2022-01-19 07:36 & \checkmark &  & 970 & $0.79 \pm 0.12$ \\
 & 2022-02-27 05:41 & \checkmark &  & 1940 & $0.33 \pm 0.04$ \\
 & 2022-03-31 03:01 & \checkmark &  & 1940 & $0.50 \pm 0.05$ \\
 & 2023-04-20 00:38 & \checkmark &  & 2184 & $1.02 \pm 0.13$ \\
 & 2024-03-19 02:15 &  & \checkmark & 2184 & $0.56 \pm 0.11$ \\
 & 2024-05-27 03:04 &  & \checkmark & 2184 & $0.91 \pm 0.10$ \\
 & 2024-06-12 00:37 &  & \checkmark & 2184 & $0.29 \pm 0.03$ \\
 & 2025-07-05 00:10 &  & \checkmark & 2544 & $0.75 \pm 0.06$ \\
 & 2025-07-18 22:53 &  & \checkmark & 2544 & $0.66 \pm 0.04$ \\
\addlinespace
3903785286749087744 & 2022-01-19 08:03 & \checkmark &  & 1100 & $0.64 \pm 0.06$ \\
 & 2022-03-02 03:40 & \checkmark &  & 2200 & $0.56 \pm 0.10$ \\
 & 2022-03-04 04:34 & \checkmark &  & 2200 & $0.52 \pm 0.15$ \\
 & 2022-04-03 03:18 & \checkmark &  & 2200 & $0.58 \pm 0.07$ \\
 & 2023-04-19 23:54 & \checkmark &  & 2399 & $0.94 \pm 0.11$ \\
 & 2024-01-27 06:23 &  & \checkmark & 2399 & $0.48 \pm 0.07$ \\
 & 2024-02-17 05:48 &  & \checkmark & 2399 & $0.42 \pm 0.04$ \\
 & 2024-03-19 02:58 &  & \checkmark & 2399 & $0.67 \pm 0.11$ \\
\addlinespace
3911409918850971392 & 2022-01-13 05:43 & \checkmark &  & 350 & $0.50 \pm 0.03$ \\
 & 2022-02-28 09:03 & \checkmark &  & 700 & $0.32 \pm 0.02$ \\
 & 2022-03-31 01:36 & \checkmark &  & 700 & $0.62 \pm 0.05$ \\
 & 2023-04-19 00:24 & \checkmark &  & 779 & $0.59 \pm 0.05$ \\
 & 2023-06-21 01:42 & \checkmark &  & 779 & $0.63 \pm 0.03$ \\
 & 2023-12-11 08:00 &  & \checkmark & 779 & $1.04 \pm 0.11$ \\
 & 2024-01-22 05:21 &  & \checkmark & 779 & $0.57 \pm 0.04$ \\
 & 2025-07-04 00:11 &  & \checkmark & 2330 & $0.65 \pm 0.09$ \\
\addlinespace
3918910714160355584 & 2022-01-19 06:25 & \checkmark &  & 810 & $0.54 \pm 0.04$ \\
 & 2022-02-27 05:11 & \checkmark &  & 1620 & $0.29 \pm 0.05$ \\
 & 2022-03-30 03:22 & \checkmark &  & 1620 & $0.76 \pm 0.10$ \\
 & 2023-04-18 23:50 & \checkmark &  & 1811 & $0.65 \pm 0.04$ \\
 & 2024-01-23 05:56 &  & \checkmark & 1811 & $0.36 \pm 0.03$ \\
 & 2024-02-16 05:05 &  & \checkmark & 1811 & $0.66 \pm 0.07$ \\
\addlinespace
3932181166672666240 & 2022-02-14 06:15 & \checkmark &  & 670 & $0.37 \pm 0.04$ \\
 & 2023-06-21 01:58 & \checkmark &  & 1501 & $0.68 \pm 0.06$ \\
 & 2024-02-17 04:28 &  & \checkmark & 1501 & $0.53 \pm 0.04$ \\
 & 2024-03-07 08:46 &  & \checkmark & 1501 & $0.62 \pm 0.16$ \\
 & 2024-06-12 01:45 &  & \checkmark & 3622 & $0.34 \pm 0.03$ \\
 & 2025-05-01 23:05 &  & \checkmark & 3702 & $0.46 \pm 0.04$ \\
\addlinespace
3937811525200355328 & 2022-02-14 06:00 & \checkmark &  & 670 & $0.42 \pm 0.04$ \\
 & 2023-04-11 04:50 & \checkmark &  & 1501 & $0.54 \pm 0.05$ \\
 & 2024-01-19 08:13 &  & \checkmark & 1501 & $0.72 \pm 0.09$ \\
 & 2024-02-03 07:22 &  & \checkmark & 1501 & $0.59 \pm 0.05$ \\
 & 2024-03-30 04:37 &  & \checkmark & 3002 & $0.88 \pm 0.12$ \\
\addlinespace
3940893456293439488 & 2022-02-16 06:18 & \checkmark &  & 460 & $0.76 \pm 0.03$ \\
 & 2022-02-17 06:14 & \checkmark &  & 460 & $0.50 \pm 0.04$ \\
 & 2023-04-11 05:19 & \checkmark &  & 1032 & $0.68 \pm 0.06$ \\
 & 2023-07-21 00:18 & \checkmark &  & 1032 & $0.42 \pm 0.02$ \\
 & 2024-01-22 07:03 &  & \checkmark & 1032 & $0.33 \pm 0.03$ \\
 & 2024-02-16 05:43 &  & \checkmark & 1032 & $0.60 \pm 0.09$ \\
 & 2024-03-05 08:00 &  & \checkmark & 1032 & $0.48 \pm 0.04$ \\
 & 2024-04-01 06:09 &  & \checkmark & 2064 & $0.54 \pm 0.06$ \\
 & 2025-04-14 05:56 &  & \checkmark & 2746 & $0.77 \pm 0.10$ \\
\addlinespace
3945839407488324224 & 2022-02-16 06:00 & \checkmark &  & 810 & $0.69 \pm 0.06$ \\
 & 2024-02-17 05:05 &  & \checkmark & 1811 & $0.50 \pm 0.04$ \\
 & 2024-03-18 03:34 &  & \checkmark & 1811 & $0.69 \pm 0.12$ \\
 & 2024-05-14 01:24 &  & \checkmark & 3622 & $0.36 \pm 0.05$ \\
 & 2024-05-14 02:31 &  & \checkmark & 1811 & $0.38 \pm 0.03$ \\
\addlinespace
3946715718255175168 & 2022-02-16 06:30 & \checkmark &  & 970 & $0.78 \pm 0.13$ \\
 & 2023-04-18 00:36 & \checkmark &  & 2184 & $0.52 \pm 0.03$ \\
 & 2024-02-01 06:18 &  & \checkmark & 2184 & $0.52 \pm 0.06$ \\
 & 2024-03-18 02:25 &  & \checkmark & 2184 & $0.56 \pm 0.05$ \\
 & 2024-04-17 04:23 &  & \checkmark & 4368 & -- \\
 & 2024-05-14 00:39 &  & \checkmark & 2184 & -- \\
\addlinespace
3947104223817128320 & 2022-02-09 08:57 & \checkmark &  & 350 & $0.85 \pm 0.07$ \\
 & 2023-04-11 05:40 & \checkmark &  & 779 & $0.72 \pm 0.08$ \\
 & 2024-01-22 06:44 &  & \checkmark & 779 & $0.32 \pm 0.02$ \\
 & 2024-02-16 06:09 &  & \checkmark & 779 & $0.53 \pm 0.08$ \\
 & 2024-06-19 00:43 &  & \checkmark & 1558 & $0.58 \pm 0.03$ \\
 & 2025-05-30 00:07 &  & \checkmark & 2330 & $0.59 \pm 0.08$ \\
\addlinespace
3952789626645417216 & 2022-01-19 06:43 & \checkmark &  & 670 & $0.68 \pm 0.07$ \\
 & 2023-04-11 05:57 & \checkmark &  & 1501 & $0.70 \pm 0.10$ \\
 & 2023-04-19 00:57 & \checkmark &  & 1501 & $0.75 \pm 0.07$ \\
 & 2024-01-23 06:33 &  & \checkmark & 1501 & $0.34 \pm 0.02$ \\
 & 2024-02-16 06:28 &  & \checkmark & 1501 & $0.44 \pm 0.04$ \\
 & 2024-03-09 07:08 &  & \checkmark & 1501 & $0.36 \pm 0.03$ \\
 & 2025-05-24 01:20 &  & \checkmark & 3702 & $0.72 \pm 0.08$ \\
\addlinespace
3954703017395718016 & 2022-02-05 08:06 & \checkmark &  & 460 & $0.60 \pm 0.04$ \\
 & 2023-04-19 01:26 & \checkmark &  & 1032 & $0.78 \pm 0.13$ \\
 & 2024-01-23 07:06 &  & \checkmark & 1032 & $0.39 \pm 0.06$ \\
 & 2024-03-18 03:10 &  & \checkmark & 1032 & $0.77 \pm 0.05$ \\
 & 2024-03-19 03:46 &  & \checkmark & 1032 & $0.81 \pm 0.13$ \\
 & 2024-04-11 02:50 &  & \checkmark & 2064 & $0.59 \pm 0.05$ \\
 & 2024-06-12 01:21 &  & \checkmark & 1032 & $0.37 \pm 0.06$ \\
\addlinespace
3980392449448341888 & 2022-01-13 06:07 & \checkmark &  & 320 & $0.63 \pm 0.03$ \\
 & 2023-04-19 00:42 & \checkmark &  & 709 & $0.49 \pm 0.02$ \\
 & 2023-06-07 23:27 & \checkmark &  & 709 & $0.57 \pm 0.03$ \\
 & 2023-12-22 07:54 &  & \checkmark & 709 & $0.60 \pm 0.02$ \\
 & 2024-01-23 05:37 &  & \checkmark & 709 & $0.38 \pm 0.02$ \\
 & 2024-02-16 04:18 &  & \checkmark & 709 & $0.76 \pm 0.08$ \\
 & 2025-05-10 01:02 &  & \checkmark & 2187 & $0.76 \pm 0.11$ \\
\addlinespace
3995373707693411072 & 2022-01-08 06:51 & \checkmark &  & 970 & $0.49 \pm 0.07$ \\
 & 2022-02-27 04:35 & \checkmark &  & 1940 & $0.32 \pm 0.03$ \\
 & 2022-03-30 02:46 & \checkmark &  & 1940 & $0.68 \pm 0.10$ \\
 & 2022-03-30 04:08 & \checkmark &  & 1940 & $0.52 \pm 0.06$ \\
 & 2023-04-17 23:55 & \checkmark &  & 2184 & $0.50 \pm 0.05$ \\
 & 2023-06-08 00:19 & \checkmark &  & 2184 & $0.66 \pm 0.07$ \\
 & 2024-01-22 05:41 &  & \checkmark & 2184 & $0.50 \pm 0.04$ \\
 & 2024-02-11 07:46 &  & \checkmark & 2184 & $0.56 \pm 0.04$ \\
 & 2024-03-05 05:02 &  & \checkmark & 2184 & $0.44 \pm 0.08$ \\
 & 2025-04-13 01:24 &  & \checkmark & 2544 & $0.53 \pm 0.06$ \\
 & 2025-04-13 02:13 &  & \checkmark & 2644 & $0.51 \pm 0.06$ \\
\addlinespace
4007351482424776576 & 2022-01-19 06:57 & \checkmark &  & 1100 & $0.80 \pm 0.10$ \\
\addlinespace
4017904350214299520 & 2022-01-05 07:59 & \checkmark &  & 350 & $0.66 \pm 0.06$ \\
 & 2022-02-27 04:19 & \checkmark &  & 700 & $0.32 \pm 0.07$ \\
 & 2022-03-30 02:31 & \checkmark &  & 700 & $0.61 \pm 0.05$ \\
 & 2023-04-06 02:32 & \checkmark &  & 779 & $0.57 \pm 0.02$ \\
 & 2024-01-22 06:24 &  & \checkmark & 779 & $0.31 \pm 0.01$ \\
 & 2024-02-08 08:28 &  & \checkmark & 779 & $0.39 \pm 0.03$ \\
 & 2024-03-09 06:47 &  & \checkmark & 779 & $0.32 \pm 0.03$ \\
 & 2025-04-27 02:09 &  & \checkmark & 2330 & $0.36 \pm 0.06$ \\
\addlinespace
4029722313506309760 & 2022-02-01 08:19 & \checkmark &  & 970 & $1.21 \pm 0.07$ \\
 & 2022-02-03 08:25 & \checkmark &  & 970 & $0.95 \pm 0.03$ \\
\addlinespace
4031060457812768768 & 2022-02-03 07:53 & \checkmark &  & 1200 & $1.10 \pm 0.13$ \\
 & 2023-05-02 01:25 & \checkmark &  & 2635 & $0.77 \pm 0.14$ \\
 & 2024-01-23 07:29 &  & \checkmark & 2635 & $0.36 \pm 0.05$ \\
 & 2025-04-07 02:40 &  & \checkmark & 3000 & $0.44 \pm 0.04$ \\
 & 2025-04-07 03:37 &  & \checkmark & 3000 & $0.50 \pm 0.05$ \\
\addlinespace
4264320784544595328 & 2022-03-28 09:15 & \checkmark &  & 200 & $0.49 \pm 0.03$ \\
 & 2022-03-28 09:25 & \checkmark &  & 200 & $0.48 \pm 0.04$ \\
 & 2023-04-08 09:26 & \checkmark &  & 444 & $0.44 \pm 0.01$ \\
 & 2023-05-15 08:30 & \checkmark &  & 444 & $0.45 \pm 0.03$ \\
 & 2023-05-30 09:27 & \checkmark &  & 444 & $0.94 \pm 0.11$ \\
 & 2023-06-22 06:47 & \checkmark &  & 444 & $0.46 \pm 0.03$ \\
 & 2023-10-01 01:33 &  & \checkmark & 444 & $0.43 \pm 0.01$ \\
 & 2023-10-16 00:47 &  & \checkmark & 444 & $1.09 \pm 0.34$ \\
 & 2023-10-17 01:15 &  & \checkmark & 444 & $0.60 \pm 0.03$ \\
\addlinespace
4772734377062372224 & 2021-10-22 07:55 & \checkmark &  & 240 & -- \\
 & 2022-03-02 00:47 & \checkmark &  & 480 & $0.55 \pm 0.04$ \\
 & 2022-03-31 01:22 & \checkmark &  & 480 & $0.66 \pm 0.05$ \\
 & 2025-04-06 23:49 &  & \checkmark & 1831 & $0.60 \pm 0.06$ \\
 & 2025-04-30 23:20 &  & \checkmark & 1831 & $0.70 \pm 0.10$ \\
\addlinespace
4934640384830032512 & 2021-11-09 07:02 & \checkmark &  & 420 & -- \\
 & 2023-06-06 09:26 & \checkmark &  & 939 & $0.50 \pm 0.03$ \\
 & 2023-07-14 06:52 & \checkmark &  & 939 & $1.09 \pm 0.11$ \\
 & 2023-07-15 07:03 & \checkmark &  & 939 & $0.78 \pm 0.06$ \\
 & 2023-09-30 06:18 &  & \checkmark & 939 & $0.83 \pm 0.11$ \\
 & 2023-10-25 04:20 &  & \checkmark & 939 & $0.71 \pm 0.02$ \\
 & 2023-11-13 05:05 &  & \checkmark & 939 & $0.88 \pm 0.08$ \\
 & 2025-06-17 08:53 &  & \checkmark & 2558 & $0.81 \pm 0.11$ \\
\addlinespace
4935140284663674368 & 2021-11-14 01:42 & \checkmark &  & 1340 & $0.56 \pm 0.06$ \\
 & 2025-07-12 09:05 &  & \checkmark & 3262 & $0.97 \pm 0.15$ \\
 & 2025-07-15 08:04 &  & \checkmark & 3262 & $0.45 \pm 0.03$ \\
 & 2025-08-12 08:26 &  & \checkmark & 3262 & $0.56 \pm 0.05$ \\
 & 2025-08-13 05:07 &  & \checkmark & 3262 & $0.84 \pm 0.09$ \\
\addlinespace
4999449100568563584 & 2021-11-12 05:55 & \checkmark &  & 740 & $0.51 \pm 0.02$ \\
 & 2023-06-06 09:49 & \checkmark &  & 1649 & $0.55 \pm 0.04$ \\
 & 2023-07-09 06:02 & \checkmark &  & 1649 & $0.75 \pm 0.13$ \\
 & 2023-09-30 06:43 &  & \checkmark & 1649 & $0.86 \pm 0.07$ \\
 & 2023-10-25 01:51 &  & \checkmark & 1649 & $0.61 \pm 0.05$ \\
 & 2023-11-13 02:50 &  & \checkmark & 1649 & $0.86 \pm 0.10$ \\
 & 2025-06-10 08:29 &  & \checkmark & 4002 & $0.55 \pm 0.08$ \\
\addlinespace
5001167366660075904 & 2021-11-14 01:22 & \checkmark &  & 810 & $0.61 \pm 0.08$ \\
\addlinespace
5006585382004810752 & 2021-11-16 02:56 & \checkmark &  & 970 & $0.63 \pm 0.04$ \\
 & 2025-07-13 09:20 &  & \checkmark & 2544 & $1.97 \pm 0.37$ \\
 & 2025-07-16 06:32 &  & \checkmark & 2544 & $0.63 \pm 0.04$ \\
 & 2025-08-08 06:21 &  & \checkmark & 2544 & $1.04 \pm 0.09$ \\
 & 2025-08-12 07:29 &  & \checkmark & 2544 & $0.50 \pm 0.06$ \\
\addlinespace
5014635662546524416 & 2021-11-14 02:07 & \checkmark &  & 970 & $0.42 \pm 0.03$ \\
 & 2025-07-18 07:08 &  & \checkmark & 2544 & $0.51 \pm 0.04$ \\
 & 2025-08-13 06:08 &  & \checkmark & 2544 & $0.45 \pm 0.08$ \\
 & 2025-08-14 06:36 &  & \checkmark & 2544 & $0.58 \pm 0.08$ \\
\addlinespace
5015490635916742912 & 2021-11-16 07:04 & \checkmark &  & 460 & $0.88 \pm 0.16$ \\
\addlinespace
5035014766969503488 & 2021-12-16 03:28 & \checkmark &  & 460 & $0.47 \pm 0.04$ \\
 & 2025-07-23 07:36 &  & \checkmark & 2746 & $0.46 \pm 0.08$ \\
 & 2025-08-13 09:04 &  & \checkmark & 2746 & $0.60 \pm 0.09$ \\
\addlinespace
5040226172911337344 & 2021-12-16 04:08 & \checkmark &  & 610 & $0.42 \pm 0.02$ \\
\addlinespace
5040747242639330432 & 2021-12-16 03:42 & \checkmark &  & 1240 & $0.46 \pm 0.03$ \\
\addlinespace
5066899264145265792 & 2021-11-14 07:57 & \checkmark &  & 350 & $0.35 \pm 0.01$ \\
 & 2023-06-09 09:46 & \checkmark &  & 779 & $0.57 \pm 0.05$ \\
 & 2023-07-22 06:42 & \checkmark &  & 779 & $0.56 \pm 0.04$ \\
 & 2023-09-30 07:19 &  & \checkmark & 779 & $0.88 \pm 0.06$ \\
 & 2023-10-25 05:05 &  & \checkmark & 779 & $0.71 \pm 0.03$ \\
 & 2023-11-15 03:52 &  & \checkmark & 779 & $0.71 \pm 0.05$ \\
 & 2025-07-25 06:28 &  & \checkmark & 2230 & $0.63 \pm 0.06$ \\
\addlinespace
5117683953885263360 & 2021-11-13 00:45 & \checkmark &  & 200 & $0.69 \pm 0.01$ \\
\addlinespace
5142271095466719744 & 2021-12-16 04:51 & \checkmark &  & 970 & $0.46 \pm 0.05$ \\
\addlinespace
5142451621532291840 & 2021-11-10 06:59 & \checkmark &  & 1440 & $1.05 \pm 0.16$ \\
 & 2021-12-16 04:24 & \checkmark &  & 1440 & $0.45 \pm 0.03$ \\
 & 2023-07-15 07:22 & \checkmark &  & 3179 & $0.73 \pm 0.05$ \\
 & 2023-07-31 08:00 & \checkmark &  & 3179 & $0.81 \pm 0.06$ \\
 & 2023-10-02 07:28 &  & \checkmark & 3179 & $0.88 \pm 0.10$ \\
 & 2023-10-24 03:51 &  & \checkmark & 3179 & $0.52 \pm 0.06$ \\
 & 2025-07-18 07:58 &  & \checkmark & 3549 & $0.55 \pm 0.07$ \\
 & 2025-07-18 09:04 &  & \checkmark & 3549 & $0.61 \pm 0.07$ \\
\addlinespace
5144335875224357888 & 2021-12-06 05:08 & \checkmark &  & 1300 & $0.69 \pm 0.08$ \\
 & 2025-08-03 08:44 &  & \checkmark & 3262 & $0.82 \pm 0.13$ \\
 & 2025-09-02 03:52 &  & \checkmark & 3262 & $0.58 \pm 0.05$ \\
\addlinespace
5295916025702362368 & 2021-10-22 08:11 & \checkmark &  & 240 & $0.70 \pm 0.04$ \\
\addlinespace
5548976735730236928 & 2021-10-22 08:03 & \checkmark &  & 240 & $0.65 \pm 0.07$ \\
 & 2025-04-07 00:27 &  & \checkmark & 1731 & $0.61 \pm 0.09$ \\
 & 2025-04-29 23:06 &  & \checkmark & 1731 & $0.33 \pm 0.04$ \\
\end{longtable*}
\normalsize

\clearpage

\section{Fabry-Perot vs On-sky observations}
\label{sec:FPbias}
Throughout the five observing periods, two distinct observing techniques were utilized, illuminating the B fiber with either (1) the Fabry-Perot source for a simultaneous calibration measurement or (2) the sky - respectively shorthanded as ``FP'' observations and ``Sky'' observations. ``FP'' observations provide a simultaneous measurement of a wavelength calibration source that can be used to measure the wavelength drift between the time of observation and the time the wavelength calibrations are taken the following morning.

\begin{figure}[h]
    \centering
    {\includegraphics[width=0.49\textwidth]{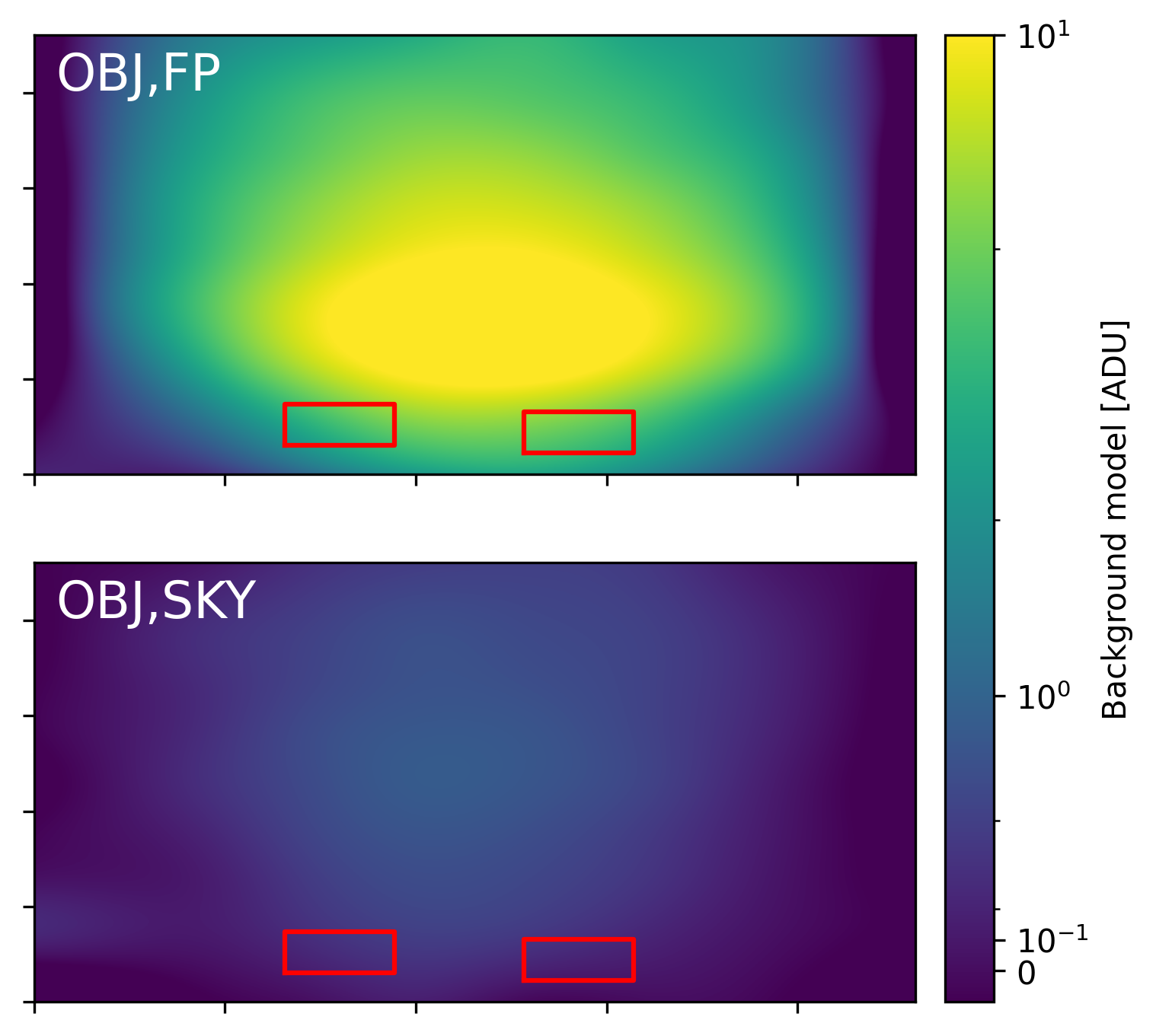}}%
    {\includegraphics[width=0.49\textwidth]{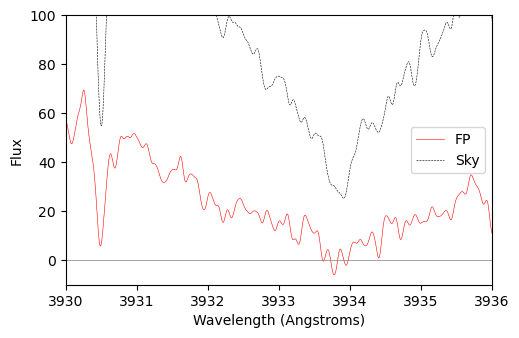}}%
    \caption{{\bf (left)} A figure representing the difference in the background levels for ``FP'' and ``SKY'' observations, modeled by the pipeline from the raw exposures. The red regions show the locations of the Ca II H/K lines. A small error in the modeling of this background is more than enough to cause the cores of the lines to go negative. For the ``Sky'' observations the background is not actually modeled from the raw exposure. Instead, the background is estimated from the sky (B) fiber. {\bf (right)} Overlay of observed flux in the Ca II K lines from our ``FP'' and ``Sky'' observations for Gaia DR3 4999449100568563584. Note that the flux is reported negative under the FP observation due to an overestimate of the detector background subtracted by the pipeline. \label{fig:background}}
\end{figure}

We initially used the ``FP'' observation approach during the first phase of the campaign. However, in analyzing the results from the first two semesters utilizing these ``FP'' observations, the $\log R'_{\rm HK}$ measurements made by the pipeline were significantly lower than expected (c.f. \citealp{2018A&A...616A.108B}), and sometimes not calculable. This was caused by an overestimate of the 2D detector background (Fig. \ref{fig:background}) which, when subtracted from the image before the extraction of the spectrum, led to an over-subtraction in the core of the Ca H \& K lines, sometimes with a negative flux. In subsequent periods, we moved to the ``Sky'' observation technique, where the B fiber was observing the blank sky. Measurements of the Fabry-Perot source were instead made in both the A and B fibers immediately after each observation, before the telescope was moved to its next observation.

To recover the $\log R'_{\rm HK}$ measurements affected in our ``FP'' observations, we used the ``Sky'' observations of the same stars to measure and correct the over-subtraction of the detector background map. For each object with both ``FP'' and ``Sky'' observations, we constructed a reference spectrum from an average of each ``Sky'' epoch shifted to the rest frame of the star. Then, for each spectrum from an ``FP'' observation shifted to the same rest frame, we determined the additive offset, multiplicative scaling factor, and linear gradient that provided the best match between the observation and the reference spectrum between 3885\AA~and 4010\AA.

\section{Radial velocity time series}
\begin{figure*}
    \includegraphics[width=0.8\textwidth]{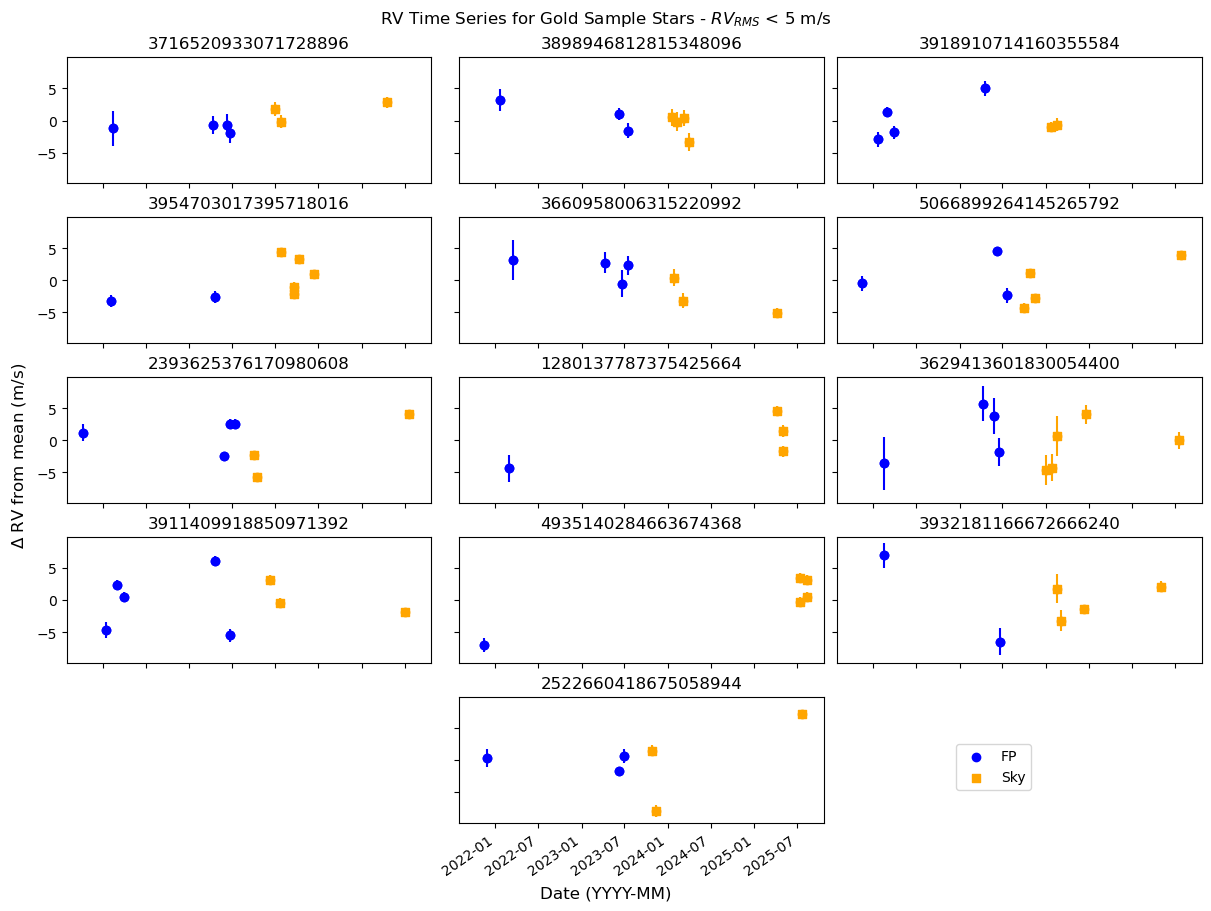}
\caption{RV epochs for sources with $RV_{\rm RMS} < 5$ m/s (above) and $5 < RV_{\rm RMS} < 10$ m/s (below)}
\label{fig:goldsampleRVtimeseries1}    
\end{figure*}

\begin{figure*}
    \includegraphics[width=0.8\textwidth]{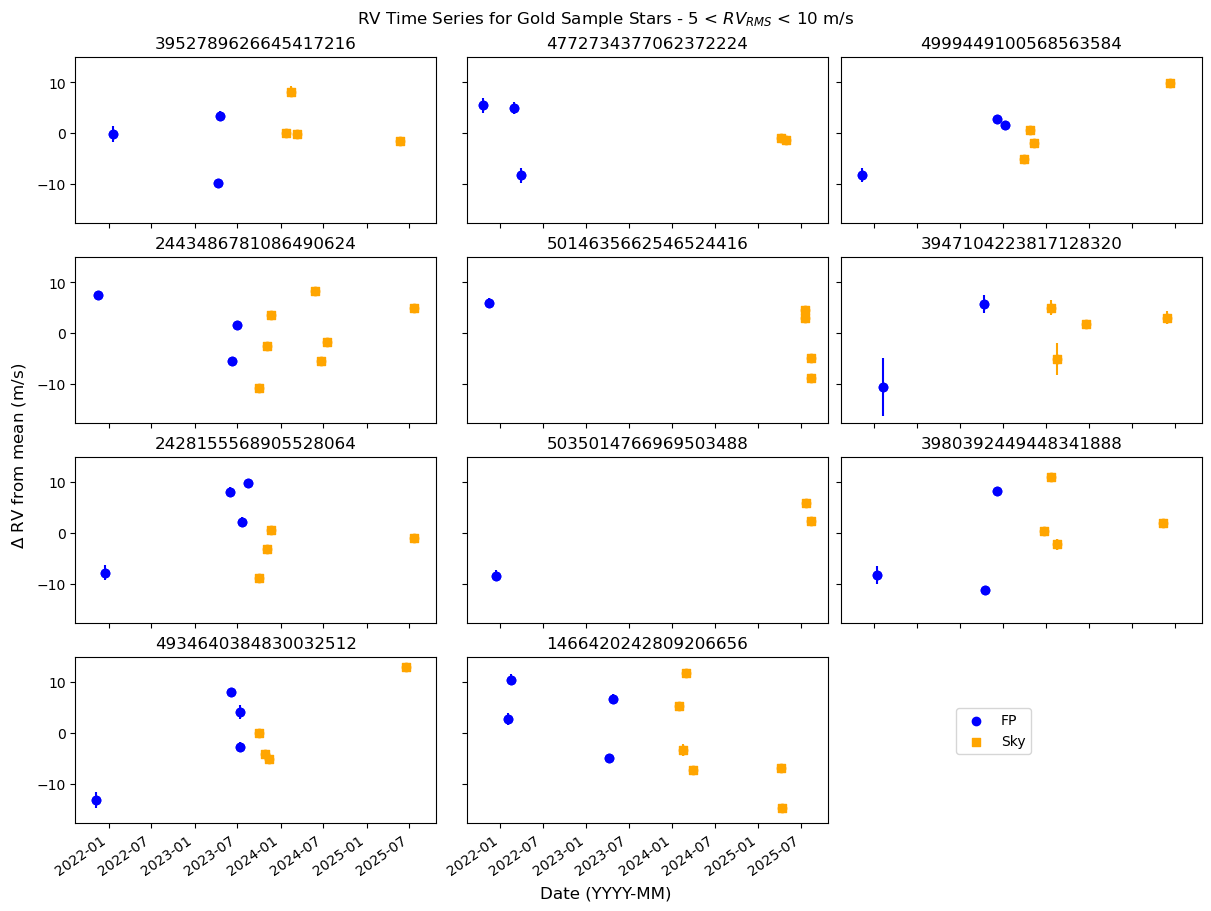}
\label{fig:goldsampleRVtimeseries2}      
\end{figure*}

\begin{figure*}
    \includegraphics[width=0.8\textwidth]{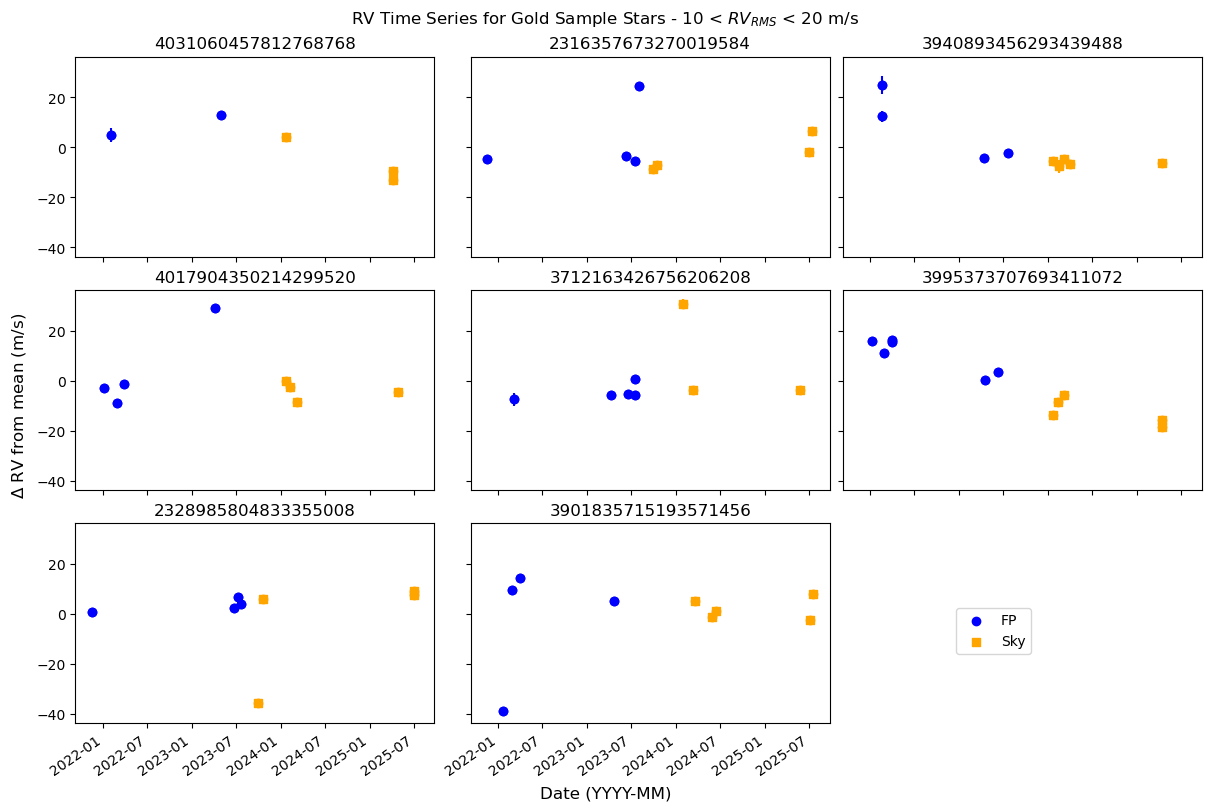}
\caption{RV epochs for sources with $ 10 < RV_{\rm RMS} < 20$ m/s (above) and $RV_{\rm RMS} > 20$ m/s (below)}
\label{fig:goldsampleRVtimeseries3}      
\end{figure*}

Figures \ref{fig:goldsampleRVtimeseries1} - \ref{fig:goldsampleRVtimeseries3} show the RV epochs over time for our current gold sample stars; the change in the RV relative to the mean is shown here.  These figures are separated into bins that have $RV_{\rm RMS}$ $<$ 5 m/s, $5<RV_{\rm RMS} < 10$ m/s, $ 10 <RV_{\rm RMS} <$ 20 m/s, and $>$ 20 m/s.  For the set of 13 stars with $RV_{\rm rms} < 5$ m/s, the mean $RV_{\rm RMS}$ = 3.48 m/s; for the 11 stars with $ 10 <RV_{\rm RMS} <$ 20 m/s, the mean $RV_{\rm RMS}$ = 6.84 m/s, and the set of 10 stars with $ 20 <RV_{\rm RMS} <$ 30 have mean $RV_{\rm RMS}$ = 12.62 m/s.

\begin{figure*}
    \includegraphics[width=0.8\textwidth]{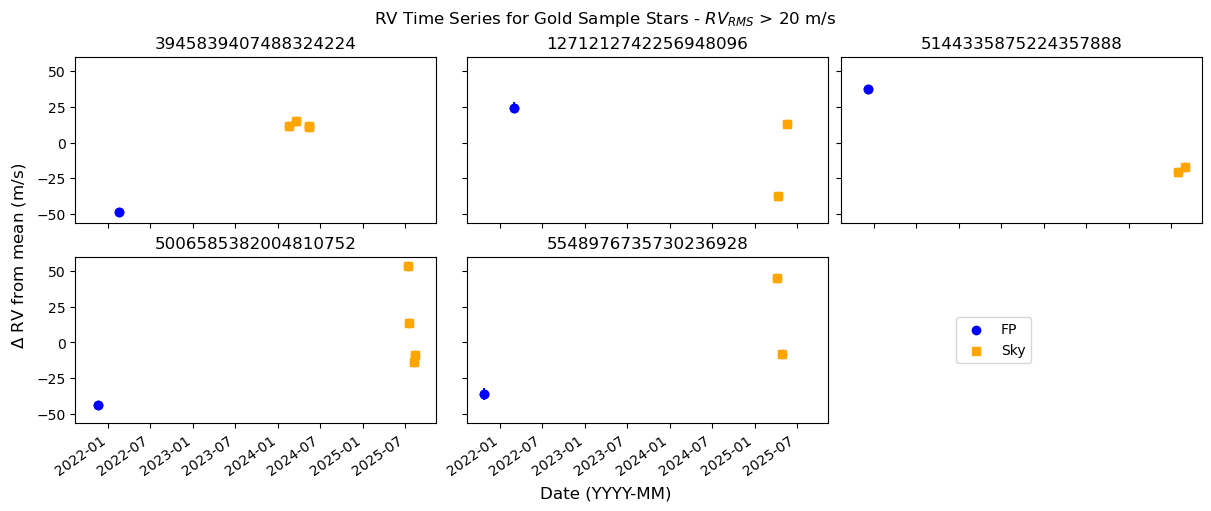}
\label{fig:goldsampleRVtimeseries4}      
\end{figure*}

\bibliography{combined}
\end{document}